\documentclass{phdthesisbo}

\usepackage[numbers]{natbib}
\usepackage{pstcol}
\usepackage{graphicx,epsf}
\newcommand{\Bibliography}{\thispagestyle{plain}%
 \bibliography{phdeg}%
 \addcontentsline{toc}{chapter}{\numberline{}Bibliography}%
 \markboth{\scshape Bibliography}{\scshape Bibliography}
}

\begin{document}

\vspace*{1.2 cm}
{\noindent \centering {\large \underline {PhD Thesis}} \par}
\vspace{5mm}
{\noindent \centering \textbf{\Huge Finite Size Effects in Integrable}\Huge \par}
\vspace{1mm}
{\noindent \centering \textbf{\Huge Quantum Field Theory:}\Huge \par}
\vspace{1mm}
{\noindent \centering \textbf{\Large the Sine-Gordon model with boundaries}\Large \par}
\vspace{15mm}
{\noindent \centering \textbf{\large Marco Bellacosa} \par}
\vspace{10mm}
{\noindent \centering {\large \emph {Department of Physics University of Bologna}} \par}
\vspace{2mm}
{\noindent \centering {\large \emph {Via Irnerio 46, I-40126 Bologna, Italy}} \par}
\vspace{2mm}
{\noindent \centering {\large \texttt{bellaco@bo.infn.it}} \par}
\thispagestyle{empty}
\newpage
\vspace*{15 cm}
\thispagestyle{empty}

\newpage

\pagenumbering{roman}

\tableofcontents

\newpage

\pagenumbering{arabic}

\chapter*{Introduction}
\addcontentsline{toc}{chapter}{\numberline{}Introduction}
\markboth{\scshape Introduction}{\scshape Introduction}

\begin{figure}
\begin{center}
\includegraphics[angle=0, width=0.7\textwidth]{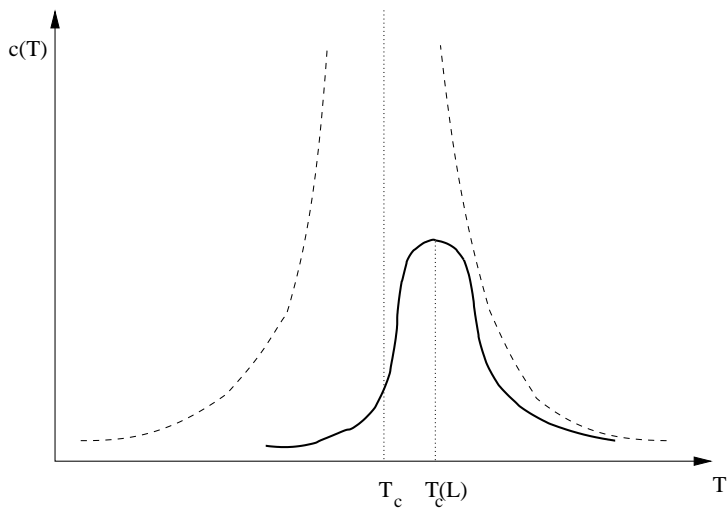}
\caption{\emph{Specific heat as a function of the temperature for a system in a finite volume \(L\)}.\label{fig:3int}}
\end{center}
\end{figure}

Finite size effects play an important role in modern Statistical Mechanics and Quantum Field Theory. In fact, they give a description of the special behaviour of observables of a statistical model like specific heat, magnetic susceptibility or correlation length due to the influences of the presence of a finite geometry. As an example, consider the specific heat \(c(T)\) of some statistical system defined in a finite domain with size \(L\). Although the specific heat of the system in an infinite volume shows a divergence \(c_{\infty}(T)\sim\left(T-T_{c}\right)^{-\alpha}\) at the critical temperature \(T_{c}\), this singularity is rounded when the system size is finite. The curves \(c(T)\) versus \(T\) show a maximum at some value of \(T\) which is shifted away from \(T_{c}\) (see Fig. (\ref{fig:3int})). One can take the temperature of the finite size maximum as an estimate \(T_{c}(L)\) of the true critical temperature \(T_{c}(\infty)=T_{c}\) and only in a finite interval around this temperature the finite size effects are relevant, out of this interval these finite size effects are negligible because the correlation length of the system is not longer comparable with the size. These phenomena are also found to hold true for many condensed matter systems. For example, many experimental mesurements have been done in physical systems as thin epitaxial layers and multi-layers of magnetic substance on non-magnetic substrates (see \cite{hen} and references therein). 

An important fact is that critical quantities, as the specific heat, have a scaling behaviour, i.e. a variation as functions of \(L\), that is \textbf{fixed} by the critical exponents of the system in the infinite size geometry. Moreover, the behaviour of the scaling functions (see later) is deeply related to the conformal field theory describing the critical points of statistical systems (see \cite{car3} and \cite{hen,priv} for a review).

\begin{figure}
\begin{center}
\includegraphics[angle=0, width=0.7\textwidth]{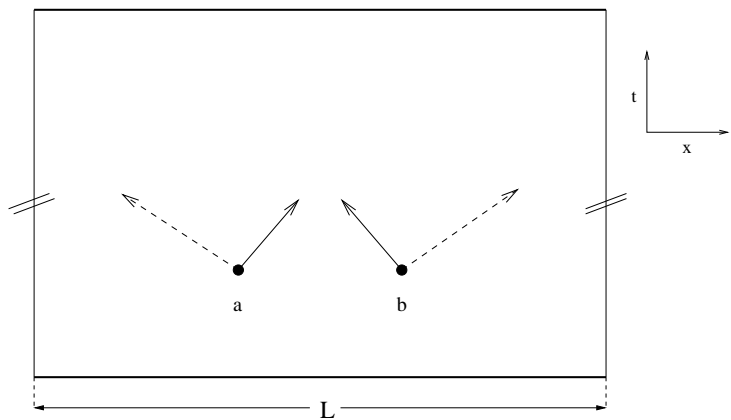}
\caption{\emph{Particles \(a\) and \(b\) interacting in two different directions in a \(1+1\) dimensional space-time with finite space size \(L\) and periodic boundary conditions}.\label{fig:1int}}
\end{center}
\end{figure}

From the point of view of quantum field theories defined in a finite geometry, interesting phenomena appear as well. Take, for instance, a \(1+1\) dimensional space-time with periodic boundary conditions in the space direction and an infinitely extended time direction, in fact, this geometry corresponds to a cylinder: there will be some Casimir effect changing, for example, the energy of a two body interaction, because two particles can interact in two possible directions along the space direction due to the periodic boundary conditions (Fig. (\ref{fig:1int})).

\begin{figure}
\begin{center}
\includegraphics[angle=0, width=0.7\textwidth]{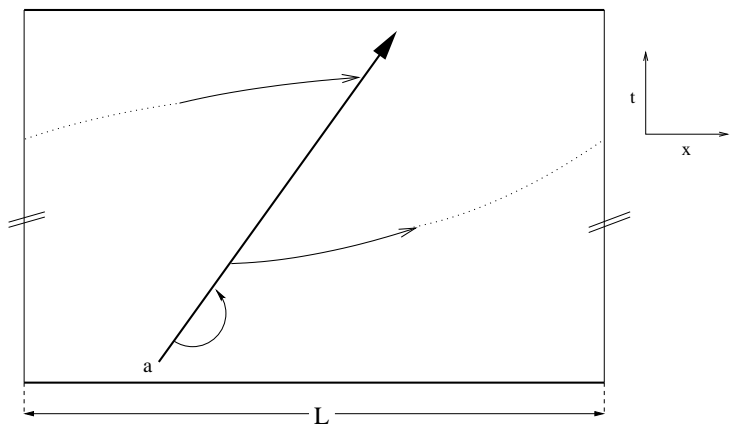}
\caption{\emph{Virtual emissions from the particle \(a\). Due to periodic boundary conditions in the space direction, they can travel along the whole space coming back to the particle from the left}.\label{fig:2int}}
\end{center}
\end{figure}

New radiative corrections to the self-energy of a propagating particle should appear as well. While in an infinite space-time only local virtual emissions are allowed, on a cylindrical geometry one can also conceive virtual emissions traveling around the whole cylinder before coming back to the bare particle. Although these virtual emissions are exponentially depressed with respect to the usual local emissions, if the cylinder circumference \(L\) (the size of space direction in Fig. (\ref{fig:2int})) is small enough, there is a mesurable probability that such virtual objects travel around the world and come back from the other side, due to the Heisenberg principle. The corrections to the self-energy can be estimated. 

Useful applications of such results show up in lattice calculations, for example many progresses have been made in some two dimensional quantum field theories sharing phenomena with QCD and other interesting four dimensional theories. In a particular class of two-dimensional quantum field theories, the so called \emph{Integrable} quantum field theory (IQFT), there are methods to exactly calculate the dependence of physical quantities on the finite size \(L\).

In principle, the behaviour of such systems in finite geometries can be known. This will be the main topic of the present thesis. Among all possible methods to compute finite size effects, we will propose the \emph{Nonlinear Integral Equation} approach to study the \emph{sine-Gordon} model in a finite volume with Dirichlet boundary conditions. This approach will be developed through a lattice regularization of the model in order to describe the ground state, i.e. the Casimir energy, and some excited states scaling functions to give predictions on the spectrum in relation to the space size \(L\). It will be shown that in the ultraviolet limit the Nonlinear Integral Equation allows to reproduce the correct conformal field theory living in the ultraviolet renormalization group flow fixed point of the model and to classify conformal states in relation to physical states of sine-Gordon. 

The thesis is organized following the scheme: in Chapter 1 some known fundamental facts about critical phenomena and conformal field theories are introduced. Particular relevance is reserved to the relations between statistical mechanical systems at criticality and conformal field theory, to conformal field theory defined in a cylindrical geometry and to conformal theories with boundaries. Chapter 2 contains a general survey of IQFTs and boundary IQFTs. The \(S\) matrix approach and the finite size scaling in relation to conformal invariant theories are introduced. Chapter 3 is devoted to introduce basic material on the sine-Gordon model with Dirichlet boundary conditions and the inhomogeneous XXZ spin chain with boundaries. A description of double row transfer matrix, its diagonalization and the derivation of Bethe Ansatz equations is also provided. Finally, in Chapter 4 the Nonlinear Integral Equation for our model is derived while the analysis of the continuum theory is proposed in Chapter 5, where the finite size effects of boundary sine-Gordon model are also discussed taking into account excited states.

\newpage
\subsection*{Acknowledgments}
I have the pleasure to express my gratitude to F. Ravanini for advice and collaboration. I remain deeply indebted to him for the scientific and human dialogue in the past few years. I thank C. Ahn for many useful discussions and for enjoyable collaboration. I would like to acknowledge E. Corrigan for kind hospitality at the Department of Mathematics at the University of York where part of results presented here has been achieved. I am also grateful to C. Dunning for discussions and for her great kindness, to D. Fioravanti and R. Soldati for advice and support during my PhD. Finally, I thank my parents and all people who always provided support and friendship. 

\begin{quote}
\raggedleft \emph{Bologna, March 2004} \hfill M.B.
\end{quote}

\chapter{Scaling and conformal field theory}
This Chapter presents a summary of critical phenomena in statistical mechanical systems, conformal field theory and their mathematical and physical relations.
       
\section{Critical phenomena}
A statistical mechanical system is said to be \emph{critical} when the correlation length \(\xi\), i.e. the distance over which the order parameters are statistically correlated, diverges: \(\xi\rightarrow \infty\). This is peculiar of continuous (second order) phase transitions, which consist in a change of macroscopic properties of the system when some parameters, for example the temperature, are varied. The special points in the phases diagram characterizing a second order phase transition are called second order critical points. In particular, a diverging correlation length makes no more relevant the length scales, i.e. every phenomenon at length scale \(\rho\le\xi\) is on average independent of the value of \(\rho\). Then the \emph{scale invariance} emerges. As a consequence, the microscopic details, such as the precise structure of the interactions, are not longer discernible. Therefore, the behaviour of the observables in a region close to second order critical points, which typically can be described by critical exponents, becomes independent of microscopic properties. The independence of quantities like critical exponents from microscopic aspects of a system is referred as \emph{universality}. Phenomenologically, the various critical behaviours of systems near their critical points can be organized in universality classes, which depend only on the space dimensionality and on underlying simmetries. One would like to be able to define a systematic classification of the possible universality classes, a problem in general not yet achieved. For two-dimensional systems, however, \emph{conformal invariance} allows a satisfying description of universality classes. Indeed, the use of conformal invariance to describe statistical mechanical systems at criticality is motivated by an argument, due to Polyakov \cite{pol}, which states that local scale invariant field theories are conformally invariant. Hence, universality classes of critical behaviours can be identified with conformal field theory (CFT), i.e. a conformally invariant quantum field theory.

\subsection{Scale invariance and scaling}
First of all let us introduce the concept of critical exponents. Given a statistical mechanical system, one can find several macroscopic "quantities" deeply describing the system, such as the temperature or the magnetic field, if any. In particular one defines the "reduced quantity", which describes the deviation of the system from critical value, e.g. for the temperature one defines the reduced temperature as

\begin{equation}
t\equiv\frac{T-T_{c}}{T_{c}}\label{eq:1}
\end{equation}
where \(T_{c}\) is the critical temperature, i.e. the temperature of the critical point. Given the basic quantity of a statistical system as the partition function

\begin{equation}
\mathcal{Z}=\sum_{\textrm{states}}\textrm{exp}\left(-\frac{1}{T}H\right)\label{eq:2}
\end{equation}
with \(H\) the Hamiltonian, averages \(\left\langle X\right\rangle\) for several thermodynamic observables can be obtained from it as functions \(X(t)\) of the reduced temperature \(t\). Away from criticality it is known that the functions \(X(t)\) follow exponential law behaviours \(X(t)\sim \left|t\right|^{y_{\pm}}\), where the critical exponents \(y_{\pm}\)'s are defined as

\begin{equation}
y_{\pm}\equiv\lim_{t\rightarrow\pm 0}\frac{\textrm{ln}X(t)}{\textrm{ln}\left|t\right|}\label{eq:3}
\end{equation}
In most cases \(y_{+}=y_{-}\), the exponents are the same for \(t>0\) and \(t<0\). The most common critical exponents of statistical mechanical systems are those associated to the specific heat \(C\), to the spontaneous magnetization \(M\) and to the susceptibility \(\chi\). They are defined as follows

\begin{equation}
C\sim\left|t\right|^{-\alpha}\;\;\;,\;\;\;M\sim\left|-t\right|^{\beta}\;\;\;,\;\;\;\chi\sim\left|t\right|^{-\gamma}\label{eq:4}
\end{equation}
We are mostly interested in the critical exponent \(\nu\) of the correlation length and the large distance behaviour of the two-point correlation function \(G(\mathbf{r})\) defined usually as

\begin{equation}
\xi\sim\left|t\right|^{-\nu}\;\;\;\;\;,\;\;\;\;\;G(\mathbf{r})\sim \mathbf{r}^{2-d-\eta}\;\;\;\textrm{at}\;\;t=0 \label{eq:5}
\end{equation}
Critical exponents describe phenomenogically the behaviour of statistical systems close to criticality. It is not simple in general to give an exaustive description of critical exponents, while in two-dimensional critical phenomena the conformal invariance of correlation functions at criticality allows one to find the exponents \textbf{exactly}. 

In fact the critical exponents can be related to each other by use of the \emph{scaling hypothesis}, which states that the free energy density near the critical point is a homogeneous function of some parameters, such as the external magnetic field \(h\) and the reduced temperature \(t\), called in this framework \emph{scaling fields}. That is there exist parameters \(a_{t}\) and \(a_{h}\) such that

\begin{equation}
f(t,h)=\lambda^{-d}f\left(\lambda^{a_{t}}t,\lambda^{a_{h}}h\right)\label{eq:6}
\end{equation}
where \(\lambda^{-d}\) is a dilatation factor describing a scale transformation of parameters \(t\) and \(h\), \(a_{t}\) and \(a_{h}\) are called renormalization group (RG) eigenvalues. From this important properties the critical behaviour of scaling fields can be derived. To be generic, one can take a scaling field \(\phi\) and the associated RG eingenvalue \(u\). Consider the scaling relation \(\Phi=\lambda^{u}\phi\) for the free energy \(f\), where the critical point corresponds to \(\phi=\Phi=0\). If \(u>0\), repeated dilatations will move the system away from criticality, given an arbitrarily small initial perturbation \(\phi\ne 0\) near the criticality. Such a scaling field is called relevant. If \(u<0\), repeated dilatations will leave the system around the critical point. In this case the scaling field is called irrelevant. Finally, if \(u=0\), the scaling field is said marginal. It follows that the behaviour of the system is completely determined by the relevant scaling fields. In particular, for the free energy the scaling hypothesis implies the homogeneity relation

\begin{equation}
f(t,h)=\left|
\frac{t}{t_{0}}\right|^{d/a_{t}}\mathcal{G}_{\pm}\left(\left(\frac{h}{h_{0}}\right)\left(\frac{t}{t_{0}}\right)^{-a_{h}/a_{t}}\right)\label{eq:7}
\end{equation}
where \(\mathcal{G}_{\pm}\) is the scaling function and the index \(\pm\) refers \(t>0\) or \(t<0\) respectively. The scaling function is universal, i.e. independent of the irrelevant scaling fields.  

Evidently, because the most relevant thermodynamic quantities of a statistical system can be obtained from the free energy density as derivatives with respect to \(t\) and \(h\), the scaling hypothesis gives a tool to calculate all critical exponents as functions of \(a_{t}\) and \(a_{h}\): 

\begin{equation}
\alpha=2-\frac{d}{a_{t}}\;\;\;,\;\;\;\beta=\frac{d-a_{h}}{a_{t}}\;\;\;,\;\;\;\gamma=\frac{2a_{h}-d}{a_{t}}\label{eq:8}
\end{equation}
For the scaling of a two-point correlation function a scale transformation imposes

\begin{equation}
G(\mathbf{r},t,h)=\lambda^{-2x}G\left(\frac{\mathbf{r}}{\lambda},\lambda^{a_{t}}t,\lambda^{a_{h}}h\right)\label{eq:9}
\end{equation}
where \(x\) is called the scaling dimension, then, eliminating the dilatation factor fixing \(\lambda^{a_{t}}t=K\) and taking for simplicity \(h=0\), one has the homogeneus relation

\begin{equation}
G(\mathbf{r},t,0)=t^{2x_{t}/a_{t}}G\left(\frac{\mathbf{r}}{(K/t)^{1/a_{t}}},K,0\right)\label{eq:10}
\end{equation}
Close to criticality, \(G(\mathbf{r})\) decays exponentially with \(\left|\mathbf{r}\right|\), so one can identify a correlation length \(\xi\), which shows the scaling

\begin{equation}
\xi\sim t^{1/a_{t}}\;\;\;\textrm{therefore}\;\;\;\nu=1/a_{t}
\end{equation}\label{eq:11}
Finally, considering the limit \(t\rightarrow 0\), one gets

\begin{equation}
\eta=2x_{t}+2-d\label{eq:12}
\end{equation}
Observe that, given the two last relations, the critical exponents \(\alpha\;,\beta\;\textrm{and}\;\gamma\) can be found by substitution in the formulae (\ref{eq:8}). Let us mention that in the particular case of \(t=h=0\), a spatial dilatation (scale transformation) \(\mathbf{r}\rightarrow \mathbf{r}^{\prime}=\mathbf{r}/\lambda\) gives the following relation for two-point correlation functions

\begin{equation}
G(\mathbf{r})=\lambda^{-2x}G(\mathbf{r}/\lambda)\label{eq:13}
\end{equation}
At least heuristically, for a field theoretic interpretation of this, it appears natural to assume that under scale transformations any field obeys the scaling behaviour 

\begin{equation}
\phi_{i}(\mathbf{r})=\lambda^{-x_{i}}\phi_{i}(\mathbf{r}/\lambda)\label{eq:14}
\end{equation}
with \(x_{i}\) scaling dimension of the field \(\phi_{i}\). The scaling hypothesis and all the procedures illustrated can be motivated and rigorously derived by the \emph{renormalization group theory}. An exhaustive presentation of the renormalization group is out of the scope of this work we just present here a brief survey.  

\section{Renormalization Group}
A quantum field theory is in general not invariant under scale transformations, so it is a very important task to determine the behaviour of a theory when scale transformations are performed: this corresponds to study the renormalization group (RG) of the theory. One way to investigate the RG is to introduce the Space of Actions of Wilson \cite{wil,ma}. It corresponds to suppose that the action of a given theory \(\mathcal{A}=\int\textrm{d}^{d}x\mathcal{L}\) depends on a set of fields and their derivatives and on the coupling constants \(\mathbf{g}=\left(g_{1},\;.\;.\;.g_{n}\right)\). The central idea is that one can see the variation of the Lagrangean density under scale transformations \(\mathbf{x}\rightarrow \mathbf{x}^{\prime}=\mathbf{x}+\mathbf{x}\textrm{d}t\) as a transformation of the couplings \(\mathbf{g}\rightarrow\mathbf{g}^{\prime}\). Scale transformations induce a motion on the \(n\)-dimensional space of coupling constants. Such a space is usually called Space of Actions. A trajectory in this space is thus a function \(\mathbf{g}(t)\) of the scale parameter \(t\). Under a scale transformation the Lagrangean density changes as \(\mathcal{L}(t+\textrm{d}t)=\mathcal{L}(t)+\partial_{\mu}J^{\mu}\) where \(J^{\mu}=x_{\nu}T^{\mu\nu}\) is the conserved current and \(T^{\mu\nu}\) is the energy-momentum tensor. If the system is scale invariant the quantity \(J^{\mu}\) is conserved and then \(T^{\mu}_{\mu}\equiv\Theta=0\), of course this situation does not appear for a general \(\mathbf{g}(t)\). Let us introduce the \(\beta\)-function

\begin{equation}
\beta_{i}(\mathbf{g})=\frac{\textrm{d}g_{i}}{\textrm{d}t}\label{eq:15}
\end{equation}
which describes the variation of coupling constants under a scaling of the parameter \(t\). Given the \(\beta\)-function for a certain initial value of couplings \(\mathbf{g}_{0}\), as \(t\) increases there are different behaviours

\begin{itemize}
\item if \(\beta\left(\mathbf{g}_{0}\right)>0\), then \(\mathbf{g}(t)\) is increasing in a neighbourhood of \(\mathbf{g}_{0}\);

\item if \(\beta\left(\mathbf{g}_{0}\right)<0\), then \(\mathbf{g}(t)\) is decreasing in a neighbourhood of \(\mathbf{g}_{0}\).
\end{itemize}
There exist special points \(\mathbf{g}^{*}\) called \emph{fixed points} where \(\beta\left(\mathbf{g}^{*}\right)=0\) and indipendently of the exact value of \(\mathbf{g}_{0}\) in a given region, the asymptotic value of \(\mathbf{g}(t)\) for \(t\rightarrow \pm\infty\) depends on the values of \(\mathbf{g}\) closest to \(\mathbf{g}^{*}\). In particular, one can distinguish two types of fixed points: infrared (IR) fixed points, if they are reached from \(\mathbf{g}_{0}\) for \(t\rightarrow +\infty\); ultraviolet (UV) fixed points, if they are reached for \(t\rightarrow -\infty\). Usually the trajectories \(\mathbf{g}(t)\) are referred as \emph{Renormalization Group flows}. The control of \(\beta\)-functions allows one to reconstruct the behaviour of the theory around the starting point \(\mathbf{g}_{0}\) and hence a description of the tangent space to the space of actions. In the tangent space the variation of the Lagrangean can be described by combinations of fields

\begin{equation}
\Phi_{i}=\frac{\partial \mathcal{L}}{\partial g_{i}}\label{eq:16}
\end{equation}
In particular one has that the trace of the energy-momentum tensor is a field living on the tangent space

\begin{equation}
T^{\mu}_{\mu}(\mathbf{x})=\Theta(\mathbf{x})=\sum_{i}\frac{\partial \mathcal{L}}{\partial g_{i}}\frac{\textrm{d}g_{i}}{\textrm{d}t}=\sum_{i}\beta_{i}(\mathbf{g})\Phi_{i}(\mathbf{g})\label{eq:17}
\end{equation}
Therefore \(\Theta(\mathbf{x})=0\) iff \(\beta_{i}(\mathbf{g})=0\;\forall\; i\in\mathbb{N}\). This means that the fixed points of a quantum field theory are scale invariant. This is a very important result.

In the renormalization theory it is possible to give a description  of the variation of \(N\) point correlation functions along the RG flow as well. What one gets is the so-called Callan-Symanzyk equation. We do not illustrate the general derivation (for a nice procedure see \cite{zam1}) and focus on the two-dimensional case. Given a set of \(N\) fields \(\mathcal{A}_{i}(\mathbf{x})\) and their \(N\) point corralation function \(\left\langle X\right\rangle=\left\langle\mathcal{A}_{1}(\mathbf{x}).\;.\;.\;.\mathcal{A}_{N}(\mathbf{x})\right\rangle\), any field changes under a scale transformation as

\begin{equation}
\delta\mathcal{A}_{j}(\mathbf{x})=\textrm{d}t\left(x^{\mu}_{j}\frac{\partial_{\mu}}{\partial x_{j}^{\mu}}+D_{j}\right)\mathcal{A}_{j}(\mathbf{x})\label{eq:18}
\end{equation}
where \(D_{j}\) is the classical dimension. One can demonstrate that for a two-dimensional quantum field theory the correlation function obeys the following equation along the RG flow

\begin{equation}
\left[\sum_{j}\left(x_{j}^{\mu}\frac{\partial_{\mu}}{\partial x_{j}^{\mu}}+\Gamma_{j}(\mathbf{g})\right)-\sum_{i}\beta_{i}(\mathbf{g})\frac{\partial}{\partial g_{i}}\right]\left\langle X\right\rangle=0\label{eq:19}
\end{equation}
where \(\Gamma_{j}(\mathbf{g})\mathcal{A}_{j}(\mathbf{x})=\left(D_{j}+\sum_{i}\beta_{i}(\mathbf{g})\frac{\partial}{\partial g_{i}}\right)\mathcal{A}_{j}(\mathbf{x})\) is the anomalous dimension of the field  \(\mathcal{A}_{j}\). Moreover, the dimension of the field \(\Phi_{i}(\mathbf{x})\) is

\begin{equation}
\Gamma(\mathbf{g})\Phi_{i}(\mathbf{x})=\sum_{j}\left(2\delta_{i}^{j}-\frac{\partial\beta_{j}(\mathbf{g})}{\partial g_{i}}\right)\Phi_{j}(\mathbf{x})\label{eq:20}
\end{equation}
and, in particular, \(\Gamma(\mathbf{g})\Theta(\mathbf{x})=2\Theta(\mathbf{x})\). Therefore, considering also the Lorentz invariance, it follows that all the components of the energy-momentum tensor have the same anomalous dimension: \(2\).

\subsection{Phenomenological renormalization}
Consider now a statistical mechanical system, say a spin chain, defined in a finite (one dimensional) lattice with dimension \(L\) in which, for simplicity, \(h=0\). According to finite size scaling, for any change of scale, the correlation length obeys the homogeneous relation

\begin{equation}
\xi^{-1}(t,0)=L^{-1}\mathcal{G}\left(L^{1/\nu}t,0\right)\label{eq:21}
\end{equation}
Given two lattice sizes \(L\) and \(L^{\prime}\) and temperatures \(t\) and \(t^{\prime}\) such that it is satisfied the condition

\begin{equation}
\frac{\xi\left(t,L\right)}{L}=\frac{\xi\left(t^{\prime},L^{\prime}\right)}{L^{\prime}}\label{eq:22}
\end{equation}
one can describe the relation between \(t\) and \(t^{\prime}\) as

\begin{equation}
t^{\prime}=t\left(\frac{L}{L^{\prime}}\right)^{1/\nu}\label{eq:23}
\end{equation}
These relations can be reinterpreted \cite{night} as a RG mapping from the two temperatures

\begin{equation}
t\rightarrow t^{\prime}=\mathcal{R}_{L,b}(t)\label{eq:24}
\end{equation}
where \(b=L/L^{\prime}\) is the scale factor. This procedure can be considered exact only in the limit \(L,L^{\prime}\rightarrow \infty\), but the definition of the RG mapping is considered to make sense also for \(L,L^{\prime}\) finite and large enough. If it is found a fixed point \(t^{*}\left(L,L^{\prime}\right)\) of the RG flow, i.e.

\begin{equation}
\frac{\xi\left(t^{*},L\right)}{L}=\frac{\xi\left(t^{*},L^{\prime}\right)}{L^{\prime}}\label{eq:25}
\end{equation}
one expects that \(t^{*}\left(L,L^{\prime}\right)\rightarrow 0\) as \(L,L^{\prime}\rightarrow \infty\). The solution can be derived in several ways, for example graphically. The important thing is that the RG approach (as it has been done in the previous discussion) allows one to relate critical (scale invariant) points to fixed points and then the change of physical quantities as scales are continuously varied. From another point of view, one can use the fact that the function \(\mathcal{G}(t,h)\) is universal and obtain that

\begin{equation}
\mathcal{G}(0,0)=\frac{L}{\xi\left(t,L\right)}\label{eq:26}
\end{equation}
Therefore if the value \(\mathcal{G}(0,0)\) is known, one can estimate \(t^{*}\) and  give an exact description of critical exponents.  

\section{Conformal invariance}
It has been said that a CFT is a quantum field theory endowed with covariance properties under conformal transformations (see \cite{bpz,difra}), i.e. special coordinate transformations in \(d\) dimensions which keep the metric invariant up to a scale factor

\[
g^{\prime}_{\mu\nu}\left(\mathbf{r}^{\prime}\right)=\Omega(\mathbf{r})g_{\mu\nu}\left(\mathbf{r}\right) \]
The set of conformal transformations forms a group: the \emph{conformal group}. In \(d>2\) this group is finite dimensional and it is generated by global rotations \(\mathbf{r}\rightarrow \mathbf{r}^{\prime}=\Lambda\mathbf{r}\), global translations \(\mathbf{r}\rightarrow \mathbf{r}^{\prime}=\mathbf{r}+\mathbf{a}\), global dilatations \(\mathbf{r}\rightarrow \mathbf{r}^{\prime}=b\mathbf{r}\) and the so called special conformal transformations 

\[
\mathbf{r}\rightarrow \mathbf{r}^{\prime}=\frac{\mathbf{r}+\mathbf{a}r^{2}}{1+2\mathbf{a}\mathbf{r}+a^{2}r^{2}} \]
We restrict to the two dimensional case and in particular to the Euclidean plane where the conformal transformations are realized by any analytic change of coordinates \(z\rightarrow \xi(z),\;\bar{z}\rightarrow \bar{\xi}(\bar{z})\). In fact, one can demonstrate \cite{bpz} that the contributions of the complex variables \(z\) and \(\bar{z}\) decouple and that they can be regarded as independent variables. One can identify a conformal theory as a quantum field theory described by the correlation functions of a set of local scaling operators \(\mathcal{A}(z,\bar{z})\) with following properties

\begin{itemize}
\item If \(\mathcal{A}(z,\bar{z})\) is local, all derivatives of \(\mathcal{A}\) are local. The set of operators \(\left\{\mathcal{A}(z,\bar{z})\right\}\) is in general infinite;

\item there exists a subset \(\left\{\Phi(z,\bar{z})\right\}\subset\left\{\mathcal{A}(z,\bar{z})\right\}\) of local operators, called quasi-primary operators, which transform covariantly under projective conformal transformations \(z\rightarrow w(z)=(az+b)/(cz+d)\) as

\[
\Phi(z,\bar{z})\rightarrow \left(\frac{\partial w}{\partial z}\right)^{-\Delta}\left(\frac{\partial \bar{w}}{\partial \bar{z}}\right)^{-\bar{\Delta}}\Phi(z,\bar{z})\;,\;\;\;\Delta,\;\bar{\Delta}\in\mathbb{R}\; ; \]

\item any local operator can be written as a linear combination of quasi-primary operators and their derivatives;

\item the vacuum of the theory is invariant under projective conformal transformations.
\end{itemize}
Under any conformal change of variables, the fields of a CFT obey some general relations. In particular \emph{primary fields}, which are a subset of quasi-primary fields, transform according to

\begin{equation}
\phi(w,\bar{w})=\left(\frac{\partial w}{\partial z}\right)^{-\Delta} \left(\frac{\partial \bar{w}}{\partial \bar{z}}\right)^{-\bar{\Delta}}\phi(z,\bar{z})\label{eq:27}
\end{equation}
where the real numbers \(\left(\Delta,\bar{\Delta}\right)\) are called \emph{conformal dimensions} of the field \(\phi\). This relation implies that it is very simple, in principle, to calculate exactly the correlation functions among primary fields using the fact that

\begin{align}
&\left\langle\phi_{1}\left(w_{1},\bar{w}_{1}\right)\;.\;.\;.\;\phi_{n}\left(w_{n},\bar{w}_{n}\right)\right\rangle= \nonumber \\
&=\prod_{i=1}^{n}\left(\frac{\partial w}{\partial z}\right)^{-\Delta_{i}}\left(\frac{\partial\bar{w}}{\partial\bar{z}}\right)^{-\bar{\Delta}_{i}}\left\langle\phi_{1}\left(z_{1},\bar{z}_{1}\right)\;.\;.\;.\;\phi_{n}\left(z_{n},\bar{z}_{n}\right)\right\rangle\label{eq:28}
\end{align}
For instance, the two point correlation function explicitly reads

\begin{equation}
\left\langle\phi_{1}\left(z_{1},\bar{z}_{1}\right)\phi_{2}\left(z_{2},\bar{z}_{2}\right)\right\rangle=
\delta_{12}z_{12}^{2\Delta}\bar{z}_{12}^{2\bar{\Delta}}\label{eq:29}
\end{equation}
where \(z_{ij}=z_{i}-z_{j}\), \(\Delta_{1}=\Delta_{2}\equiv\Delta\), \(\bar{\Delta}_{1}=\bar{\Delta}_{2}\equiv\bar{\Delta}\). The correlation vanishes if the conformal dimensions of the two fields are different. It is evident that the expression of the two point function allows one to identify the critical exponent \(\eta\) given in Eq. (\ref{eq:12}) as

\begin{equation}
\eta=2\Delta+2-d=2\Delta\label{eq:30}
\end{equation}
In this scheme, i.e. studying the conformal properties of a massless field theory living in a scale invariant second order critical point, one finds exactly the critical exponents of the system. Similar analysis can be done for three and four point functions.  

At this point it is useful to introduce the energy-momentum tensor \(T_{\mu\nu}\). It is defined classically considering the variation of the action under coordinates transformations, in quantum theory one makes use of Ward identities. It has some very important properties: the symmetric property \(T_{12}=T_{21}\) as a result of rotational invariance, and the tracelessness \(T_{11}+T_{22}=0\) as a result of scale invariance. In addition the energy-momentum tensor is always conserved, i.e. \(\partial_{\mu}T_{\mu\nu}=0 \). In general one introduces complex components

\begin{equation} \left\{\begin{array}{c}
T_{zz}=\frac{1}{4}\left(T_{11}-T_{22}-2iT_{12}\right)  \\
T_{\bar{z}\bar{z}}=\frac{1}{4}\left(T_{11}-T_{22}+2iT_{12}\right)\end{array}\right.\label{eq:31}
\end{equation}
while \(T_{z\bar{z}}=T_{\bar{z}z}=\frac{1}{4}\left(T_{11}+T_{22}\right)=0\) because of tracelessness. The \(zz\) component of the energy-momentum tensor is analytic as a consequence of tracelessness and conservation and denoted by \(T(z)\). The \(\bar{z}\bar{z}\) component \(\bar{T}(\bar{z})\) is antianalytic. \(T(z)\) is the generator of infinitesimal conformal transformations \(z\rightarrow w=z+\epsilon(z)\) and likewise \(\bar{T}(\bar{z})\) for \(\bar{z}\rightarrow \bar{w}=\bar{z}+\bar{\epsilon}(\bar{z})\), in the sense that under this change of variables a correlation function changes as

\begin{align}
&\delta\left\langle\phi_{1}\left(z_{1},\bar{z}_{1}\right).\;.\;.\;\phi_{n}\left(z_{n},\bar{z}_{n}\right)\right\rangle=\nonumber \\
&=\frac{1}{2\pi i}\oint_{\mathcal{C}}\textrm{d}x\epsilon(w)T(w)\left\langle \phi_{1}\left(z_{1},\bar{z}_{1}\right).\;.\;.\;\phi_{n}\left(z_{n},\bar{z}_{n}\right)\right\rangle+\textrm{c.c.} \nonumber
\end{align}
with an integration contour \(\mathcal{C}\) encircling all points \(z_{1}.\;.\;.\;z_{n}\). The change of \(T(z)\) itself has the form

\begin{equation}
T(w)=\left(\frac{\partial w}{\partial z}\right)^{-2}\left[T(z)-\frac{c}{12}\left\{w,z\right\}\right]\label{eq:32}
\end{equation}
where \(\left\{w,z\right\}\) denotes the schwarzian derivative

\[
\left\{w,z\right\}\equiv\frac{\frac{\partial^{3} w}{\partial z^{3}}}{\frac{\partial w}{\partial z}}-\frac{3}{2}\left(\frac{\frac{\partial^{2} w}{\partial z^{2}}}{\frac{\partial w}{\partial z}}\right)^{2} \]
with the constant term \(c\) called \emph{central charge}. In simple words the energy-momentum tensor \(T(z)\) transforms under analytical changes of coordinates as a primary field with conformal dimensions \((2,0)\) up to the schwarzian anomaly. Assuming that correlation functions are well defined in the complex plane with a finite number of possible singularities, there exist short distance expansions of products of fields usually called operator product expansion (OPE)

\begin{equation}
T(w)\phi(z,\bar{z})=\frac{\Delta\phi(z,\bar{z})}{(w-z)^{2}}+\frac{\partial \phi(z,\bar{z})}{w-z}+\textrm{regular terms}\label{eq:33}
\end{equation}

\begin{equation}
T(w)T(z)=\frac{c/2}{(w-z)^{4}}+\frac{2T(z)}{(w-z)^{2}}+\frac{\partial T(z)}{w-z}+\textrm{regular terms}\label{eq:34}
\end{equation}
that describe the singular behaviour of correlation functions \(\left\langle T(w)\phi(z,\bar{z}).\;.\;.\right\rangle\) and \(\left\langle T(w)T(z).\;.\;.\right\rangle\) as \(z\rightarrow w\).  

\subsection{Virasoro Algebra}
\begin{figure}
\begin{center}
\includegraphics[angle=0, width=0.7\textwidth]{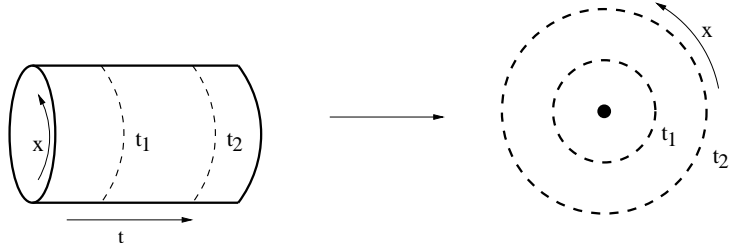}
\caption{\emph{Mapping from an infinite space-time strip to the complex plane}.\label{fig:1}}
\end{center}
\end{figure}

The picture given up to now is the one dealing with correlation functions, but, as usual in quantum field theory, it is possible also to deal with an operator formalism that describes the system by states, i.e. as vectors in a Hilbert space. In CFT it is useful to work with radial quantization in the complex plane, that is surfaces of equal time are circles centered at the origin and the Hamiltonian is the dilatation operator. It is denoted by the mapping \(z=e^{2\pi(t-ix)/L}\), \(\bar{z}=e^{2\pi(t+ix)/L}\) with \(t\) euclidean time (see Fig. (\ref{fig:1})).

On any circle there is a description of the system in terms of a Hilbert space of states on which field operators act.

At this point one can expand the energy-momentum tensor (remember that it is a conserved current) on its Laurent modes

\begin{equation}
T(z)=\sum_{n=-\infty}^{+\infty} z^{-n-2}L_{n}\label{eq:35}
\end{equation}
and, given a field \(\mathcal{A}(z,\bar{z})\), putting this expansion into the OPE provides a definition of what \(L_{n}\mathcal{A}(z)\) is. For instance \(L_{0}\mathcal{A}(z)=\Delta\mathcal{A}(z)\), \(L_{-1}\mathcal{A}(z)=\partial\mathcal{A}(z)\), etc. In general one does not expect fields with negative dimensions, then, for every field, \(L_{n}\mathcal{A}(z)\) must vanish for \(n\) positive large enough. For example, the primary fields have the highest singularity behaving as 
\(1/z^{2}\), they satisfy \(L_{n}\mathcal{A}(z)=0,\;n>0\). Intuitively the \(L_{n}\) are operators acting on the space of fields of the theory, then it is tempting to ask oneself what the algebra of these operators is. Commuatators among operators \(L_{n}\) can be calculated considering the products \(L_{n}L_{m}\mathcal{A}\) and \(L_{m}L_{n}\mathcal{A}\). Making use of the OPE among energy-momentum tensor and fields and after integrations over closed contours via the Cauchy theorem, one has

\begin{equation}
\left[L_{n},L_{m}\right]=(n-m)L_{n+m}+\frac{c}{12}n\left(n^{2}-1\right)\delta_{n+m,0}\label{eq:36}
\end{equation}
This is the celebrated \emph{Virasoro Algebra} that is an infinite dimensional Lie algebra in which the central charge plays the role of the coefficient of the central extension. Observe that there exist similar commutation rules for the antiholomorphic part of the theory and, in particular, any commutator between operators \(L_{n}\) and \(\bar{L}_{m}\) vanishes because the holomorphic and antiholomorphic parts of the fields are decoupled.

\subsection{Representation theory of the Virasoro algebra}
A very important use of the Virasoro algebra is to provide one with a natural structure to organize all the states, and then all the fields, of a given CFT. The representations which one shall consider are the highest weight representations or \emph{Verma} modules. They are characterized by a real number \(\Delta\). A Verma module \(\mathcal{V}_{c}(\Delta)\) is generated from a highest weight vector of a CFT with central charge \(c\) denoted \(\left|\Delta\right\rangle\) satisfying

\begin{equation}
L_{0}\left|\Delta\right\rangle=\Delta\left|\Delta\right\rangle\;\;\;\;,\;\;\;\;L_{n}\left|\Delta\right\rangle=0,\;\forall\; n\in \mathbb{N}\label{eq:37}
\end{equation}
by the action of the operators \(L_{-n},\; n>0\). Among these eigenstates there is the vacuum \(\left|0\right\rangle\) and

\begin{equation}
\mathcal{V}_{c}(\Delta)=\textrm{Span}\left\{L_{-k_{1}},L_{-k_{2}},.\;.\;.\;L_{-k_{r}}\right\},\;\;\;\;1\le k_{1}\le k_{2}\le .\;.\;.\le k_{r}\label{eq:38}
\end{equation}
The whole set of Hilbert states is thus organized into products of representations of the holomorphic and antiholomorphic Virasoro algebras, for which primary fields are highest weights. In compact form this is usually expressed as

\begin{equation}
\mathcal{H}=\bigoplus_{\Delta,\bar{\Delta}} \mathcal{N}_{\Delta,\bar{\Delta}}\mathcal{V}_{c}(\Delta)\otimes\bar{\mathcal{V}}_{c}(\bar{\Delta})\label{eq:39}
\end{equation}
Such a representation has the important property of being graded for the action of the Virasoro generator \(L_{0}\). This means that the spectrum of \(L_{0}\) in \(\mathcal{V}_{c}(\Delta)\) is \(\left\{\Delta,\Delta+1, \Delta+2,.\;.\;.\;\right\}\). The numbers \(\mathcal{N}_{\Delta,\bar{\Delta}}\) count the multiplicity of each representation in \(\mathcal{H}\), this implies they must always be non negative integers and \(\mathcal{N}_{\Delta,\bar{\Delta}}=0\) if a certain representation \(\mathcal{V}_{c}(\Delta)\otimes\bar{\mathcal{V}}_{c}(\bar{\Delta})\) does not appear, but they are not completely fixed by conformal invariance. Constraints on them come from other physical requirements such as locality \cite{dot1,dot2} or modular invariace \cite{car3}. The situation is then quite similar to the case of angular momentum in ordinary quantum mechanics where the space of states can be organized in terms of representations of the angular momentum algebra, of course here we have an algebra with an infinite number of generators. Note that the operators \(L_{0},\;L_{1},\;L_{-1}\) form a subalgebra of the Virasoro algebra and, in particular, \(L_{0}+\bar{L}_{0}\) is the generator of dilatations: this gives a representation of the Hamiltonian operator in term of Virasoro generators as \(H\sim L_{0}+\bar{L}_{0}\). 

The correspondence between the primary fields and heighest weight vectors is given by

\begin{equation}
\left|\Delta_{i}\right\rangle=\phi_{\Delta_{i}}(0)\left|0\right\rangle\label{eq:40}
\end{equation}
All other fields of the theory associated with states \(L_{-n}\left|\Delta\right\rangle\) can be obtained from primary fields as \emph{descendants}, in the sense that \(\phi^{-n}(z)=\left(L_{-n}\phi\right)(z)\) with conformal dimensions \(\Delta+n\). They form, together with primary fields, the so-called \emph{conformal families} \(\left[\phi_{\Delta_{i}}\right]\). Any conformal family forms a representation of the Virasoro algebra, because under conformal transformations each member of the family is mapped into a representative of the same family.

Given a module \(\mathcal{V}_{c}(\Delta)\) it could be interesting to ask oneself whether this module is unitary, i.e. characterized by the absence of negative-norm states and whether it is reducible or not, in the sense that it contains a singular vector (usually called \emph{null vector}), i.e. a vector satisfying the axioms (\ref{eq:37}) of a highest weight vector. If so, from this vector a highest weight module descends, which is a submodule of \(\mathcal{V}_{c}(\Delta)\) too. The Verma module can contain several such submodules with non trivial intersections. One constructs the irreducible representation by quotienting out the submodule(s) of \(\mathcal{V}_{c}(\Delta)\). Of course, one has to define systematically the ranges of \(c\) and \(\Delta\) where unitary irreducible representations of the Virasoro algebra are possible. It has been shown in \cite{fri} that unitarity is only possible in the following two cases:

\begin{itemize}
\item for \(c\ge 1\): all representations are unitary and unitarity only implies \(\Delta\ge 0\). However, the number of primary states is infinite.
\item for \(0\le c<1\): the following set of \(\mathcal{V}_{c}(\Delta)\) is unitary 
\end{itemize}
\begin{equation}
c=1-\frac{6}{x(x+1)}\;\;\;\;,\;\;\;\;\Delta=\frac{(r(x+1)-sx)^{2}-1}{4x(x+1)}\label{eq:41}
\end{equation}
with \(x\in \mathbb{Q}\) and \(r,s\in \mathbb{N}\) s.t. \(1\le r\le s<x\). CFTs with central charge and conformal dimensions in the set (\ref{eq:41}) have a finite number of primary states and consequentely a finite number of conformal families. They are called \emph{minimal models}. The structure of the Verma modules defined by primary fields \(\phi_{(r,s)}\) is encoded in the corresponding Virasoro characters in terms of which the partition function is exactly known

\begin{equation}
\chi_{r,s}(q)=q^{-c/24}\textrm{Tr}\;q^{L_{0}}=q^{-c/24+\Delta_{r,s}}\sum_{n=0}d_{r,s}(n)q^{n}\label{eq:42}
\end{equation}
where \(d_{r,s}(n)\) is the number of linearly independent states of the representation \(\left\{\phi_{(r,s)}\right\}\) at level \(n\). The meaning of the variable \(q\) will be clear in the next sections. 

\subsection{Correlation functions}

All fields \(\mathcal{A}(z,\bar{z})\) of a CFT obey a closed algebra under OPE

\begin{equation}
\mathcal{A}_{i}(z,\bar{z})\mathcal{A}_{j}(0,0)=\sum_{k}B_{ij}^{k}\mathcal{A}_{k}(0,0)\label{eq:43}
\end{equation}
where the indices \(i,j,k\) run over all fields of the theory. Conformal invariance reduces the problem to the computation of the OPE algebra among primary fields, indeed correlators among descendant fields can be reduced to correlators among primaries because of Eq. (\ref{eq:33}). The OPE algebra among primary fields reads

\begin{equation}
\phi_{i}(z,\bar{z})\phi_{j}(0,0)=\sum_{k}C_{ij}^{k}z^{\Delta_{k}-\Delta_{i}-\Delta_{j}}\bar{z}^{\bar{\Delta}_{k}-\bar{\Delta}_{i}-\bar{\Delta}_{j}}
\left[\phi_{k}(0,0)\right]
\end{equation}
where \(i,j,k\) count only primaries and \(\left[\phi_{k}(0,0)\right]\) denotes, as usual, the conformal families with ancestor the primary \(\phi_{k}\). If the constants \(C_{ij}^{k}\) are known, it is possible to reduce all the correlators to the two and three point functions. The two point function has been written in Eq. (\ref{eq:29}) and for the three point function one has, making use of the projective invariance only,

\begin{equation}
\left\langle\phi_{1}\left(z_{1},\bar{z}_{1}\right)\phi_{2}\left(z_{2},\bar{z}_{2}\right)\phi_{3}\left(z_{3},\bar{z}_{3}\right)\right\rangle=C_{12}^{3}z_{12}^{\gamma_{12}}z_{13}^{\gamma_{13}}z_{23}^{\gamma_{23}}\bar{z}_{12}^{\bar{\gamma}_{12}}\bar{z}_{13}^{\bar{\gamma}_{13}}\bar{z}_{23}^{\bar{\gamma}_{23}}\label{eq:45}
\end{equation}
where \(\gamma_{ij}=\Delta_{k}-\Delta_{i}-\Delta_{j}\) with \(i\ne j\ne k\) and \(i,j,k=1,2,3\). Therefore, almost in principle, all correlation functions are known. If the \(C_{ij}^{k}\)'s are unknown, constraints come from the requirement of associativity of the OPE among fields, which corresponds to impose duality properties to the four point functions.

\section{CFT on a cylinder}
It has been observed that in two dimensions the conformal transformations are represented by analytical transformations, so they are in general infinite in number. Among all possible anlytical transformations there is one which will be very important in next Chapters. It is the \emph{strip geometry} defined by 

\begin{equation}
z\rightarrow w=\frac{L}{2\pi}\ln (z)\label{eq:46}
\end{equation}
which maps all the plane onto a strip of width \(L\). The negative part of the real axis corresponds to both edges of the strip, which has therefore the topology of a cylinder (periodic boundary conditions in the space direction). Under this particular transformation Eq. (\ref{eq:32}) can be rewritten as

\begin{equation}
T_{cyl}(w)=\left(\frac{2\pi}{L}\right)^{2}\left(T(z)z^{2}-\frac{c}{24}\right)\label{eq:47}
\end{equation}
As a consequence, the expectation value of \(T\) in the cylinder is always different from zero; for instance, if \(\left\langle T(z)\right\rangle=0\),

\begin{equation}
\left\langle T_{cyl}(w)\right\rangle_{cyl}=-\frac{\pi^{2}c}{6L}\label{eq:48}
\end{equation}
The physical meaning of the apparence of the central charge \(c\), also known as conformal anomaly, is the way a specific system reacts to macroscopic length scales introduced, e.g. boundary conditions. It gives a direct mesure of the variation of the energy, or free energy, due to finite size effects. Due to the correspondence between the free energy of two-dimensional systems and the ground state energy of one-dimensional quantum systems, one can realize the analogy with Casimir effect in quantum electrodynamics (there \(\hslash\) plays the role of \(c\)). In the scheme of a CFT on a cylinder, the Hamiltonian reads

\begin{equation}
H=\frac{2\pi}{L}\left(L_{0}+\bar{L}_{0}-\frac{c}{12}\right)\label{eq:49}
\end{equation}
Another quantity of interest is the two point function of a primary field \(\phi\left(z,\bar{z}\right)\) with conformal dimensions \(\left(\Delta,\bar{\Delta}\right)\). In order to write it down on the cylinder one needs to use the covariance relation (\ref{eq:27}) of primary fields with the mapping (\ref{eq:46}) and gets

\begin{align}
&\left\langle\phi\left(w_{1},\bar{w}_{1}\right)\phi\left(w_{2},\bar{w}_{2}\right)\right\rangle= \nonumber \\
&=\left(\frac{2\pi}{L}\right)^{4\Delta}\left[4 \sinh \frac{\pi\left(w_{1}-w_{2}\right)}{L}\sinh \frac{\pi\left(\bar{w}_{1}-\bar{w}_{2}\right)}{L}\right]^{-2\Delta}\label{eq:50}
\end{align}
where it has been supposed, for simplicity, \(\Delta=\bar{\Delta}\). Observe that for \(\left|w_{1}-w_{2}\right|\) and \(\left|\bar{w}_{1}-\bar{w}_{2}\right|\) much smaller than \(L\) the effect of the finite size disappears and one recovers the infinite plane result. In the opposite case one can easily see that the correlator decays exponentially with a correlation length 

\begin{equation}
\xi=\frac{L}{4\pi\Delta}\label{eq:51}
\end{equation}
The apparence of this correlation length is due to the presence of the size scale \(L\). In a more general way, one takes a set of scaling operators \(\phi_{i}\) with scaling dimensions \(x_{i}\). It has been observed that they scale as in (\ref{eq:14}) and for any two point correlation function a correlation length \(\xi_{i}\) can be introduced. Therefore, at criticality, making use of conformal invariance, it is simple to identify exactly the scaling dimensions of fields (in the special case of scalar primary fields: \(x_{i}=2\Delta_{i}\)). Moreover, with respect to the formulae (\ref{eq:13}) and (\ref{eq:26}), one can show that for an infinitely long strip of finite width \(L\) and with periodic boundary conditions the scaling functions of fields \(\phi_{i}\) at the RG fixed points (critical points) are exactly known

\begin{equation}
\mathcal{G}(0,0)=4\pi x_{i}=4\pi \Delta_{i}\;.\label{eq:52}
\end{equation}
CFT gives an \emph{a posteriori} confirmation of the scaling hypothesis and, from a physical point of view, what which one is more interested in, i.e. the quantitative relations among universal scaling amplitudes, critical exponents and correlation functions. 

\subsection{Modular invariance}
Up to now it has been assumed that the CFTs live in the whole complex plane or in the cylinder described in the previous section. It could be useful to study these theories on a torus geometry. The motivation to do that is the possibility of extracting constraints on the content of the theories coming from the interactions of the holomorphic and antiholomorphic sectors revealed by the so called \emph{modular transformations}. In this scheme the Hamiltonian and the momentum operators propagate states along the two different directions of the torus and the spectrum is embodied in the partition function.

A torus may be defined by two linearly independent vectors on the plane and identifying points that differ by an integer combination of these vectors. On the complex plane these lattice vectors can be represented by two complex numbers \(\omega_{1}\) and  \(\omega_{2}\). The relevant parameter is the ratio \(\tau=\omega_{1}/\omega_{2}\). Regarding the torus as a cylinder of finite length whose ends are glued together, one has an expression for Hamiltonian as the one given in Eq. (\ref{eq:49}), i.e. \(H=(2\pi /L)(L_{0}+\bar{L}_{0}-c/12)\) and for the momentum operator one has \(P=(2\pi i/L)(L_{0}-\bar{L}_{0})\). Choosing then \(\omega_{1}\) real and equal to \(L\), it is possible to write the partition function as

\begin{equation}
\mathcal{Z}(\tau)=\textrm{Tr}\;\left(q^{L_{0}-c/24}\bar{q}^{\bar{L}_{0}-c/24}\right)\label{eq:53}
\end{equation}
where 

\begin{equation}
q=e^{2\pi i \tau}\;\;\;\;,\;\;\;\;\bar{q}=e^{-2\pi i \bar{\tau}}\label{eq:54}
\end{equation}
For a CFT it is necessary to have the partition function \(\mathcal{Z}(\tau)\) invariant under the modular group \(SL\left(2,\mathbb{Z}\right)\) generated by

\begin{equation}
\mathcal{T}\;:\;\tau\rightarrow \tau+1\;\;\;,\;\;\;\mathcal{S}\;:\;\tau\rightarrow -\frac{1}{\tau}\label{eq:55}
\end{equation}
This invariance imposes constraints on the operator content of the theory. For example, given the Virasoro algebra character functions \(\chi_{\Delta}(q)\), the partition function is \cite{car3,car1}

\begin{equation}
\mathcal{Z}(q,\bar{q})=\sum_{\Delta,\bar{\Delta}}\mathcal{N}_{\Delta,\bar{\Delta}}\chi_{\Delta}(q)\chi_{\bar{\Delta}}(\bar{q})\label{eq:56}
\end{equation}
with \(\mathcal{N}_{\Delta,\bar{\Delta}}\) non negative integers characterizing the operator content of the theory, therefore its class of universality. It is possible to demonstrate that the requirement of modular invariance fixes the numbers \(\mathcal{N}_{\Delta,\bar{\Delta}}\). Moreover, the modular invariant partition functions may be classified, which corresponds in a certain sense to a classification of CFTs themselves. It has also been shown that there is a deep correspondence with the ADE classification of simply laced finite dimensional Lie algebras \cite{capp1,capp2}. See \cite{difra} and references therein for more details.

\subsection{Free boson}
It is now necessary to give a brief summary of the free boson theory which is a \(c=1\) CFT of a scalar field \(\phi(x,t)\) compactified on a circle of radius \(R\). The Lagrangean is

\begin{equation}
\mathcal{L}=\frac{1}{8\pi}\int_{0}^{L}\textrm{d}x\partial_{\mu}\phi(x,t)\partial^{\mu}\phi(x,t)\label{eq:57}
\end{equation}
where \(L \) is the spatial volume, i.e. the theory is defined on a cylinder of circumference \(L\). The field is quasi-periodic in the space direction, that is \(\phi(x+L,t)=\phi(x,t)+2\pi mR\) with \(m\in\mathbb{Z}\). The integer \(m\) specifies a topological class of configurations obeying the above periodicity condition. Therefore one can characterize several sectors labelled by a pair \((n,m)\), where \(\frac{n}{R}\) is the eigenvalue of the total momentum and \(m\) (winding number) is the eigenvalue of the topological charge, which are defined respectively by

\[
\Pi_{0}=\frac{1}{4\pi}\int_{0}^{L}\textrm{d}x\partial_{t}\phi(x,t)\;\;\;,\;\;\;Q=\frac{1}{2\pi R}\int_{0}^{L}\textrm{d}x \partial_{x}\phi(x,t) \]
Adopting the variables \(z\) and \(\bar{z}\), the field is expanded in modes as it follows

\begin{equation}
\phi(z,\bar{z})=\phi_{0}-ip_{+}\log z-ip_{-}\log \bar{z}+i\sum_{k\ne 0}\frac{1}{k}\left(\alpha_{k}z^{-k}+\bar{\alpha}_{k}\bar{z}^{-k}\right)\label{eq:58}
\end{equation}
where \(\phi_{0}\) is the boson zero mode, \(p_{\pm}=\left(\frac{n}{R}\pm\frac{1}{2}mR\right)\) and

\begin{equation}
\left[\alpha_{n},\alpha_{m}\right]=\left[\bar{\alpha}_{n},\bar{\alpha}_{m}\right]=n\delta_{n+m,0}\;\;\;,\;\;\;\left[\alpha_{n},\bar{\alpha}_{m}\right]=0\label{eq:59}
\end{equation}
The Virasoro generators take the form

\begin{equation}
L_{n}=\frac{1}{2}\sum_{k=-\infty}^{\infty}\alpha_{n-k}\alpha_{k}\;\;\;,\;\;\;\bar{L}_{n}=\frac{1}{2}\sum_{k=-\infty}^{\infty}\bar{\alpha}_{n-k}\bar{\alpha}_{k}\label{eq:60}
\end{equation}
The products above have to be considered as normal ordered. The modes \(\alpha_{n}\) and \(\bar{\alpha}_{n}\) are annihilation operators for \(n>0\) and creation operators for \(n<0\). For any sector, the whole space of fields is obtained starting from the highest weight states \(\left|n,m\right\rangle\) created from the vacuum by the vertex operators

\[
V_{(n,m)}(z,\bar{z})=e^{i\left[p_{+}\phi(z)+p_{-}\phi(\bar{z})\right]}\]
of conformal dimensions \(\Delta_{(n,m)}=p_{+}^{2}/2\) and \(\bar{\Delta}_{(n,m)}=p_{-}^{2}/2\), with successive application of creation operators on the states. Schematically one has

\begin{equation}
\mathcal{H}=\bigoplus_{n,m} \mathcal{H}_{n,m}\otimes \bar{\mathcal{H}}_{n,m}\label{eq:61}
\end{equation}
where \(\mathcal{H}_{n,m}\) and \(\bar{\mathcal{H}}_{n,m}\) are representations of Heisenberg algebra. Of course the states \(\left|n,m\right\rangle\) are also Virasoro highest weights, therefore one could build up the whole Hilbert space by acting with the \(L_{n}\)'s, but this would be more complicated.

The boson Hamiltonian is expressed in terms of Virasoro operators as in Eq. (\ref{eq:49}) with central charge \(c=1\).

\section{Boundary conformal field theory}
\begin{figure}
\begin{center}
\includegraphics[angle=0, width=0.7\textwidth]{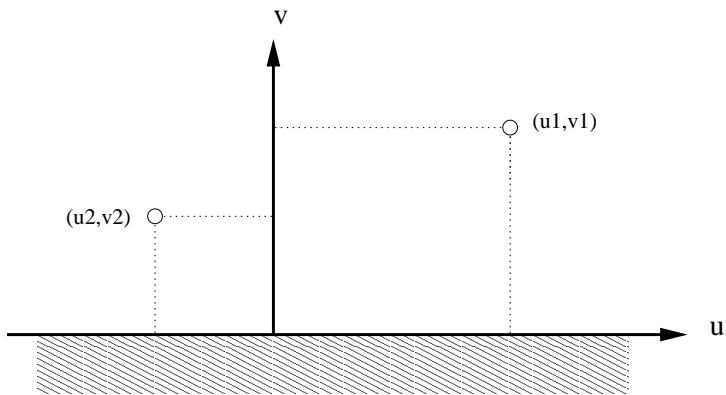}
\caption{\emph{Semi-infinite geometry. The physically accessible half-plane is the upper one}.\label{fig:2}}
\end{center}
\end{figure}

We want to consider now a quantum field theory defined only in the half upper plane \(v>0\) (see Fig. (\ref{fig:2})). One would like that the presence of the boundary on the axis \(u\) does not break several symmetries of the theory. In particular one expects that it is invariant under global rotations, dilatations and translations preserving the boundary and that these invariances should be implemented locally, i.e. conformal invariance realized in presence of boundary \cite{car4}.

A condition for boundary conformal invariance is that \(T_{12}=T_{21}\) on the axis \(u\), which means that there is no flux of energy through the boundary. As a consequence the holomorphic and antiholomorphic components of the energy-momentum tensor are not independent, but \(T(z)=\bar{T}(\bar{z})\) for \(\mathfrak{Im}z=0\). So in the region \(\mathfrak{Im}z<0\) one can define the energy-momentum tensor

\begin{equation}
T(z)=\bar{T}(\bar{z})\;,\;\;\;\;\mathfrak{Im}z<0\label{eq:62}
\end{equation}
Therefore in a boundary conformal field theory (BCFT), instead of having holomorphic and antiholomorphic parts of the fields in the half plane, it is possible to describe the whole theory with the holomorphic (antiholomorphic) part of the fields in the plane. For instance, in radial quantization, the Hamiltonian becomes 

\begin{figure}
\begin{center}
\includegraphics[angle=0, width=0.7\textwidth]{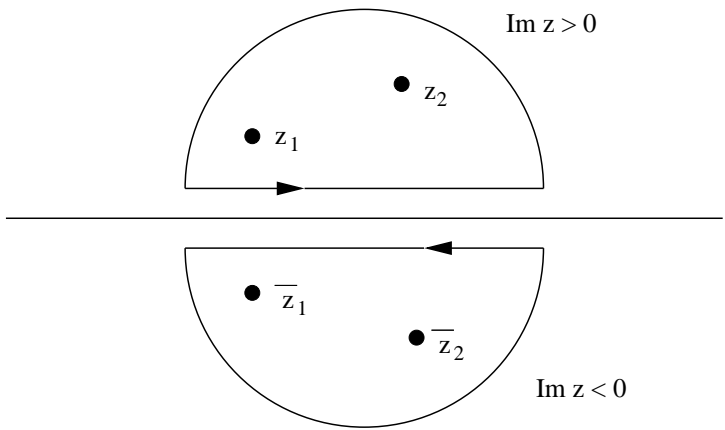}
\caption{\emph{Geometry of the integration contour for the boundary case}.\label{fig:3}}
\end{center}
\end{figure}

\begin{equation}
H\sim\oint_{C}T(z)\textrm{d}z+\textrm{c.c.}\label{eq:63}
\end{equation}
which means that the Hamiltonian generates the motion just outwards from the origin in the upper half plane. Because of the definition (\ref{eq:62}), the contour in Fig. (\ref{fig:3}) becomes a closed contour integral of \(T\) only: therefore the Hilbert space of a BCFT is described by a sum of representations of only one Virasoro algebra: 

\begin{equation}
\mathcal{H}=\bigoplus_{\Delta}\mathcal{N}_{\Delta}\mathcal{V}_{c}(\Delta)\label{eq:64}
\end{equation}
Such a system is naturally represented mapping the half complex plane onto a strip of width \(L\) with some boundary conditions through the coordinates transformation \(w=(L/ \pi)\ln(z)\). Note that, because of the single Virasoro algebra, the Hamiltonian in the cylinder now reads

\begin{equation}
H=\frac{\pi}{L}\left(L_{0}-\frac{c}{24}\right).\label{eq:65}      
\end{equation}
The space depends on the boundary conditions, it is organized into a vector space by a single Virasoro algebra strictly related to the boundary. Physically this implies that the same observables are associated to different representations of Virasoro algebra in the bulk and boundary cases.

\subsection{Partition function and boundary states}
\begin{figure}
\begin{center}
\includegraphics[angle=0, width=0.7\textwidth]{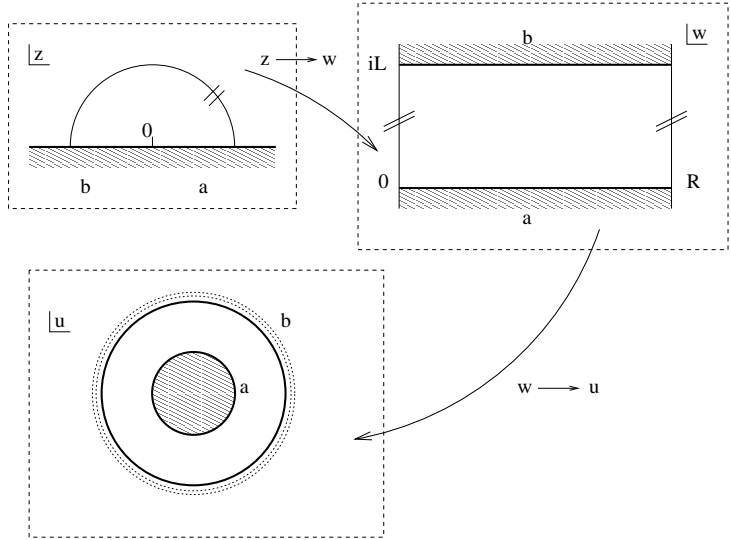}
\caption{\emph{The same domain in different coordinates: the upper half complex plane punctured at the origin, the infinite strip with coordinate \(w=(L/2\pi)\log z\), the circular annulus with coordinate \(u=\textrm{exp}(-2\pi iw/R)\)}.\label{fig:4}}
\end{center}
\end{figure}

To study boundary conditions preserving conformal invariance and boundary states, it is useful to consider the BCFT on a cylinder within two equivalent quantization schemes, one in which the time flows around the cylinder, another one in which the time flows along that. In the first case the Hamiltonian depends on the boundary conditions \(a\) and \(b\) on the edges of the cylinder. In the second case, the boundary conditions are embodied in initial and final boundary states \(\left|a\right\rangle\) and \(\left|b\right\rangle\) and the Hamiltonian is obtained from the whole complex plane. It is convenient to introduce a partition function

\begin{equation}
\mathcal{Z}_{ab}=\textrm{Tr}\;\textrm{exp}\left(-\frac{\pi R}{L}H_{ab}\right)=\textrm{Tr}\;q^{H_{ab}}\label{eq:66}
\end{equation}
with \(R\) and \(L\) respectively the circumference and the length of the cylinder, \(q=e^{2\pi i\tau}\), \(\tau=iR/2L\). In particular the cylinder can be obtained first mapping the upper half plane into an infinite strip, \(w=\frac{L}{\pi}\log z\), and then imposing periodic boundary conditions in the \(R\) direction. The conformal invariance imposes that the spectrum of the Hamiltonian \(H_{ab}\) falls into representations of the Virasoro algebra. If one calls \(\mathcal{N}^{i}_{ab}\) the number of copies of representations in the spectrum, the partition function can be written as

\begin{equation}
\mathcal{Z}_{ab}(q)=\sum_{i}\mathcal{N}_{ab}^{i}\chi_{i}(q)\label{eq:67}
\end{equation}
where the \(\chi_{i}(q)\)'s are the Virasoro characters of the representation \(i\). Interchanging now the roles of \(L\) and \(R\), that corresponds to the modular transformation \(\tau\rightarrow -1/\tau\), it is possible to regard the partition function as a trace of a Hamiltonian generating translation along the spatial coordinate. This can be done via the coordinate transformation \(u=\textrm{exp}(-2\pi iw/R)\), which is a plane distinct from the upper half \(z\)-plane (see Fig. (\ref{fig:4}) for a pictorial description of these transformations).

The Hamiltonian reads now

\begin{equation}
\widetilde{H}=\frac{2\pi}{R}\left(L_{0}^{u}+\bar{L}_{0}^{u}-\frac{c}{12}\right)\label{eq:68}
\end{equation}
On the \(u\)-plane the boundary conditions are imposed propagating states from the initial boundary state \(\left|a\right\rangle\) to the final one 
\(\left|b\right\rangle\). The partition function becomes

\begin{equation}
\mathcal{Z}_{ab}(q)=\left\langle a\right|e^{L\bar{H}}\left|b\right\rangle=\left\langle a\right|\left(\bar{q}^{1/2}\right)^{\left(L_{0}^{u}+\bar{L}_{0}^{u}-c/12\right)}\left|b\right\rangle\label{eq:69}
\end{equation}
where \(\bar{q}=e^{-2\pi i/\tau}\). Under these coordinates transformations, the condition (\ref{eq:62}) implies on the energy-momentum tensor
\[
u^{2}T(u)=\bar{u}^{2}\bar{T}(\bar{u})\;\;\;,\;\;\;\left|u\right|=1,\;e^{2\pi L/R} \]
This gives a strong condition on the boundary states \(\left|a\right\rangle\) and \(\left|b\right\rangle\)

\begin{equation}
\left(L_{n}-\bar{L}_{-n}\right)\left|a\right\rangle=0\;\;\;,\;\;\;\left(L_{n}-\bar{L}_{-n}\right)\left|b\right\rangle=0\; .\label{eq:70}
\end{equation}
A solution to these equations is given by the Ishibashi states \cite{ishi}

\begin{equation}
\left|j\right\rangle=\sum_{N}\left|j;N\right\rangle\otimes\left|\bar{j};N\right\rangle\;.\label{eq:71}
\end{equation}
where \(\left|j;N\right\rangle\) and \(\left|\bar{j},N\right\rangle\) are orthonormal basis of Virasoro representations j (\(N\) labels the different states within that module). Every boundary state should be a linear combination of vectors (\ref{eq:71}). For the partition function one has

\begin{equation}
\mathcal{Z}_{ab}(q)=\sum_{j}\left\langle a\right|j\left\rangle\right\langle j\left|b\right\rangle \chi_{j}(\bar{q})\label{eq:72}
\end{equation}
Making use of the modular properties of the characters and comparing the expressions (\ref{eq:67}) and (\ref{eq:72}) it is also possible to give an exact valuation of numbers \(\mathcal{N}_{ab}^{i}\).
 
In the case of the free boson, the boundary states have to satisfy the stronger constraint (remember the form of Virasoro generators in the free boson scheme)

\[
\left(\alpha_{n}\pm\bar{\alpha}_{n}\right)\left|a\right\rangle=0\;\;\;,\;\;\; \left(\alpha_{n}\pm\bar{\alpha}_{n}\right)\left|b\right\rangle=0\]corresponding to Neumann and Dirichlet boundary conditions (\(T_{12}=0\)). In particular the negative sign can be solved by states of type

\[
\left|a\right\rangle\propto \textrm{exp}\left[-\sum_{n=1}^{\infty}\frac{\alpha_{-n}\bar{\alpha}_{-n}}{n}\right]\left|0,k\right\rangle \]
therefore, any boundary state is a linear combination of vectors of this type. If one looks at this system on the cylinder and considers two boundary fields \(\Phi_{+}\) and \(\Phi_{-}\) acting on boundaries of both sides of the cylinder, one can calculate the partition function of the system explicitly (here we consider Dirichlet boundary conditions) and find

\begin{equation}
\mathcal{Z}_{DD}=\propto \frac{1}{\eta(q)}\sum_{n}q^{\frac{1}{2\pi}\left(\Phi_{+}-\Phi_{-}+2\pi nr\right)^{2}}\label{eq:73}   
\end{equation}
where \(4\sqrt{\pi}r=R\) and \(\eta(q)=q^{1/24}\prod_{n=1}^{\infty}\left(1-q^{n}\right)\) is the Dedekind function. 

For Neumann boundary conditions, or, equivalently, Dirichlet boundary conditions on the dual fields \(\widetilde{\Phi}_{\pm}\),

\begin{equation}
\mathcal{Z}_{NN}\propto \frac{1}{\eta(q)}\sum_{n}q^{\frac{1}{2\pi}\left(\widetilde{\Phi}_{+}-\widetilde{\Phi}_{-}+ n/r\right)^{2}}\;.\label{eq:74}
\end{equation}
These expressions have simple interpretations: one has to sum over all the sectors where the difference of hights between the two sides of the cylinder is \(\Phi_{+}-\Phi_{-}+2\pi nr\) (Dirichlet) and \(\widetilde{\Phi}_{+}-\widetilde{\Phi}_{-}+ n/r\) (Neumann). For each such sectors the partition function corresponds to the product of a basic partition function with hights equal on both sides, times the exponential of a classical action. The boundary states are, respectively,

\[
\left|B_{D}\left(\Phi_{\pm}\right)\right\rangle\propto \sum_{k=-\infty}^{\infty}\textrm{exp}\left(-\frac{ik\Phi_{\pm}}{r}\right)\textrm{exp}\left[-\sum_{n=1}^{\infty}\frac{\alpha_{-n}\bar{\alpha}_{-n}}{n}\right]\left|0,k\right\rangle \]

\[ 
\left|B_{N}\left(\widetilde{\Phi}_{\pm}\right)\right\rangle\propto \sum_{l=-\infty}^{\infty}\textrm{exp}\left(-2\pi il\widetilde{\Phi}_{\pm}r\right)\textrm{exp}\left[\sum_{n=1}^{\infty}\frac{\alpha_{-n}\bar{\alpha}_{-n}}{n}\right]\left|l,0\right\rangle \]
Details on this subject can be found in \cite{car2}.

\chapter{Integrable theories and finite size effects}
In this Chapter the main properties of integrable field theories are illustrated and the physical phenomena coming from finite size are introduced.

\section{Near the critical point}
It has been noticed that the analysis of the universality classes of two dimensional statistical models includes the study of CFT living on the RG fixed points or, equivalently, from a statistical point of view, second order critical points and the description of the behaviour around them. The way a CFT behaves in a neighbourhood of RG fixed points can be observed considering that any RG trajectory flowing away from such fixed points can be described, as briefly illustrated in the previous Chapter, by combinations of relevant fields \(\Phi_{i}\), i.e. conformal fields characteristic of the given CFT \cite{zam3}. The off-critical theory is described by the action

\begin{equation}
\mathcal{A}=\mathcal{A}_{CFT}+\sum_{i}\lambda_{i}\int\textrm{d}^{2}x\;\Phi_{i}(x)\label{eq:75}
\end{equation}
where \(\mathcal{A}_{CFT}\) corresponds to the conformally invariant action and the \(\lambda_{i}\)'s are coupling constants. The Lorentz invariance in two dimensions is equivalent to rotational invariance once a Wick rotation is performed. Therefore, to have Lorentz invariance, one has to require the fields \(\Phi_{i}\) to be scalars under two-dimensional rotations, as a result their spin is \(s=\Delta_{i}-\bar{\Delta}_{i}=0\), which implies that \(\Delta_{i}=\bar{\Delta}_{i}\) and the anomalous dimensions are \(2\Delta_{i}\). The relevant operators \(\Phi_{i}\) in (\ref{eq:75}) do not affect the short distance behaviour of the theory because they are superrenormalizable, but they modify it at large distance scales. In particular, one can expect that through the RG flow the system either can reach another fixed point, in case described by another CFT, or go to a non critical point, hence it will correspond to a massive quantum field theory. In this scheme, the knowledge of the properties of the RG flow and of the theories linked by that is one of the main aims. For an unitary theory there exists a theorem, due to Zamolodchikov, which allows one to have an exact function describing the RG flow and, in the fixed points of such a flow, coinciding with the central charge of the underlying CFT.

\subsection{The c-theorem}
The \(c\)-theorem \cite{zam2} states that, given an unitary \(1+1\) dimensional quantum field theory endowed with rotational invariance and conservation of the energy-momentum tensor, there exists a function \(C(\lambda_{i})\) of the couplings \(\lambda_{i}\) which is non-increasing along the RG flow and stationary only at the RG fixed points where it coincides with the central charge \(c\) of the corresponding CFT. One can demonstrate the theorem in a very simple way. Let \(T\), \(\Theta\) and \(\bar{T}\) be respectively the components with spin \(2\), \(0\) and \(-2\) of the energy-momentum tensor. The two-point functions among them can be written as

\[
\left\langle T(z,\bar{z})T(0,0)\right\rangle=\frac{F(mz\bar{z})}{z^{4}} \]

\[
\left\langle T(z,\bar{z})\Theta(0,0)\right\rangle=\frac{G(mz\bar{z})}{z^{3}\bar{z}} \]

\[
\left\langle \Theta(z,\bar{z})\Theta(0,0)\right\rangle=\frac{H(mz\bar{z})}{z^{2}\bar{z}^{2}} \]
where \(m\) is a mass scale parameter and \(F\), \(G\) and \(H\) some scalar functions. As a result of the energy-momentum tensor conservation one deduces differential equations for the functions \(F\), \(G\) and \(H\)

\begin{equation}
\dot{F}+\frac{1}{4}\left(\dot{G}-3G\right)=0\;\;\;\;,\;\;\;\;\dot{G}-G+\frac{1}{4}\left(\dot{H}-2H\right)=0\label{eq:76}
\end{equation}
where \(\dot{F}=\textrm{d}F(x)/\textrm{d}\log x\). Defining the scalar function

\begin{equation}
C\equiv 2F-G-\frac{3}{8}H\label{eq:77}
\end{equation}
it follows

\begin{equation}
\dot{C}=-\frac{3}{4}H\label{eq:78}
\end{equation}
The unitarity conditon of the quantum field theory imposes that \(H\) is a positive quantity, thus \(C\) is not increasing. At the critical point the energy-momentum tensor is traceless: \(\Theta=0\). As a consequence, at fixed points, \(G=H=0\) and \(F=\frac{c}{2}\) and hence the function \(C\) reduces to the central charge \(c\). Moreover it is possible to extract a sum rule for the total change in the function \(C\) from short to long distances. In fact one can consider a CFT simply perturbed by one relevant field \(\Phi\) with coupling \(\lambda\) and scaling dimension \(x=2\Delta\). The trace of the energy-momentum tensor follows by direct calculation 

\begin{equation}
\Theta(x)=2\pi \lambda\left(2-x\right)\Phi(x)\label{eq:79}
\end{equation}
and using the \(c\)-theorem, the total change of \(C\) is

\begin{equation}
\delta C=3\pi\lambda^{2}\left(2-x\right)\int \textrm{d}x^{2}\left|x\right|^{2}\left\langle\Phi(x)\Phi(0)\right\rangle\label{eq:80}
\end{equation}
which means that one has an exact computation of the total change of the central charge along two different fixed points of the RG trajectory. If the CFT \(\mathcal{A}_{CFT}^{1}\), from which the RG flow starts by application of a relevant scalar operator \(\Phi\), is known, in principle the CFT \(\mathcal{A}_{CFT}^{2}\) in which the RG flow ends is also known. 

\section{Integrable models}

An integrable quantum field theory (IQFT) is characterized by the existence of an infinite number of conservation laws, i.e. by an infinite set of conserved currents. In two dimensional systems one can use holomorphic and antiholomorphic indices and, in particular, as noticed before, an important characteristic of CFT in two dimensions is the decoupling of \(z\) and \(\bar{z}\) dependence. So, in a CFT one can take as current any independent operator in the conformal block of the energy-momentum tensor \(T(z,\bar{z})\) and, as a result of the conservation laws

\begin{equation}
\partial_{\bar{z}}T(z,\bar{z})=\partial_{z}\bar{T}(z,\bar{z})=0,\label{eq:81}
\end{equation}
it follows that the CFT's possess an infinite set of conserved charges, i.e. they are integrable. It can be demonstrated that, given a polynomial \(T_{n}(z)\) of the energy-momentum tensor \(T(z)\) at level \(n\) (and, equivalently, for \(\bar{T}(\bar{z})\)) there are infinite integrals of motion of the form

\begin{equation}
I_{n+1}=\oint \textrm{d}z\;T_{n}(z)\label{eq:82}
\end{equation}
which are non-trivial for \(n\) even. In the deformed theory defined in (\ref{eq:75}), this set of conserved quantities - together with the decoupled variables \(z\) and \(\bar{z}\) - is in general lost. Anyway, there exist the so-called \emph{integrable} deformations of CFT in which an infinite number of conserved quantities survives and the theory can be treated non-perturbatively.

\subsection{Conservation laws}
For integrable theories originated by a perturbation of a CFT, the integrals of motion can be interpreted as deformations of the conformal conservation laws. For simplicity let us consider just one perturbing field \(\Phi\).

In fact, if the equation \(\partial_{\bar{z}}T_{n}(z)=0\) for the linear combination \(T_{n}(z)\) of fields in the conformal block of the energy-momentum tensor introduced above holds at criticality, this is not more valid off-criticality. The equation is deformed by several contributions

\begin{equation}
\partial_{\bar{z}}T_{n}(z)=\lambda Q_{n-1}^{(1)}+\lambda^{2}Q_{n-1}^{(2)}+\;.\;.\;.\label{eq:83}
\end{equation}
It is reasonably correct to think that the fields \(Q_{n-1}^{(k)}\) have a limit in the unperturbed CFT, defined by some combination of fields in the Virasoro algebra of the CFT, with conformal dimensions \(\left(\Delta^{(k)},\bar{\Delta}^{(k)}\right)\). The following relations hold

\begin{equation}
\Delta^{(k)}=n-k(1-\Delta)\;\;\;\;,\;\;\;\;\bar{\Delta}^{(k)}=1-k(1-\Delta)\label{eq:84}   
\end{equation}
with \(\Delta\) the conformal dimension of the perturbing field \(\Phi\). This requires that the series (\ref{eq:83}) contains a finite number of fields \(Q_{n-1}^{(k)}\), as the conformal dimensions of a CFT are bounded by below. In the simplest case there is only one field with conformal dimensions given in (\ref{eq:84}), hence one can consider that only the field \(Q_{n-1}^{(1)}\) of first order in \(\lambda\) appears, it has to be a secondary of the field \(\Phi\) with spin \(s=n-1\) and conformal dimensions \(\left(n-1+\Delta,\Delta\right)\). It can be demonstrated that

\begin{equation}
\partial_{\bar{z}}T_{n}-\partial_{z}\Theta_{n-2}=0\label{eq:85}
\end{equation}
where \(\Theta_{n-2}\) is a field in the set of secondary descendant fields of \(\Phi\). This is an argument of the existence of perturbed conserved currents in IQFT. A simple example to show how to calculate exactly the conserved charges can be done just considering a minimal model \(\mathfrak{M}_{r,s}\) perturbed by a primary scalar field \(\Phi_{r,s}(z,\bar{z})\). The action of the system is, following Eq. (\ref{eq:75}),

\begin{equation}
\mathcal{A}=\mathcal{A}_{\mathfrak{M}}+\lambda\int\textrm{d}^{2}z\;\Phi_{r,s}(z,\bar{z})\label{eq:86}
\end{equation}
The simplest conserved current which one can work with is the energy-momentum tensor \(T(z)\). Its OPE with the perturbing field is known from Eq. (\ref{eq:33}) and explicitly reads 

\begin{equation}
T(z,\bar{z})\Phi_{r,s}(w,\bar{w})=\frac{\Delta_{r,s}\Phi_{r,s}(w,\bar{w})}{(z-w)^{2}}+\frac{\partial_{w}\Phi_{r,s}(w,\bar{w})}{z-w}+.\;.\;.\label{eq:87}
\end{equation}
Considering now that the Ward identities can be written in terms of the conformal ones as

\begin{align}
&\left\langle T(z,\bar{z}).\;.\;.\right\rangle=\left\langle T(z,\bar{z}).\;.\;.\right\rangle_{\mathfrak{M}}+\nonumber \\
&+\lambda\int\textrm{d}w\textrm{d}\bar{w}\left\langle T(z,\bar{z})\Phi_{r,s}(w,\bar{w}).\;.\;.\right\rangle+\mathcal{O}\left(\lambda^{2}\right),\label{eq:88}
\end{align}
one easily arrives to the perturbed counterpart of the conformal conservation law at the first order in \(\lambda\)

\begin{equation}
\partial_{\bar{z}}T(z,\bar{z})=\lambda\left[(1-\Delta)\partial_{z}\Phi_{r,s}(z,\bar{z})\right]\label{eq:89}
\end{equation}
Therefore, identifying \(\Theta(z,\bar{z})=\lambda(1-\Delta)\Phi_{r,s}(z,\bar{z})\), the first integral of motion is

\begin{equation}
I_{1}=\oint\textrm{d}z\;T(z,\bar{z})+\oint\textrm{d}\bar{z}\;\Theta(z,\bar{z})\label{eq:90} 
\end{equation}
It is also possible to give several arguments to find higher conserved quantities \(I_{n}\) \cite{zam3}.

\subsection{S-matrix approach}
In the analysis of a massive IQFT the scattering description has a very important meaning. In general one can assume to have massive particles distinguished by some label \(a\), with mass \(\mathcal{M}_{a}\). Their momenta \(p^{\mu}_{a}\) can be written in terms of a rapidity variable \(\theta\): \(E=\mathcal{M}\cosh \theta\), \(P=\mathcal{M}\sinh \theta\). In any scattering process there are the "in-states", corresponding physically to a set of particles incoming from the infinite past, arranged by decreasing order of rapidities, formally described by \(\left|\theta_{1},\;.\;.\;.\;,\theta_{N}\right\rangle^{a_{1},...,a_{N}}_{IN}\). At infinite time in the future, there are "out-states" \(\left|\theta_{1}^{\prime},\;.\;.\;.\;,\theta_{N}^{\prime}\right\rangle^{a_{1}^{\prime},...,a_{N}^{\prime}}_{OUT}\) describing in general a different set of particles arranged by increasing order of different rapidities. The "in" and "out" states form a complete set of states of a local quantum field theory and they are connected by the operator called \(S\) matrix.

\begin{figure}
\begin{center}
\includegraphics[angle=0, width=0.7\textwidth]{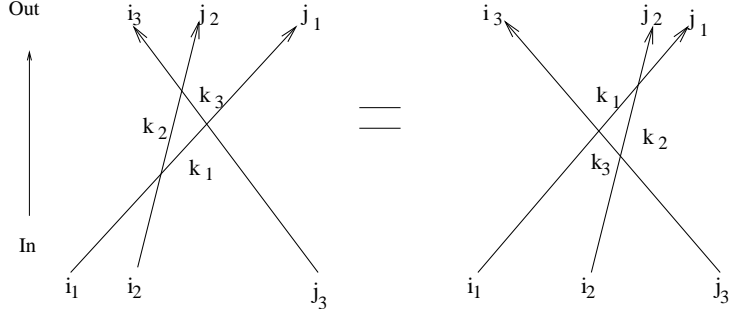}
\caption{\emph{Factorization of the scattering}.\label{fig:5}}
\end{center}
\end{figure}

In a IQFT, the existence of infinitely many conserved quantities has very important consequences \cite{zamzam}. In any scattering process it results that:
\begin{itemize}
\item the number of particles is conserved, more precisely the number of particles of same mass is conserved;

\item the set of final momenta coincides with the set of initial momenta, \(\left\{p_{1},p_{2},\;.\;.\;.p_{n}\right\}=\left\{p^{\prime}_{1},p^{\prime}_{2},\;.\;.\;.p^{\prime}_{n}\right\}\), such that the scattering processes which take place are purely elastic.
\end{itemize}
From this it follows that the \(S\) matrix factorizes into a product of 2-body scattering \(S\) matrices. To verify this, let \(A_{i}(\theta)\) be a set of non-commutative operators which describe the corresponding particles. They are regarded as generators of the infinite-dimensional algebra given by all possible products of type \(A_{a_{1}}\left(\theta_{1}\right)A_{a_{2}}\left(\theta_{2}\right).\;.\;.A_{a_{n}}\left(\theta_{n}\right)\). The incoming and outgoing asymptotic states correspond to a decreasing and increasing arrangement of the rapidities \(\theta_{i}\). In this approach the \(S\) matrix is an operator such that

\begin{equation}
A_{i}\left(\theta_{1}\right)A_{j}\left(\theta_{2}\right)=S_{ij}^{kl}\left(\theta_{12}\right)A_{l}\left(\theta_{2}\right)A_{k}\left(\theta_{1}\right)\label{eq:91}
\end{equation}
where the relativistic invariance constrains the \(S\) matrix to depend on the difference of rapidities \(\theta_{ij}=\theta_{i}-\theta_{j}\). These commutation relations are required to be compatible with the algebraic associativity, which translates to the \emph{Yang-Baxter equation} (YBE)

\begin{equation}
S_{i_{1}i_{2}}^{k_{1}k_{2}}\left(\theta_{12}\right)S_{k_{1}k_{3}}^{j_{1}j_{3}}\left(\theta_{13}\right)S_{k_{2}i_{3}}^{j_{2}k_{3}}\left(\theta_{23}\right)=S_{i_{2}j_{3}}^{k_{2}k_{3}}\left(\theta_{23}\right)S_{i_{1}k_{3}}^{k_{1}i_{3}}\left(\theta_{13}\right)S_{k_{1}i_{2}}^{j_{1}j_{2}}\left(\theta_{12}\right)\label{eq:92}
\end{equation}
It can be described by the following physical argument. One can consider just a particle, say \(a_{1}\), with its total momentum \(p^{\mu}_{1}\). Since the \(S\) matrix conserves momenta, one can think to conjugate it by the operator \(\textrm{exp}(ip_{1}^{\mu}x_{\mu})\), in fact it does commute with this operator, therefore this operation does not change anything. But it is a non-trivial physical action: conjugating the \(S\) matrix with \(\textrm{exp}(ip_{1}^{\mu}x_{\mu})\) changes the space-time coordinates of the particle \(a_{1}\). In particular, choosing appropriately \(p_{1}\), one should be able to arrange for \(a_{1}\) to scatter with the other particles only when they are separated, therefore, the complete scattering process can occour as a succession of two particle scatterings. So, proceeding inductively for any particle, the \(S\) matrix factorizes. Of course, for reasons of consistency, the scattering must be "associative", i.e. the scattering of three particles has to be decomposed into three pairwise scattering with a result indipendent of which particular decomposition is used. From this the YBE follows and the Eq. (\ref{eq:92}) can be understood looking at the Fig. (\ref{fig:5}).

\begin{figure}
\begin{center}
\includegraphics[angle=0, width=0.7\textwidth]{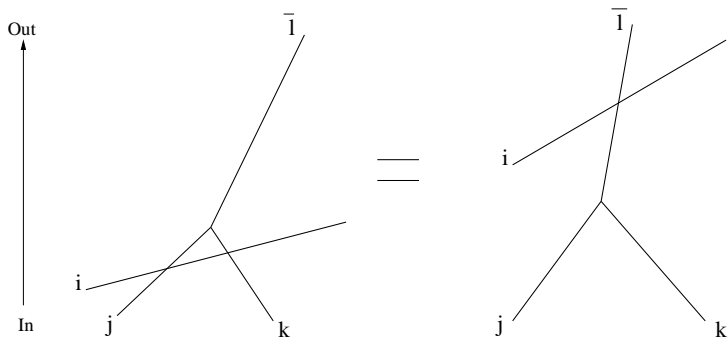}
\caption{\emph{Bootstrap relation}.\label{fig:6}}
\end{center}
\end{figure}

The YBE is the fundamental constraint of the algebraic approach to IQFT's. The \(S\) matrix is not only characterized by the fact that it has to solve the YBE; there are several other requirements, as unitarity and crossing symmetry

\begin{equation}
\sum_{n,m}S_{ij}^{nm}(\theta)S_{nm}^{kl}(-\theta)=\delta_{i}^{k}\delta_{j}^{l}\;\;\;,\;\;\;S_{ik}^{lj}(\theta)=S_{ij}^{kl}(i\pi-\theta)\label{eq:93}
\end{equation}
There exists also a consistent bootstrap principle, in the sense that there is a consistent set of couplings among the particles, signalled by poles in the \(S\) matrix at certain imaginary relative rapidities. These may be used to relate the \(S\) matrix elements to each other and, in a certain sense, to give a physical description of the particle system. In fact the \(S\) matrices are analytic functions in the complex plane of the Mandelstam variable \(s=\left(p_{1}^{\mu}+p_{2}^{\mu}\right)^{2}\) with branch cut singularities at \(\left(\mathcal{M}_{1}-\mathcal{M}_{2}\right)^{2}\) and \(\left(\mathcal{M}_{1}+\mathcal{M}_{2}\right)^{2}\), hence \(S_{ij}^{kl}(\theta)\) are meromorphic functions of the rapidity. Usually the bound states are associated to simple poles, and in the bootstrap approach they are associated with some of the particles appearing in the asymptotic states. In  particular, if \(\theta=i\alpha_{ij}^{n}\) is a pole in the scattering process of particles \(a_{i}\), \(a_{j}\), tha mass of the bound state \(a_{n}\) is \(\mathcal{M}_{n}^{2}=\mathcal{M}_{i}^{2}+\mathcal{M}_{j}^{2}+2\mathcal{M}_{i}\mathcal{M}_{j}\cos \alpha_{ij}^{n}\). If the system has a non degenerate mass spectrum or presents a spectrum with degenerate particles which can be distinguished by their higher conserved charge eigenvalues, the \(S\) matrix is diagonal, therefore the YBE is satisfied trivially and the unitarity and crossing relations simply become \(S_{ij}(\theta)S_{ij}(-\theta)=1\), \(S_{ij}(i\pi-\theta)=S_{\bar{i}j}(\theta)\) where \(\bar{i}\) denotes the antiparticle. In this case the bootstrap principle is encoded in the following functional equations

\begin{equation}
S_{i\bar{l}}(\theta)=S_{ij}\left(\theta+\bar{\alpha}_{jl}^{k}\right)S_{ik}\left(\theta_{d}-\bar{\alpha}_{lk}^{j}\right)\label{eq:94}
\end{equation}
where \(\bar{\alpha}_{ij}^{k}=\pi-\alpha_{ij}^{k}\). A pictorial representation of the bootstrap is illustrated in Fig. (\ref{fig:6}).

We conclude this section mentioning that there are also arguments to find the spin of the conserved quantities of an IQFT, by solving some particular equations encoding conserved charges eigenvalues, their spins and the angles \(\alpha_{ij}^{k}\) themselves with the help of special sets of resonance angles \(\alpha_{ij}^{k}\).

\section{Integrable theory with boundaries}
At this point it is very useful to give a general idea of integrability when the two dimensional field theory is restricted to a half-line, or to a segment of the line (in space direction). It is clear that field theory defined in finite space with non trivial boundary conditions turns out to have more constraints than theories in infinite space or with (anti)-periodic boundary conditions.

For example, suppose an integrable field theory describes a collection of distinguishable particles. It is not trivial to establish what is the spectrum of particle energies if the system is enclosed on the interval \(\left[-L,L\right]\). Or, if the theory is confined to the space region \(x<0\), one might expect that the boundary at \(x=0\) affects the particles approaching from the left. In fact, one has the almost minimal effect of reversing all the momentum-like conserved quantities and the preservation of all energy-like conserved quantities. It might be expected that the "out" state consisting of a single particle is proportional to the "in" state with the momentum reversed.

The principal idea to solve the problem of integrability in the presence of the boundary (suppose, for the moment, it is at \(x=0\)) is that particle states continue to be eigenstates of energy-like conserved charges and any initial state containing a single particle moving towards the boundary will evolve into a final state of a single particle moving away from the boundary \cite{cher}, i.e.

\begin{equation}
\left|a,\theta\right\rangle_{OUT}=K_{ab}\left|b,-\theta\right\rangle_{IN}\label{eq:95}
\end{equation}
where the states \(a\) and \(b\) correspond to multiplets of particles and \(K_{ab}\) is a matrix which may mix the particles as a result of the reflection from the boundary. The first task in a boundary field theory is to determine the reflection matrices \(K\) for a given boundary condition.

\begin{figure}
\begin{center}
\includegraphics[angle=0, width=0.7\textwidth]{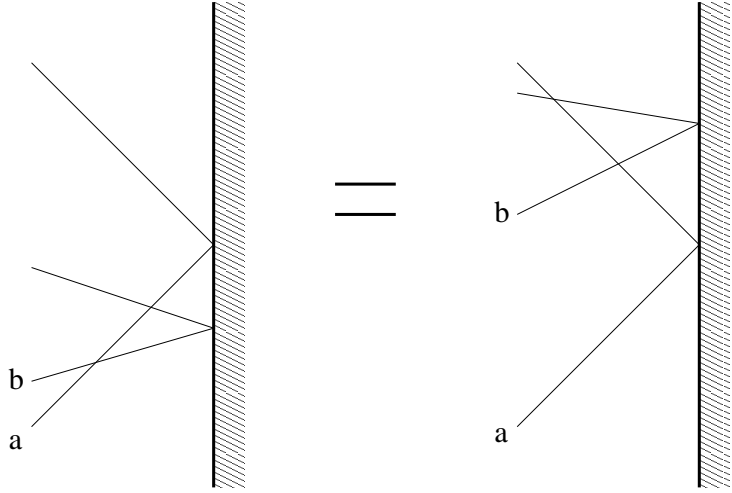}
\caption{\emph{Boundary Yang Baxter equation}.\label{fig:7}}
\end{center}
\end{figure}

In what follows one assumes that the boundary does not affect particles moving towards it until they are very "close" to it, therefore for all particles "far" from the boundary all the results of integrable theories hold. A first reasonable way to find out reflection matrices is to suppose that, given an "in" state with several particles, the order of the individual scatterings from each other and reflections from the boundary is irrelevant. If it is also supposed that these events take place factorizably, one obtains a boundary Yang Baxter equation (BYBE):

\begin{align}
&K_{a}\left(\theta_{a}\right)S_{ab}\left(\theta_{b}+\theta_{a}\right)K_{b}\left(\theta_{b}\right)S_{ab}\left(\theta_{b}-\theta_{a}\right)= \qquad\qquad\nonumber \\
&\qquad\qquad=S_{ab}\left(\theta_{b}-\theta_{a}\right)K_{b}\left(\theta_{b}\right)S_{ab}\left(\theta_{b}+\theta_{a}\right)K_{a}\left(\theta_{a}\right)\label{eq:96}
\end{align}
(see the Fig. (\ref{fig:7}) for a pictorial representation).

\begin{figure}
\begin{center}
\includegraphics[angle=0, width=0.7\textwidth]{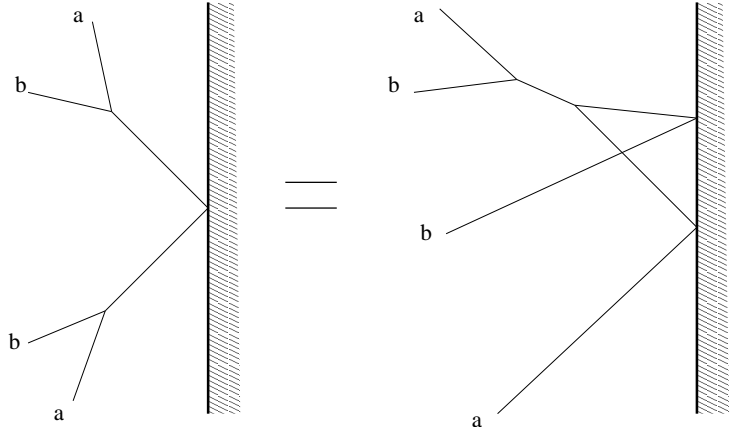}
\caption{\emph{Reflection bootstrap equation}.\label{fig:8}}
\end{center}
\end{figure}

Assuming now that the family of all masses and couplings remains the same in the presence of the boundary, the boostrap implies relations between the various reflection factors. Therefore, for example, if at special values of difference of rapidities \(\theta_{ab}\) the particles \(a\) and \(b\) form a bound state \(c\), one might think of either \(a\), \(b\) separately reflecting from the boundary in advance or after the bound state forms, or the particle \(c\) reflects from the boundary. The reflection bootstrap reads

\begin{equation}
K_{c}\left(\theta_{c}\right)=K_{a}\left(\theta_{a}\right)S_{ab}\left(\theta_{b}+\theta_{a}\right)K_{b}\left(\theta_{b}\right)\label{eq:97}
\end{equation}
where \(\theta_{a}=\theta_{c}-i\bar{\theta}_{ac}^{b}\) and \(\theta_{b}=\theta_{c}+i\bar{\theta}_{bc}^{a}\) (see also Fig. (\ref{fig:8})).

\begin{figure}
\begin{center}
\includegraphics[angle=0, width=0.7\textwidth]{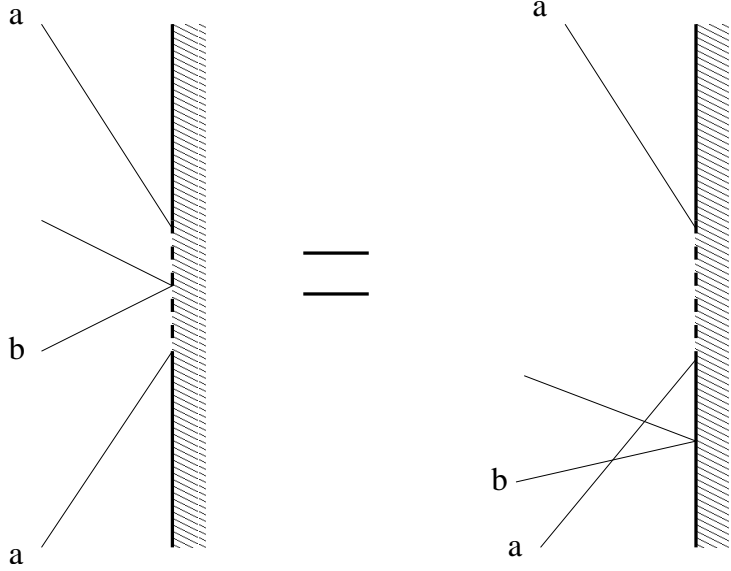}
\caption{\emph{Boundary bound state bootstrap. The dashed line on the boundary represents pictorially an excitation of the boundary itself, i.e. the boundary bound state}.\label{fig:9}}
\end{center}
\end{figure}

One might also think to the possibility of bound states involving a particle and the boundary, i.e. the boundary can be excited \cite{ghoshzam,corr1,fring1}, this should be indicated by the presence of poles in \(K_{a}\left(\theta_{a}\right)\). For instance, a particle \(a\) and a boundary, that for simplicity here we denote with \(\alpha\), have, say, a reflection factor \(K_{a}^{\alpha}\): a pole at \(\theta_{a}=i\psi_{a,\alpha}^{\beta}\) can be interpreted as a boundary bound state \(\beta\). A two particle state \(a\), \(b\) could also have the pole \(\theta_{a}\), in this case the particle \(b\) should be regarded as either reflecting from the boundary \(\beta\), or reflecting from the boundary state \(\alpha\), therefore it scatters twice with the particle \(a\), once before and once after its reflection from the boundary state \(\alpha\). This implies a factorization assumption that algebraically is

\begin{equation}
K_{b}^{\beta}\left(\theta_{b}\right)=S_{ab}\left(\theta_{b}+i\psi_{a,\alpha}^{\beta}\right)K_{b}^{\alpha}\left(\theta_{b}\right)S_{ab}\left(\theta_{b}-i\psi_{a,\alpha}^{\beta}\right)\label{eq:98}
\end{equation}
Pictorially, the boundary bound state bootstrap is represented as in Fig. (\ref{fig:9}). Finally one can also have unitarity and crossing relation as in the case without boundary

\begin{equation}
K_{a}(\theta)=K_{a}^{-1}(-\theta)\;\;\;\;,\;\;\;\;K_{a}\left(\theta-i\frac{\pi}{2}\right)K_{\bar{a}}\left(\theta+i\frac{\pi}{2}\right)S_{\bar{a}a}(2\theta)=1\label{eq:99}
\end{equation}
In terms of lagrangean formulation of a quantum field theory with boundary, one has to consider that the first step is to take a Lagrangean evidently modified due to the presence of the boundary. 

For example, one can take for semplicity a field theory of a scalar field \(\phi\) defined on the negative half line with a boundary at \(x=0\). If \(\mathcal{L}_{0}\) is the Lagrangean of the corresponding IQFT without boundary, the modified Lagrangean might take the form

\begin{equation}
\mathcal{L}_{B}=\textrm{Step}(-x)\mathcal{L}_{0}-\delta(x)\mathcal{B}(\phi)\label{eq:100}
\end{equation}
Therefore, the field equations are restricted to the region \(x\le 0\) with a boundary condition at \(x=0\)

\begin{equation}
\partial _{x}\phi=-\frac{\partial \mathcal{B}(\phi)}{\partial \phi}\label{eq:101}
\end{equation}
The main problem is to establish what value of \(\mathcal{B}\) is compatible with integrability. In particular, one has to take into account that the presence of the boundary removes the translational invariance, it is not expected that the momentum should be conserved. The energy should be. One can expect that energy-like charges might continue to be conserved, while momentum-like charges cannot be. The way to find integrability in presence of boundary is, essentially, to generalize the standard Lax pairs approach including the boundary. Moreover, it is also requested, in order to have a complete description of a boundary IQFT, an inverse scattering procedure. It does not exist a systematic determination of integrability in presence of boundary and the full set of reflection factors for any specific integrable model is in general unknown. Anyway, many progresses have been done in the search for integrability and knowledge of reflection factors in the boundary integrable theories as sine-Gordon and affine Toda theories \cite{fring2,fring3,sasa}.   

\section{From Integrable models to critical points}
A link between IQFT coming from perturbations of a given CFT and factorizable scattering theories can be done by studying the finite size effects associated to theories due to the finiteness of space regions. What one usually works with is the scaling properties of a quantum field theory and, in particular, the scaling functions and their behaviour along the RG flow. For example, it can be useful to deal with special quantities, as the energy of a system, whose scaling in particular space regions of definition is well known. It has been noticed that, in a cylinder, a CFT develops a sort of Casimir effect due to the finiteness of the spatial extension \(L\), therefore, using formulae (\ref{eq:48}) and (\ref{eq:49}), it follows that for a conformal state \(\left|i\right\rangle\) the energy eigenvalues behave as

\begin{equation}
E_{i}=-\frac{\pi c_{i}}{6L}\;\;\;\;\textrm{with}\;\;\;\;c_{i}=c-12\left(\Delta_{i}+\bar{\Delta}_{i}\right)\label{eq:102}
\end{equation}
Of course, off-criticality this dependence of the energy on \(L\) might change. However, all the variations of the dependence can be put in a dimensionless function such that it reproduces (\ref{eq:102}) in the limit \(L\rightarrow 0\) because the dependence \(1/L\) is fixed by dimensional reasons. This allows to introduce the dimensionless parameter \(l=\mathcal{M}L\) and the scaling function \(c_{i}(l)\) which describe the behaviour of the energy state \(\left|i\right\rangle\) in the regions \(l\rightarrow 0\) and \(l\rightarrow \infty\) through a sort of generalized version of the Eq. (\ref{eq:102})

\begin{equation}   
E_{i}(l)=-\frac{\pi c_{i}(l)}{6L}\label{eq:103}
\end{equation}
with the request that \(c_{i}(l)\rightarrow c-12\left(\Delta_{i}+\bar{\Delta}_{i}\right)\) for \(l\rightarrow 0\). The presence of a mass parameter \(\mathcal{M}\) has not to be surprising because the function \(c_{i}(l)\) is describing the system flowing away from the critical point, it is thus natural to introduce a mass scale. Moreover, in this scheme, the IR and UV limits take a stronger physical meaning as well. The limit \(l\rightarrow 0\) (UV) can be also interpreted, taking the spatial extension \(L\) fixed, as a limit for very small masses, i.e. a very large energy scale system in which the specific masses have no more significance. This situation, in fact, is usually referred as UV (high energies) limit in quantum field theory. On the other hand, the limit \(l\rightarrow \infty\) can be seen as a physical limit where the masses are very large, acquiring, thus, a great relevance, i.e. one realizes a theory with several scattering massive particles. Let us notice that for a BCFT the relation (\ref{eq:102}) becomes 

\begin{equation}
E_{i}=-\frac{\pi c_{i}}{24L}\;\;\;\;\textrm{with}\;\;\;\;c_{i}=c-24\Delta_{i}\;.\label{eq:104}
\end{equation}
One way to compute the scaling functions is to make use of perturbation theory, and in our case of conformal perturbation theory. However, it is valid only in a neighbourood of a fixed point and, only sometimes, it allows to make contact between different fixed points, i.e. CFT's. The real aim is to link CFT to massive scattering quantum field theory, which are evidently not conformally invariant. This leads to the application of non-perturbative methods. These methods are essentially:

\begin{itemize}
\item the \emph{Truncated Conformal Space Approach} (TCSA) \cite{yu} consisting in diagonalizing the truncated Hamiltonian numerically. Even if it is non-perturbative and applicable also to non-intergrable systems, it is affected by numerical errors and it does not give any analytic control of the scaling functions;
\item the \emph{Thermodynamic Bethe Ansatz} (TBA) \cite{zam4} based on the idea of implementing the thermodynamics of a statistical system of particles interacting via a given \(S\) matrix. In particular, the TBA allows to calculate exactly the free energy of the system and other thermodynamical quantities as the entropy. Once the role of space and time have been exchanged, the finite size effects of the vacuum energy and first excited states \cite{dt1,dt2} energy appear and can be studied analytically;
\item the \emph{Nonlinear Integral Equation} (NLIE) \cite{ddv1,ddv2,kp,kbp}, which is an approach allowing to obtain scaling functions for a given IQFT starting from its definition as a quantum field theory regularized on a finite lattice.
\end{itemize}
In the next Chapter we will be introducing this last method for a particular integrable system and giving a description of its finite size scaling between the UV and IR limits.

\chapter{Boundary Sine-Gordon Theory}
The sine-Gordon model in a strip with Dirichlet boundary conditions is introduced.
Particular emphasis is given to its regularization on the lattice via an XXZ spin chain with magnetic fields at the boundaries. Bethe Ansatz equations and energy eigenvalues are derived as well.
\section{General results}
Quantum field theory with boundaries has been for many years an interesting subject of theoretical investigations and has been found to have many applications too, for example, in dissipative quantum systems \cite{exp1,exp2}. In particular, boundary IQFTs provide some useful tools as to study the structure of the space of boundary interactions, which have great significance in open String Theory \cite{exp3}, as to probe some fascinating physical phenomena in Condensed Matter Physics, for instance, the Kondo effect and the quantum Hall effect \cite{exp4,exp5}.

Let us consider the sine-Gordon model

\begin{equation}
\mathcal{A}_{SG}=\int_{-\infty}^{+\infty} \textrm{d}t\int_{-\infty}^{+\infty}\textrm{d}x \;a\left(\phi,\partial_{\mu}\phi\right)\label{eq:103pri}
\end{equation}
with

\[
a\left(\phi,\partial_{\mu} \phi\right)=\frac{1}{2}\left(\partial_{\mu}\phi \right)^{2}-\frac{m^{2}}{\beta^{2}}\cos\left( \beta \phi\right) \]
where \(\phi(x,t) \) is a scalar field and \(\beta\) a dimensionless coupling constant. The field theory in the presence of a boundary, say at \(x=0\), can be defined by the following action

\begin{equation}
\mathcal{A}=\int_{-\infty}^{+\infty} \textrm{d}t\int_{-\infty}^{0}\textrm{d}x \;a\left(\phi,\partial_{\mu}\phi\right)+\int_{-\infty}^{+\infty}\textrm{d}t\;\mathcal{B}\left(\phi(0,t),\partial_{t}\phi(0,t)\right)\label{eq:103}
\end{equation}
where the possibility of boundary degrees of freedom other than the boundary value of the bulk field \(\phi(x,t)\) has been discarded. It has been found in \cite{ghoshzam} that the boundary action preserving the integrability is of the form (\ref{eq:103}) with

\begin{equation}
\mathcal{B}\left(\phi(0,t),\partial_{t}\phi(0,t)\right)=-\mu \cos\left(\frac{\beta}{2}\left(\phi(0,t)-\phi_{0}\right)\right)\label{eq:103bis}
\end{equation}
where \(\mu\) and \(\phi_{0}\) are free parameters. The boundary reflection amplitude has been derived by using boundary Yang-Baxter equation \cite{ghoshzam,ghosh}. Exept for the case \(\mu=\infty\), the boundary value \(\phi(x=0,t)\) is not fixed in the boundary theory (\ref{eq:103}) and hence the topological charge

\begin{equation}
Q=\frac{\beta}{2\pi}\int _{-\infty}^{0}\textrm{d} x \frac{\partial}{\partial x}\phi(x,t)\label{eq:104}
\end{equation}
is not conserved. We will refer to the theory (\ref{eq:103}) with \(\mathcal{B}\left(\phi(0,t),\partial_{t}\phi(0,t)\right)\) given in (\ref{eq:103bis}) as \emph{boundary sine-Gordon} model.

Let us give a brief summary of the theory in the bulk, further details can be found in \cite{zamzam}. The bulk sine-Gordon model is known to be integrable at both the classical and quantum levels \cite{tak,kor}. The theory has an infinite number of degenerate vacua with the discrete symmetry \(\phi\rightarrow\phi+\frac{2\pi}{\beta}m\), with \(m\in\mathbb{Z}\). The spectrum consists of a soliton \(a\) and an antisoliton \(\bar{a}\) and a number of soliton - antisoliton bound states \(b_{n}\), \(n=1,2,\;.\;.\;.\;<\frac{1}{p}\), usually called \emph{breathers} with 

\begin{equation}
p=\frac{\beta^{2}}{8\pi-\beta^{2}}\label{eq:104bis}
\end{equation}
The soliton and the antisoliton have the same mass \(\mathcal{M}_{s}\), while the mass of the \(n^{\textrm{th}}\) breather is \(\mathcal{M}_{n}=2\mathcal{M}_{s}\sin\left(\frac{n\pi p}{2}\right)\). The topological charge of the soliton (antisoliton) is \(1\) (\(-1\)), while the breathers are neutral. It is always conserved. As a consequence of the integrability, there is no particle production, i.e. the scattering is factorized and, if one denotes the soliton \(S\) matrix as \(S_{cd}^{ab}(\theta)\) with \(a,b,c,d\) taking the value \(+\) (\(-\)) if the particle is a soliton (antisoliton), the non-zero amplitudes are \(S_{++}^{++}(\theta)=S_{--}^{--}(\theta)=a(\theta)\), which represent either the soliton - soliton scattering or the antisoliton - antisoliton scattering, \(S_{+-}^{+-}(\theta)=S_{-+}^{-+}(\theta)=b(\theta)\), which denotes the soliton - antisoliton transmission and finally the soliton - antisoliton reflection \(S_{+-}^{-+}(\theta)=S_{-+}^{+-}(\theta)=c(\theta)\), where the functions \(a(\theta)\), \(b(\theta)\) and \(c(\theta)\) explicitly read

\begin{equation}
\begin{cases}
a(\theta)=\sin\left(\frac{\pi-u}{p}\right)\rho(u) \\
b(\theta)=\sin\left(\frac{u}{p}\right)\rho(u) \\
c(\theta)=\sin\left(\frac{\pi}{p}\right)\rho(u) \end{cases}\label{eq:105}
\end{equation}
with \(u=-i\theta\) and \(\rho(u)\) can be written either in terms of Gamma functions as

\begin{equation}
\rho(u)=-\frac{1}{\pi}\Gamma\left(\frac{1}{p}\right)\Gamma\left(1-\frac{u}{p\pi}\right)\Gamma\left(1-\frac{1}{p}+\frac{u}{p\pi}\right)\prod_{l=1}^{\infty}\frac{F_{l}(u)F_{l}(\pi-u)}{F_{l}(0)F_{l}(\pi)}\label{eq:105bis}
\end{equation}
where

\[
F_{l}(u)=\frac{\Gamma\left(\frac{2l}{p}-\frac{u}{p\pi}\right)\Gamma\left(1+\frac{2l}{p}-\frac{u}{p\pi}\right)}{\Gamma\left(\frac{2l+1}{p}-\frac{u}{p\pi}\right)\Gamma\left(1+\frac{2l-1}{p}-\frac{u}{p\pi}\right)}\;, \]
or in terms of the Barnes' diperiodic sine function \(S_{2}\left(x\left|\;\omega_{1},\omega_{2}\right.\right)\) \cite{barnes1,barnes2}, which are meromorphic functions parametrized by the pair of quasiperiods \(\left(\omega_{1},\omega_{2}\right)\), as

\begin{equation}
\rho(u)=\left[\sin\left(\frac{u-\pi}{p}\right)\right]^{-1}\frac{S_{2}\left(\pi-u\;\left|\;p\pi,2\pi\right.\right) S_{2}\left(u\;\left|\;p\pi,2\pi\right.\right)}{S_{2}\left(\pi+u\;\left|\;p\pi,2\pi\right.\right)S_{2}\left(-u\;\left|\;p\pi,2\pi\right.\right)}.\label{eq:106}
\end{equation}
Let us mention that, due to the properties of Gamma functions, the function \(\rho(u)\) and, therefore, Eqs. (\ref{eq:105}) can be expressed in integral form. As an example, we write the soliton - soliton (antisoliton - antisoliton) scattering amplitude as

\begin{equation}
a(\theta)=\textrm{exp}\left[i\int\frac{\textrm{d}k}{2\pi k}e^{ik\theta}\frac{\sinh\frac{\pi}{2}(p-1)k}{2\sinh\frac{\pi}{2}pk\cosh\frac{\pi}{2}k}\right],\label{eq:106bis}
\end{equation}
this expression will be acquiring a great significance in next Chapters. The amplitudes \(b(\theta)\) and \(c(\theta)\) have simple poles at \(\theta=i(\pi-n\pi p)\) and \(a(\theta)\) and \(b(\theta)\) have poles at \(\theta=in\pi p\). They correpond to the creation of breathers \(b_{n}\) in the forward and cross channel respectively.

The model we are going to deal with is the special case with two boundaries, on the left and on the right of a semi-infinite strip, infinite in time direction and from \(0\) to \(L\) in space direction. The boundary conditions are a special restriction of the most general one (\ref{eq:103bis}) preserving the integrability, i.e. they are of Dirichlet types. We define the \emph{Dirichlet sine-Gordon} (DSG) model in a strip of lenght \(L\) through the Action 

\begin{equation}
\mathcal{A}_{DSG}= \frac{1}{2} \int_{-\infty}^{+\infty}\textrm{d}t\int _{0}^{L}\textrm{d}x\left[\left(\partial _{t}\phi \right)^{2}-\left(\partial _{x}\phi \right)^{2}+ \frac{m_{0}^{2}}{\beta^{2}}\cos \beta \phi \right]\label{eq:107}
\end{equation}
with the constraints fixing the value of the field at the boundaries:

\begin{equation}
\phi \left(0,t\right)\equiv \phi _{-}+\frac{2\pi}{\beta}m_{-}\;\;\;\textrm{and}\;\;\;\phi (L,t)\equiv\phi_{+}+\frac{2\pi}{\beta}m_{+}\label{eq:108}
\end{equation}
with \(m_{\pm}\in\mathbb{Z}\). In the Dirichlet diagonal case the topological charge can be written as

\begin{equation}
Q\equiv\frac{\beta}{2\pi}\int_{0}^{L}\textrm{d}x\frac{\partial}{\partial x}\phi(x,t)=\frac{\beta}{2\pi}\left(\phi_{+}-\phi_{-}\right)+m_{+}-m_{-}\label{eq:109}
\end{equation}
The model enjoies the discrete symmetry of the field \(\phi\rightarrow \phi+\frac{2\pi}{\beta}m\) and simultaneously \(\phi_{\pm}\rightarrow \phi_{\pm}+\frac{2\pi}{\beta}m\). The charge conjugation symmetry is also guaranteed provided \(\phi\rightarrow -\phi\) and \(\phi_{\pm}\rightarrow -\phi_{\pm}\) simultaneously. It sends \(Q\rightarrow -Q\), therefore, one can restrict attention to the study of positive \(Q\) and then act with this tranformation to obtain states with negative \(Q\). Then, only processes conserving the topological charge are allowed, in particular, any particle of the theory scattering off the boundaries has to conserve its topological charge. Observe that the periodicity allows to restrict the boundary parameters to the range \(0\le \phi_{\pm} \le \frac{2\pi}{\beta}\). Scattering off the boundary processes are described by the diagonal term of the most general boundary reflection amplitude found in \cite{ghoshzam}. Therefore, if one introduces the reflection factors for the solitons \(\mathcal{P}_{\pm}(u)\) (a soliton or an antisoliton incident on the boundary is reflected back unchanged) and \(\mathcal{Q}_{\pm}(u)\) (a soliton is reflected back as an antisoliton, or vice versa), one has that they take the values

\begin{equation}
\mathcal{Q}_{\pm}(u)=0\label{eq:110}
\end{equation}

\begin{equation}
\mathcal{P}_{\pm}(u)=R_{0}(u)\frac{S_{2}\left(\frac{\pi p}{2}\mp p\xi_{\pm}+\pi+u\;\left|\;p\pi,2\pi\right.\right)S_{2}\left(\frac{\pi p}{2}\mp p\xi_{\pm}-u\;\left|\;p\pi,2\pi\right.\right)}{S_{2}\left(\frac{\pi p}{2}\mp p\xi_{\pm}+\pi-u\;\left|\;p\pi,2\pi\right.\right)S_{2}\left(\frac{\pi p}{2}\mp p\xi_{\pm}+u\;\left|\;p\pi,2\pi\right.\right)}\label{eq:111}
\end{equation}
where 

\begin{equation}
R_{0}=\frac{S_{2}\left(\frac{\pi}{2}-u\;\left|\frac{\pi p}{2},2\pi\right.\right)S_{2}\left(\frac{\pi p}{2}+u\;\left|\frac{\pi p}{2},2\pi\right.\right)}{S_{2}\left(\frac{\pi}{2}+u\;\left|\frac{\pi p}{2},2\pi\right.\right)S_{2}\left(\frac{\pi p}{2}-u\;\left|\frac{\pi p}{2},2\pi\right.\right)} \label{eq:112}
\end{equation}

\begin{equation}
\xi_{\pm}=\frac{4\pi\phi_{\pm}}{\beta}\label{eq:113}
\end{equation}
and \(S_{2}\left(x\;\left|\;\omega_{1},\omega_{2}\right.\right)\) are the Barnes' diperiodic functions introduced above. It has been shown \cite{dm} that there exists a quite intricate set of poles of the diagonal reflection factors \(\mathcal{P}_{\pm}(u)\) on the imaginary axis, which correspond to boundary bound states. In particular, it has been stated that the boundary bound states exist only for particular values of the topological charge and they are represented by the set of poles 

\begin{equation}
\left\{\nu_{n_{i}},w_{m_{i}}:\;\frac{\pi}{2}>\nu_{n_{1}}>w_{m_{1}}>\nu_{n_{2}}>\;.\;.\;.>0\right\}\label{eq:114}
\end{equation}
where
\begin{equation}
\nu_{n}=p\xi_{\pm}-\frac{\pi}{2}(2n+1)p\;\;\;\textrm{with}\;\;\;n=0,1,2,\;.\;.\;.\label{eq:115}
\end{equation}

\begin{equation}
w_{m}=\pi-p\xi_{\pm}-\frac{\pi}{2}(2m-1)p\;\;\;\textrm{with}\;\;\;m=1,2,\;.\;.\;.\label{eq:116}
\end{equation}
In order to investigate all the RG flow of this theory, from the IR limit, i.e. the scattering theory limit, to the UV limit, i.e. the conformal limit, and associate states of the scattering particles interpretation to states of the \(c=1\) conformal field theory, we will construct the DSG model as the continuum limit of a lattice model by an inhomogeneus XXZ spin chain with fixed boundary magnetic fields, i.e. diagonal reflection amplitudes on the boundaries, in a Hamiltonian formalism.

\section{Boundary XXZ model}

The homogeneous antiferromagnetic XXZ spin \(\frac{1}{2}\) model in a chain of $N$ sites spaced by \(a\), coupled to magnetic fields $h_{-}$ and $h_{+}$ at the
left and right boundaries respectively, has Hamiltonian

\begin{equation}
a\mathcal{H}\left(h_{+},h_{-}\right)=-J\sum_{n=1}^{N-1} \left(\sigma^{x}_{n}\sigma^{x}_{n+1}+\sigma^{y}_{n}\sigma^{y}_{n+1}+\cos \gamma \sigma^{z}_{n}\sigma^{z}_{n+1}\right)+h_{-}\sigma_{1}^{z}+h_{+}\sigma_{N}^{z}\label{eq:117}
\end{equation}
Here $\sigma _{n}^{\alpha }$, $\alpha =x,y,z$ are Pauli matrices
acting on the $n$-th site, so that

\[
\left[\sigma _{n}^{\alpha },\sigma _{m}^{\beta }\right]=2i\varepsilon ^{\alpha \beta \gamma }\sigma _{n}^{\gamma }\delta _{nm}\]
$0\leq \gamma <\pi $, so that the isotropic case XXX is reproduced for $\gamma =0$. The parameter $\gamma $ is often referred as \emph{anisotropy} of the chain. The model (\ref{eq:117}) is invariant under global \( U(1)\) rotations of the spins (\(\sigma_{n}^{\pm}=\frac{1}{2}\left(\sigma_{n}^{x}\pm\sigma_{n}^{y}\right)\))

\[
\sigma_{n}^{\pm}\rightarrow e^{\pm i\omega}\sigma_{n}^{\pm}\;\;\;,\;\;\;\sigma_{n}^{z}\rightarrow \sigma_{n}^{z} \]
conserving the third component of the total spin 

\[
S=\frac{1}{2}\sum_{n=1}^{N} \sigma_{n}^{z} \in \frac{\mathbb{Z}}{2}\;\;\;,\;\;\;\left[\mathcal{H},S\right]=0\]
Therefore, states can be classified according to their value of \(S\), which in a chain of \(N\) sites can take all values \(S=-\frac{N}{2}, -\frac{N}{2}+1,...,\frac{N}{2}-1,\frac{N}{2}\).

The transformation 

\[
\mathcal{C}:\;\sigma_{n}^{\pm}\rightarrow \sigma_{n}^{\mp}\;\;,\;\;\sigma_{n}^{z}\rightarrow -\sigma_{n}^{z}\;\;,\;\; h_{\pm}\rightarrow -h_{\pm}\]
is another important symmetry, under which 

\[
\mathcal{C}^{-1}\mathcal{H}(h_{+},h_{-})\mathcal{C}=\mathcal{H}(-h_{+},-h_{-})\;\;\;\textrm{and}\;\;\; \mathcal{C}^{-1}S\mathcal{C}=-S\]
which means that states of a Hamiltonian with given \(h_{+}\) and \(h_{-}\) with negative spin are equivalent to states with positive \(S\) of the corresponding Hamiltonian at the same \(\gamma\) but with reversed boundary magnetic fields. In this sense one can decide to restrict to consider \(S\ge 0\) states only, as the negative ones will be obtained by just changing sign to both the boundary magnetic fields in all formulae.

\begin{figure}
\begin{center}
\includegraphics[angle=0, width=0.7\textwidth]{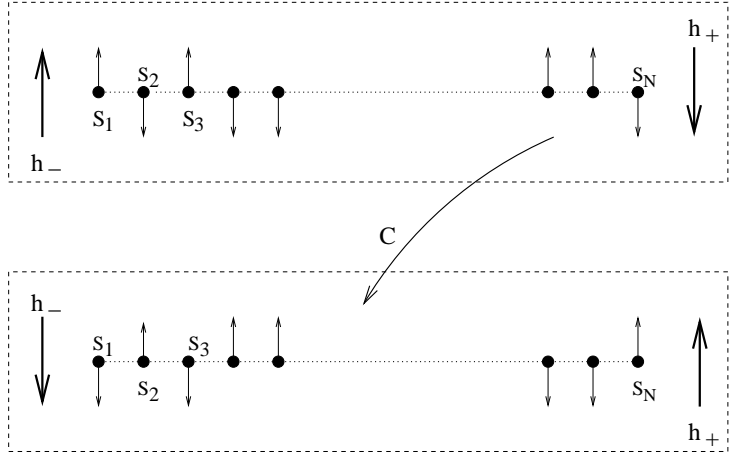}
\caption{\emph{Pictorial representation of the XXZ spin chain. The system is invariant under the transfotmation \(\mathcal{C}^{-1}\mathcal{H}(h_{+},h_{-})\mathcal{C}\) and \(\mathcal{C}^{-1}S\mathcal{C}\)}.}
\end{center}
\end{figure}
However, unlike the case of a single boundary and a semi-infinite chain, this symmetry is not enough to allow to restrict to both \(h_{\pm}>0\). Indeed it is physically evident that the case with parallel magnetic fields (\(h_{\pm}\ge 0\)) is different from the antiparallel one (\(h_{+}\ge 0\;,\;h_{-}\le 0\) for example). The only allowed inversion of magnetic fields is the \emph{simultaneous} one. This symmetry will be related to the charge conjugation of the DSG model. In what follows we will consider the antiparallel case and choose \(h_{-}\ge 0\) and \(h_{+}\le 0\).

\subsection{Double row transfer matrix}
The Hamiltonian (\ref{eq:117}), as well as its inhomogeneous generalizations, can be constructed in a transfer matrix framework. Together with the two dimensional spaces $\mathcal{V}_{n}$ on which the $\sigma _{n}^{\alpha }$'s act, consider an auxiliary space $\mathcal{V}_{0}$, isomorphic to the $\mathcal{V}_{n}$'s. Introduce the $sl(2)_{q}$ $R$-matrix

\[
R_{n,m}(u)=\frac{1+c(u)}{2}\mathbf{1}_{n}\mathbf{1}_{m}+\frac{1-c(u)}{2}\sigma _{n}^{z}\sigma _{m}^{z}+b(u)\left[\sigma _{n}^{+}\sigma _{m}^{-}+\sigma _{n}^{-}\sigma _{m}^{+}\right]\]
and the quantum Lax operator

\[
L_{n}(u)=P_{n,0}R_{n,0}(u) \]
where $P_{n,m}$ is the permutation matrix $P_{n,m}=\sigma _{n}^{+}\sigma _{m}^{-}+\sigma _{n}^{-}\sigma _{m}^{+}$.
The $u_{n}$'s are fixed parameters called inhomogeneities and

\[
b(u)=\frac{\sinh u}{\sinh (u+i\gamma )}\, \, ,\quad c(u)=\frac{i\sin \gamma }{\sinh (u+i\gamma )}\]
$R_{n,m}(u)$ satisfies the Yang-Baxter equation.

Define the \emph{monodromy matrix} 

\[
\mathbf{T}(u|\mathbf{u})=\prod _{n=1}^{N-1}L_{n}(u)=\prod _{n=1}^{N-1}R_{n,n-1}(u-u_{n})=\left(\begin{array}{cc}
 A(u) & B(u)\\
 C(u) & D(u)\end{array}\right)\]
which is a $2\times 2$ matrix in the $\mathcal{V}_{0}$ space, whose
elements are operators in the Hilbert space $\mathcal{V}=\textrm{ }\bigotimes _{n=1}^{N-1}\mathcal{V}_{n}$,
and $\mathbf{u}=(u_{1},...,u_{n})$. The trace in the $\mathcal{V}_{0}$
space of $\mathbf{T}$

\[
\mathbf{t}(u|\mathbf{u)}=\mathrm{Tr}_{0}\mathbf{T}(u|\mathbf{u)}=A(u)+B(u)\]
is well known to be an operator in $\mathcal{V}$ called \emph{row
to row transfer matrix.} As a consequence of the Yang-Baxter equation,
on a periodic chain

\[
[\mathbf{t}(u|\mathbf{u}),\mathbf{t}(v|\mathbf{u})]=0\quad ,\, \forall u,v\in \mathbb{C}\]
In the case of non trivial boundaries, however, this row to row transfer
matrix is not enough to define the integrals of motion. One has to
introduce a more complicated object that takes boundary effects into
account. Introduce also the boundary \(K\)-matrices \(K_{\pm}(u)=K\left(u,h_{\pm}\right)\) in \(\mathcal{V}_{0}\) where

\[
K(u,h)=\left(\cosh u\right)\mathbf{1}_{0}+i\left(h\sinh u\right)\sigma_{0}^{z} \]
which satisfy the boundary Yang Baxter equations. Define the $2\times 2$ matrix in $\mathcal{V}_{0}$
\[
\mathbf{U}(u|\mathbf{u})=\mathbf{T}(u|\mathbf{u)}K_{+}(u)\mathbf{T}^{-1}(-u|\mathbf{u})=\left(\begin{array}{cc}
 \mathcal{A}(u) & \mathcal{B}(u)\\
 \mathcal{C}(u) & \mathcal{D}(u)\end{array}\right)\]
In terms of this object the \emph{double row transfer matrix} can
be introduced

\[
\mathbf{d}(u|\mathbf{u})=\mathrm{Tr}_{0}K_{-}(u)\mathbf{U}(u|\mathbf{u})\]
having the fundamental integrability property that $[\mathbf{d}(u|\mathbf{u}),\mathbf{d}(v|\mathbf{u})]=0$,
$\forall u,v\in \mathbb{C}$. In the homogeneous case $\mathbf{u}=0$,
the Hamiltonian (\ref{eq:117}) is well known to be reproduced by:

\[
a\mathcal{H}=\left.\frac{\textrm{d}}{\textrm{d}u}\log \mathbf{d}(u,0)\right|_{u=0}=\sum _{n=1}^{N}\dot{L}_{n}(0)+\frac{1}{2}\dot{K}_{+}(0)+\frac{\mathrm{Tr}_{0}K_{0}(0)\dot{L}_{N}(0)}{\mathrm{Tr}K_{+}(0)}\]
where dots denote derivatives with respect to $u$, and the label
0 refers to the auxiliary space (for details see \cite{skly}).

The bare continuum limit $N\to \infty $, $a\to 0$ while $Na=L$
remains fixed, is known to give the Hamiltonian of a massless free
boson $\phi (x)$ compactified on a circle of radius $R=\frac{\sqrt{\pi }}{2(\pi -\gamma )}$ \cite{lut}.

Among the many possible deformations of the Hamiltonian (\ref{eq:117})
leading to sine-Gordon in the bare continuum limit, we choose the
one introducing alternating inhomogeneities $u_{n}=(-1)^{n}\Lambda $
in the sites of the chain. This construction has the advantage to preserve integrability of the
lattice model, thus allowing to use the Bethe Ansatz techniques needed
for our purposes. It has been known for sometime \cite{ddv4,resh}
to give a correct construction of sine-Gordon model in the bulk in
cylindrical geometry, when the appropriate scaling limit is chosen,
with periodic or twisted boundary conditions. It is then natural to
expect that the same construction in the presence of boundary magnetic
fields $h_{\pm }$ can also provide an effective tool to define the
renormalized DSG theory in a strip of lenght $L$ with fixed conditions
at the boundaries. It is worthwhile to recall that, as the homogeneous
XXZ chain is equivalent to a 6-vertex model on a square lattice, this
modified XXZ chain is also equivalent to a 6-vertex model, but --
as a consequence of the introduced inhomogeneities -- defined on a
lattice rotated by \(45^{\circ}\), i.e. on what can be thought as a Minkovski
space discretized along the light-cone directions. This is why this
construction is often referred as \emph{light cone lattice} construction
of the sine-Gordon model \cite{ddv4}.

The double row transfer matrix, in this case \cite{skly}, becomes

\begin{equation}
\mathbf{d}(u\left|\mathbf{u}\right.)=\mathbf{d}\left(u,\Lambda\right)=\hat{K}_{N}(u,h_{+})T\left(u,\Lambda\right)K_{1}\left(u,h_{-}\right)\hat{T}\left(u,\Lambda\right)\label{eq:118}
\end{equation}
where

\[
\hat{K}_{N}(u,h_{+})=\textrm{Tr}_{0}\left[R_{N,0}\left(u+(-1)^{N}\Lambda\right)K(u+i\gamma,h_{+})R_{N,0}\left(u-(-1)^{N}\Lambda\right)\right] \]

\[
T\left(u,\Lambda\right)=\prod_{n=1}^{N-1} R_{N-n+1,N-n}\left(u-(-1)^{N-n}\Lambda\right) \]

\[
\hat{T}\left(u,\Lambda\right)=\prod_{n=1}^{N-1} R_{n,n+1}\left(u+(-1)^{n}\Lambda\right) \]
It is possible to demonstrate that the transfer matrix (\ref{eq:118}) is an even function of \( \Lambda\) up to an irrelevant multiplicative function of \(\Lambda\) only, indeed, simplifying the quantity \(\hat{K}_{N}\left(u,h_{+}\right)\), one easily gets 

\begin{equation}
\hat{K}_{N}\left(u,h_{+}\right)=f\left(u,\Lambda\right)K_{N}\left(u+i\gamma,h_{+}\right)\label{eq:119}
\end{equation}
where \( f\left(u,\Lambda\right)=1+c\left(u-(-1)^{N}\Lambda\right)c\left(u+(-1)^{N}\Lambda\right) \) is evidently a numerical function. Then define

\begin{equation}
\mathbf{\tau}\left(u,\Lambda\right)=K_{N}\left(u+i\gamma,h_{+}\right)T\left(u,\Lambda\right)K_{1}\left(u,h_{-}\right)\hat T\left(u,\Lambda\right)\label{eq:120}
\end{equation}
and observe that \( \mathbf{d}\left(u,\Lambda\right)=f\left(u,\Lambda\right)\mathbf{\tau}\left(u,\Lambda\right) \). The following identity holds

\begin{equation}
\mathbf{\tau}^{-1}\left(-u,\Lambda\right)\mathbf{\tau}^{\prime}\left(-u,\Lambda\right)=\mathbf{\tau}^{-1}\left(u,\Lambda\right)\mathbf{\tau}^{\prime}\left(u,\Lambda\right)\label{eq:121}
\end{equation}
In fact, given \( \mathbf{\tau}\left(u,\Lambda\right) \) and making use of the fact that the matrix \( K_{N}\left(u+i\gamma,h_{+}\right) \) is diagonal, one has 

\begin{align}
\mathbf{\tau}\left(u,\Lambda\right) & =K_{N}\left(u+i\gamma,h_{+}\right)T\left(u,\Lambda\right)K_{1}\left(u,h_{-}\right)\hat T\left(u,\Lambda\right)\nonumber \\
&=T\left(u,\Lambda\right)K_{1}\left(u,h_{-}\right)\hat T\left(u,\Lambda\right)K_{N}\left(u+i\gamma,h_{+}\right)\label{eq:122} 
\end{align}
Now, due to the properties of \( R\) matrix and \( K\) matrices

\[
\begin{cases}
R_{i,j}(-u)=R_{i,j}^{-1}(u)\\
R_{i,j}(u)=R_{j,i}(u)\end{cases}\;,\;\;\begin{cases}
K_{j}\left(-u,h_{\pm}\right)=K_{j}^{-1}\left(u,h_{\pm}\right)\\
K_{j}\left(u,h_{\pm}\right)=K_{j}^{-1}\left(-u-2i\gamma,h_{\pm}\right)\end{cases} \]
one performs the following computation 

\begin{align}
\mathbf{\tau}\left(-u,\Lambda\right)& = K_{N}\left(-u+i\gamma,h_{+}\right)T\left(-u,\Lambda\right)K_{1}\left(-u,h_{-}\right)\hat T\left(-u,\Lambda\right)\nonumber \\
&=T\left(-u,\Lambda\right)K_{1}\left(-u,h_{-}\right)\hat T\left(-u,\Lambda\right)K_{N}\left(-u+i\gamma,h_{+}\right)\nonumber \\
&=\left[R_{N-1,N}\left(-u-(-1)^{N-1}\Lambda\right)\;.\; .\; .\;R_{1,2}\left(-u+\Lambda\right)K_{1}\left(-u,h_{+} \right)\right]\times \nonumber \\
&\times\left[R_{1,2}\left(-u-\Lambda\right)\; .\; .\; .\; R_{N-1,2N}\left(-u+(-1)^{N-1}\Lambda\right)K_{N}\left(-u+i\gamma,h_{+}\right)\right]\nonumber \\
&=\left[R_{N-1,N}^{-1}\left(u+(-1)^{N-1}\Lambda\right)\;.\; .\; .\;R_{1,2}^{-1}\left(u-\Lambda\right)K_{1}^{-1}\left(u,h_{-} \right)\right]\times \nonumber \\
&\times\left[R_{1,2}^{-1}\left(u+\Lambda\right)\; .\; .\; .\; R_{N-1,N}^{-1}\left(u-(-1)^{N-1}\Lambda\right)K_{N}^{-1}\left(u-i\gamma,h_{+}\right)\right]\nonumber \\
&=\left[R_{N-1,N}^{-1}\left(u+(-1)^{N-1}\Lambda\right)\;.\; .\; .\;R_{1,2}^{-1}\left(u-\Lambda\right)K_{1}^{-1}\left(u,h_{-} \right)\right]\times \nonumber \\
&\times\left[R_{1,2}^{-1}\left(u+\Lambda\right)\; .\; .\; .\; R_{N-1,N}^{-1}\left(u-(-1)^{N-1}\Lambda\right)K_{N}^{-1}\left(u+i\gamma,h_{+}\right)\right]\nonumber \\
&=\left[K_{N}\left(u+i\gamma,h_{+}\right)T\left(u,\Lambda\right)K_{1}\left(u,h_{-}\right)\hat T\left(u,\Lambda\right)\right]^{-1}\nonumber \\
&=\mathbf{\tau}^{-1}\left(u,\Lambda\right)\nonumber
\end{align}
and, taking the logarithm,

\begin{equation}
\log \mathbf{\tau}\left(-u,\Lambda\right)=-\log \mathbf{\tau}\left(u,\Lambda\right)\label{eq:125}
\end{equation}
which proofs the statement (\ref{eq:121}). This means that the logarithmic derivative of the operator \( \mathbf{\tau}\left(u,\Lambda\right) \) is an even function. So we have that

\begin{equation}
\frac{\textrm{d}}{\textrm{d}u}\log \mathbf{d}\left(u,\Lambda\right)=\frac{\textrm{d}}{\textrm{d}u}\log f\left(u,\Lambda\right)+\frac{\textrm{d}}{\textrm{d}u}\log \mathbf{\tau}\left(u,\Lambda\right)\label{eq:126}
\end{equation}
because \( f\left(u,\Lambda\right) \) is a numerical function. At this point all the commuting integrals of motion can be defined in terms of \(\mathbf{d}\left(u,\Lambda\right)\) as

\begin{equation}
\mathcal{Q}_{(k)}=\left.\frac{i^{k}}{a}\frac{\textrm{d}^{k}}{\textrm{d} u^{k}} \log\mathbf{d}\left(u,\Lambda\right)\right|_{u=\Lambda}\label{eq:126bis} 
\end{equation}
and, in particular, the Hamiltonian turns out to be

\begin{equation}
\mathcal{H}=\mathcal{Q}_{(1)}=\left.\frac{i}{a}\frac{\textrm{d}}{\textrm{d} u} \log\mathbf{d}\left(u,\Lambda\right)\right|_{u=\Lambda}\label{eq:126tris} 
\end{equation}
An explicit expression for our case with diagonal reflection matrices is

\begin{align}
-ia\mathcal{H} & =\sum_{n=1}^{N-1}R^{-1}_{n,n+1}\left(2\Lambda\right)\dot{R}_{n,n+1}\left(2\Lambda\right)+\nonumber \\
& +\sum_{n=\textrm{even}}^{N-1}R^{-1}_{n,n+1}\left(2\Lambda\right)\dot{R}_{n-1,n}(0)R_{n,n+1}\left(2\Lambda\right)+\nonumber \\
& +\sum_{n=\textrm{odd}}^{N-1}R^{-1}_{n,n+1}\left(2\Lambda\right)\dot{R}_{n+1,n+2}(0)R_{n,n+1}\left(2\Lambda\right)+\nonumber \\
& +K_{-}^{-1}\left(\Lambda\right)\dot{K}_{-}\left(\Lambda\right)+K_{+}^{-1}\left(\Lambda+i\gamma\right)\dot{K}_{+}\left(\Lambda+i\gamma\right) \label{eq:126quater}
\end{align}
For \(\Lambda \rightarrow 0\) the Hamiltonian (\ref{eq:117}) is reproduced.

When \(\Lambda\rightarrow \infty\), \(N\rightarrow \infty \) and \(a\rightarrow 0\), while \(L=Na\) stays fixed, contact can be made, along lines similar to those in \cite{ds}, with the lagrangean formulation of DSG model. The XXZ anysotopy $\gamma $ is related to the SG coupling $\beta $
by $\beta ^{2}=8(\pi -\gamma )$. It is often convenient to use the
parameter \(p\) defined in (\ref{eq:104bis}): $p=\frac{\pi }{\gamma }-1$, $(0<p<\infty )$.

\section{Bethe Ansatz equations}

The Bethe Ansatz equations for the boundary XXZ chain (\ref{eq:117})
have been written some years ago \cite{skly,alk}, using
an algebraic approach. It is straightforward to generalize them for Eq. (\ref{eq:126quater}) with the introduction of the alternating inhomogeneities. Eigenstates of the transfer matrix $\mathbf{d}(u,\Lambda )$ can be constructed applying repeatedly $\mathcal{B}(u_{j})$ operators to the ferromagnetic vacuum $|\Omega \rangle $

\begin{equation}
\mathcal{B}(u_{1})\;.\;.\;.\;\mathcal{B}(u_{M})|\Omega \rangle \label{eq:127}
\end{equation}
The all dinstinct numbers $u_{1},.\;.\;.\;,u_{M}$ are called \emph{roots.}
They are in number of $M$, ($M\leq N$) and must satisfy the Bethe
Ansatz equations 

\begin{align}
\left[s_{1/2}\left(\vartheta_{j}+\Theta\right)\right. & \left.s_{1/2}\left(\vartheta_{j}-\Theta\right)\right]^{N}s_{H_{+}/2}\left(\vartheta_{j}\right)s_{H_{-}/2}\left(\vartheta_{j}\right)=\nonumber \\
&=\prod_{k=1,k\ne j}^{M}s_{1}\left(\vartheta_{j}-\vartheta_{k}\right)s_{1}\left(\vartheta_{j}+\vartheta_{k}\right)\label{eq:128}
\end{align}
where

\begin{equation}
s_{\nu}(\vartheta)\equiv\frac{\sinh\frac{\gamma}{\pi}(\vartheta+i\nu\pi)}{\sinh\frac{\gamma}{\pi}(\vartheta-i\nu\pi)}\;\;\;\;\;\textrm{and}\;\;\;\;\;\vartheta=\frac{\pi}{\gamma}u\;\;,\;\;\Theta=\frac{\pi}{\gamma}\Lambda\label{eq:129}
\end{equation}
The parameter \(H_{\pm}\) is defined such that \(h_{\pm}=\sin \gamma\cot \frac{\gamma}{2}\left(H_{\pm}+1\right)\) and we choose as fundamental domain \(-p-1<H_{\pm}<p+1\). Notice that for \(H_{\pm}=0\), i.e. \(h_{\pm}=1+\cos \gamma\equiv h_{c}\), the boundary terms in the Bethe equations disappear and the theory becomes \(SL_{q}(2)\) invariant \cite{pas,ddv5}.

The antiferromagnetic vacuum turns out to be a set \(M=\frac{N}{2}\) of real roots and it exists for \(N\) even only. In the region \(0\le \gamma<\pi\) and for small enough boundary magnetic fields, this is the true ground state of the model. For \(N\) odd instead the states with lowest possible total spin have \(M=\frac{N-1}{2}\) roots and one hole. However, to deal correctly with the continuum limit one has to consider both \(N\) even and odd sectors. The symmetry of \(\left\{\vartheta_{j}\right\}\rightarrow \left\{-\vartheta_{j}\right\}\), evident from the Bethe equations, implies that only roots with positive real part are independent parameters characterizing the Bethe states. The value \(\vartheta_{j}=0\) is a solution of Eqs. (\ref{eq:128}) for any \(N\) and \(M\). However, the corresponding Bethe state would vanish, one has to subtract this unwanted root, i.e. to create a hole at \(\vartheta=0\).

The domain of roots distribution can be considered as a semistrip \(\mathbb{U}_{+}\) of the complex \(\vartheta\)-plane

\begin{align}
\mathbb{U}_{+} & =\left\{\vartheta\in\mathbb{C}:\;\mathfrak{Re} (\vartheta)>0\;,\;-\frac{\pi^{2}}{2\gamma}<\mathfrak{Im} (\vartheta)\le \frac{\pi^{2}}{2\gamma}\right.\;\;\;\;\textrm{or}\nonumber \\
&\qquad\qquad\qquad\qquad\qquad\left.\mathfrak{Re} (\vartheta)=0\;,\;0<\mathfrak{Im} (\vartheta) < \frac{\pi^{2}}{2\gamma}\right\}\label{eq:130}
\end{align}
This excludes another unwanted root at \(i\frac{\pi^{2}}{2\gamma}\) as well and considers only half of the imaginary axis, as it should for symmetry. For computational purposes, it is useful to double this strip by mirroring all the roots

\begin{equation}
\mathbb{U}=\left\{\vartheta\in\mathbb{C}:\;\mathfrak{Re} (\vartheta)\in \mathbb{R}\;,\;-\frac{\pi^{2}}{2\gamma}<\mathfrak{Im} (\vartheta)\le\frac{\pi^{2}}{2\gamma}\right\}\label{eq:131}
\end{equation}
where to each root \(\vartheta_{j}\) is associated its mirror root \(\vartheta_{-j}=-\vartheta_{j}\). Define the function

\begin{equation}
\varphi_{\nu}(\vartheta)=\pi+i\log s_{\nu}(\vartheta)\label{eq:132}
\end{equation}
with the oddity condition \(\varphi_{\nu}(-\vartheta)=-\varphi_{\nu}(\vartheta)\) fixing the fundamental branch of the logarithm. It is periodic in the imaginary axis, with period \(i\frac{\pi^{2}}{\gamma}\) and real on the real axis. We choose as fundamental peridicity the strip \(\vartheta\in\mathbb{U}\). Singularities of this function appear along the imaginary axis:

\begin{equation}
\mathfrak{Re} (\vartheta)=0\;\;\;,\;\;\;\mathfrak{Im} (\vartheta)=\pm\pi(k(p+1)-\nu)\;\;,\;\;k\in\mathbb{Z}\label{eq:133}
\end{equation}
The fundamental analyticity strip is limited to \(\left|\mathfrak{Im} (\vartheta)\right|<\pi\textrm{min}(\nu,p+1-\nu)\). In terms of the function (\ref{eq:132}) the logarithm of the Bethe Ansatz equations can be expressed as

\begin{align}
&N\left[\varphi_{1/2}\left(\vartheta_{j}+\Theta\right)+\varphi_{1/2}\left(\vartheta_{j}-\Theta\right)\right]+\varphi_{H_{+}/2}\left(\vartheta_{j}\right)+\varphi_{H_{-}/2}\left(\vartheta_{j}\right)+\nonumber \\
&+\varphi_{1}\left(\vartheta_{j}\right)+\varphi_{1}\left(2\vartheta_{j}\right)-\sum_{k=-M}^{M}\varphi_{1}\left(\vartheta_{j}-\vartheta_{k}\right)=2\pi I_{j}\;\;\;,\;\;\;I_{j}\in\mathbb{Z}\label{eq:134}
\end{align}

\subsection{Energy}

An useful result can be achieved from the diagonalization of the double row transfer matrix (\ref{eq:118}): eigenvalues of the Hamiltonian can be expressed in terms of solutions of Bethe equations (\ref{eq:128}), due to the formula (\ref{eq:126tris}). Therefore, given a Bethe state characterized by \(M\) roots in the set \(\mathbb{U}\), the corresponding energy turns out to be

\begin{equation}
E=-\frac{1}{2a}\sum_{k=1}^{M}\frac{\textrm{d}}{\textrm{d}\Theta}\left[\varphi_{1/2}\left(\Theta-\vartheta_{k}\right)+\varphi_{1/2}\left(\Theta+\vartheta_{k}\right)\right]+\frac{1}{a}\frac{\textrm{d}}{\textrm{d}\Theta}\varphi_{1/2}(\Theta)\label{eq:135}
\end{equation}

\chapter{Nonlinear Integral Equation}
In this Chapter the fundamental Nonlinear Integral Equation for the Bethe Ansatz is derived following the approach introduced in \cite{ddv1,ddv2,kp,kbp}. In the last section a general formula for the energy of the model is given, it allows to define and study scaling functions in the next Chapter.

\section{Counting function}
Define for \(\vartheta\in \mathbb{U}\) the \emph{counting function}

\begin{align}
Z_{N}(\vartheta) &=N\left[\varphi_{1/2}\left(\vartheta+\Theta\right)+\varphi_{1/2}\left(\vartheta-\Theta\right)\right]+\varphi_{H_{+}/2}(\vartheta)\nonumber \\
&+\varphi_{H_{-}/2}(\vartheta)-\sum_{k=-M}^{M}\varphi_{1}\left(\vartheta-\vartheta_{k}\right)+\varphi_{1}(\vartheta)+\varphi_{1}(2\vartheta)\label{eq:136} 
\end{align}
where the function \(\varphi_{\nu}(\vartheta)\) has been defined in the last Chapter and the values \(\vartheta_{k},\;k=-M,-M+1,\;.\;.\;.M-1,M\), are solutions of the Bethe Ansatz equations (\ref{eq:128}). It is evident that in terms of the counting function \(Z_{N}(\vartheta)\) the logarithm of the Bethe equations simply becomes 

\begin{equation}
Z_{N}\left(\vartheta_{j}\right)=2\pi I_{j}.\label{eq:137}
\end{equation}
The last two terms in Eq. (\ref{eq:136}) take care of the fact that in the second member of (\ref{eq:128}) it is not included the factor with \(k=j\) and that the root \(\vartheta_{0}=0\) is an unwanted solution and it has to be subtracted. Eq. (\ref{eq:137}) plays the role of a quantization condition for Bethe Ansatz equations roots, therefore, in the following, we will refer to the integers \(I_{j}\) as quantum numbers. 

\subsection{Bethe roots}
The analytic structure of \(Z_{N}(\vartheta)\) suggests to classify the roots and related objects of the Bethe equations as follows

\begin{itemize}
\item \emph{real roots} \(x_{k},\;k=1,2,\;.\;.\;.\; M_{R},\;x_{k}>0\): strictly positive real solutions of Eqs. (\ref{eq:128}) and (\ref{eq:137});

\item \emph{holes} \(h_{k},\;k=1,2,\;.\;.\;.\; N_{H},\;h_{k}>0\): strictly positive real soltutions of (\ref{eq:137}) that are \textbf{not} solutions of (\ref{eq:128});

\item \emph{close roots} \(c_{k},\;k=1,2,\;.\;.\;.\; M_{C}\): complex solutions of Eqs. (\ref{eq:128}) and (\ref{eq:137}) with \(\mathfrak{Re} (c_{k})\ge 0\) and imaginary part in the range \(0<\left|\mathfrak{Im} (c_{k})\right|<\textrm{min}(\gamma,\pi-\gamma)\);

\item \emph{wide roots} \(w_{k},\;k=1,2,\;.\;.\;.\; M_{W}\): complex conjugate solutions with \(\mathfrak{Re} (w_{k})\ge 0\) and \(\textrm{min}(\gamma,\pi-\gamma)<\left|\mathfrak{Im} (w_{k})\right|<\frac{\pi}{2}\).
\end{itemize}
It is also useful to define \(M_{W}^{\uparrow}\), the number of wide roots with positive imaginary part and \(M_{W}^{\downarrow}\) those with negative imaginary part. Clearly, it holds \(M_{W}^{\uparrow}+M_{W}^{\downarrow}=M_{W}\), while \(M_{W}^{\uparrow}-M_{W}^{\downarrow}=M_{SC}\) is the number of self-conjugate roots with \(\mathfrak{Im}(w_{k})=\frac{\pi}{2}\), whose complex conjugate is the root itself, due to the \(i\pi\) periodicity of Bethe Ansatz equations.

The function \(Z_{N}(\vartheta)\) for \(\vartheta\in\mathbb{R}\) is globally monotonically increasing. However, there may be points where locally \(\dot{Z}_{N}(\vartheta)<0\). In particular, one can call \emph{special} roots or holes \(s_{k}\) the objects such that \(\dot{Z}_{N}(s_{k})<0\). If such a special object appears, there must be two other objects (either roots or holes) with the same quantum number as well. This is imposed by the global increasing monotonicity of the counting function. Moreover, as two roots with the same quantum number are not allowed in the Bethe equations, only one of this triple can be a root, the others are forced to be holes. The number of special objects will be indicated as \(N_{S}\).

\begin{figure}
\begin{center}
\includegraphics[angle=0, width=0.7\textwidth]{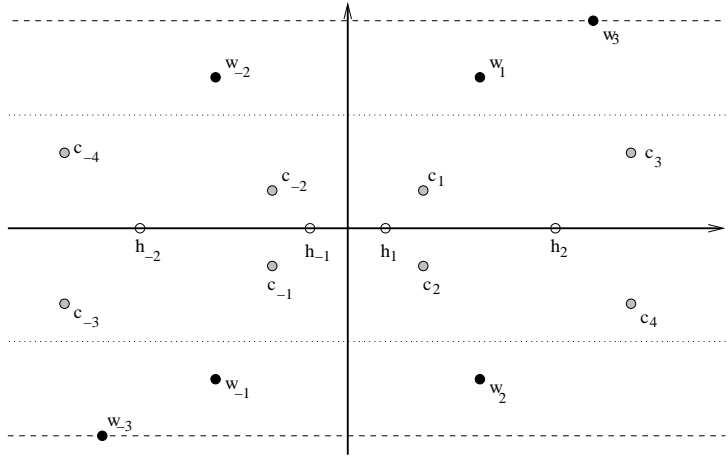}
\caption{\emph{Bethe roots and holes in the physical strip. The dashed lines denote the periodicity \(\frac{\pi^{2}}{\gamma}\) and the dotted lines the value \(\pm\pi \textrm{min}(1,p)\).}}
\end{center}
\end{figure} 

\subsection{Counting equation}
One may relate the numbers of various types of roots to the total third component of the spin of the system 

\begin{equation}
S=\frac{N}{2}-M.\label{eq:138}
\end{equation}
In order to do that, one makes use of the asymptotic values of the functions \(\phi_{\nu}(\vartheta)\)

\begin{equation}
\lim_{\vartheta\rightarrow\pm\infty}\varphi_{\nu}(\vartheta)=\pm\pi(\pi-2\nu)\label{eq;139}
\end{equation}
which allows one to express the \(\pm\infty\) asymptotics of the counting function as

\begin{align}
Z_{N}(+\infty) = & 2\pi\left(N-M\right)+\left(4M-2N\right)\gamma\nonumber \\
&-2\gamma H+4(\pi-\gamma)+2\pi\textrm{Sign}(p-1)M_{W}^{\downarrow}\label{eq:140}
\end{align}

\begin{align}
Z_{N}(-\infty) = & -2\pi\left(N-M\right)+\left(4M-2N\right)\gamma\nonumber \\
&+2\gamma H+4(\pi-\gamma)+2\pi\textrm{Sign}(p-1)M_{W}^{\uparrow}\label{eq:141}
\end{align}
where \(H=\frac{H_{+}+H_{-}}{2}\). Substituting Eq. (\ref{eq:138}) in Eqs. (\ref{eq:140}) and (\ref{eq:141}), one gets

\begin{equation}
Z_{N}(+\infty)=2\pi(2S+M)+2\pi\alpha+2\pi\textrm{Sign}(p-1)M_{W}^{\downarrow}\label{eq:142}
\end{equation}

\begin{equation}
Z_{N}(-\infty)=-2\pi(2S+M)-2\pi\alpha+2\pi\textrm{Sign}(p-1)M_{W}^{\uparrow}\label{eq:142bis}
\end{equation}
with \(\alpha=\frac{2(p-S)}{p+1}-\frac{H}{p+1}\). Define \(I_{\textrm{max}}\) to be the largest integer smaller than \(Z_{N}(+\infty)/2\pi\), and \(I_{\textrm{min}}\) the smallest integer greater than \(Z_{N}(-\infty)/2\pi\). Evidently it holds 

\[
I_{\textrm{max}}=2S+M+\left\lfloor\alpha\right\rfloor+\textrm{Sign}(p-1)M_{W}^{\downarrow} \]

\[
I_{\textrm{min}}=-2S-M-\left\lfloor\alpha\right\rfloor-\textrm{Sign}(p-1)M_{W}^{\uparrow} \]
where \(\left\lfloor\alpha\right\rfloor\) denotes the integer part of \(\alpha\). Therefore

\begin{equation}
Z_{N}(+\infty)=2\pi I_{\textrm{max}}+2\pi\left(\alpha-\left\lfloor\alpha\right\rfloor\right)\label{eq:143}
\end{equation}

\begin{equation}
Z_{N}(-\infty)=2\pi I_{\textrm{min}}-2\pi\left(\alpha-\left\lfloor\alpha\right\rfloor\right)\label{eq:144}
\end{equation}
Now, considering the total number of real Bethe solutions on the real axis, one has that 

\begin{equation}
I_{\textrm{max}}-I_{\textrm{min}}=2\left(M_{R}+N_{H}-2N_{S}\right)\label{eq:145}
\end{equation}
and then, combining Eqs. (\ref{eq:143}), (\ref{eq:144}) and (\ref{eq:145}) and substituting the values of \(I_{\textrm{max}}\) and \(I_{\textrm{min}}\),

\begin{equation}
N_{H}-2N_{S}=2S+M_{C}+\textrm{Step}(p-1)M_{W}+\left\lfloor \frac{2\left(p-S\right)}{p+1}-\frac{H}{p+1}\right\rfloor\label{eq:146}
\end{equation}
where

\[
\textrm{Step}(x)=\left\{\begin{array}{ccc} 1 & \textrm{for} & x>0 \\
0 & \textrm{for} & x<0 \end{array}\right. \]
Eq. (\ref{eq:146}) is called \emph{counting equation} and it relates the number of roots and holes appearing in a given Bethe state to the total spin of the chain.

\section{Nonlinear integral equation}
In the counting function one may separate the contributions of all types of roots and rewrite

\begin{align}
Z_{N}(\vartheta)&=N\left[\varphi_{1/2}\left(\vartheta+\Theta\right)+\varphi_{1/2}\left(\vartheta-\Theta\right)\right]+\varphi_{H_{+}/2}(\vartheta)+\varphi_{H_{-}/2}(\vartheta)\nonumber \\
&+\sum_{k=-N_{H}}^{N_{H}}\varphi_{1}\left(\vartheta-h_{k}\right)-\sum_{k=-M_{C}}^{M_{C}}\varphi_{1}\left(\vartheta-c_{k}\right)-\sum_{k=-M_{W}}^{M_{W}}\varphi_{1}\left(\vartheta-w_{k}\right)\nonumber \\
&-\sum_{k=-N_{H}-M_{R}}^{N_{H}+M_{R}}\varphi_{1}\left(\vartheta-x_{k}\right)+\varphi_{1}(\vartheta)+\varphi_{1}(2\vartheta).\label{eq:147}
\end{align}
In order to get a more explicit and useful expression for Eq. (\ref{eq:147}) one has to perform some standard manipulations \cite{ddv3,lmss,fmqr,frt1,frt2,frt3}. Let notice that the function \(Z_{N}(\vartheta)\) has a very complicated analytic structure due to the logarithmic cuts of the functions \(\varphi_{\nu}(\vartheta)\). To deal with a meromorphic function it is useful to consider the derivative \(\dot{Z}_{N}(\vartheta)\) of the counting function. The calculation is based on the fact that, assuming \(\mathbf{x}\) to be a real solution of the Bethe Eqs. (\ref{eq:128}) such that \(\dot{Z}_{N}(\mathbf{x})\ne 0\), in a small complex neighbourood of \(\mathbf{x}:\mu=\mathbf{x}+\epsilon\), it holds

\[
\frac{1}{\mu-\mathbf{x}}=\frac{-ie^{iZ_{N}(\mathbf{x}+\epsilon)}\dot{Z}_{N}(\mathbf{x}+\epsilon)}{1-e^{iZ_{N}(\mathbf{x}+\epsilon)}}+\textrm{regular terms.} \]
For any analytic function \(f(\mathbf{x})\) in an appropriate domain around the real axis, and \(\dot{\varphi}_{\nu}(\vartheta)\) is, one can apply the Cauchy theorem and obtain

\[
\dot{\varphi}_{\nu}(\mathbf{x})=\oint_{\Gamma_{\textrm{x}}}\frac{\textrm{d}\mu}{2\pi i}\frac{\dot{\varphi}_{\nu}(\mu)}{\mu-\mathbf{x}}=\oint_{\Gamma_{\textrm{x}}}\frac{\textrm{d}\mu}{2\pi i}\dot{\varphi}_{\nu}(\mu)\frac{-ie^{iZ_{N}(\mu)}\dot{Z}_{N}(\mu)}{1-e^{iZ_{N}(\mu)}} \]
where \(\Gamma_{\textrm{x}}\) is a curve enclosing \(\mathbf{x}\). The Bethe real roots are finite in number (\(N\) is a finite number), hence it is always possible to find for every root a closed encircling curve \(\Gamma_{\mathbf{x}_{i}}\). The sum of all these contours around points \(\mathbf{x}_{i}\) in the real axis may be modified to a unique curve \(\Gamma\) encircling all the real roots and avoiding the complex ones, which is possible because in the large \(N\) limit the complex roots remain finite in number. Therefore

\begin{align}
\sum_{k=-M}^{M}\dot{\varphi}_{1}\left(\vartheta-\mathbf{x}_{k}\right) & =\sum_{k=-M}^{M}\oint_{\Gamma}\frac{\textrm{d}\mu}{2\pi i} \dot{\varphi}_{1}\left(\vartheta-\mu\right)\frac{-ie^{iZ_{N}(\mu)}\dot{Z}_{N}(\mu)}{1-e^{iZ_{N}(\mu)}}=\nonumber \\
&=\int_{-\infty}^{+\infty}\frac{\textrm{d}x}{2\pi i}\dot{\varphi}_{1}\left(\vartheta-x+i\epsilon\right)\frac{-ie^{iZ_{N}(x-i\epsilon)}\dot{Z}_{N}(x-i\epsilon)}{1-e^{iZ_{N}(x-i\epsilon)}}\nonumber \\
&-\int_{-\infty}^{+\infty}\frac{\textrm{d}x}{2\pi i}\dot{\varphi}_{1}\left(\vartheta-x-i\epsilon\right)\frac{-ie^{iZ_{N}(x+i\epsilon)}\dot{Z}_{N}(x+i\epsilon)}{1-e^{iZ_{N}(x+i\epsilon)}}\label{eq:148}
\end{align}
where the curve \(\Gamma\) has been supposed to be contained in the small strip around the real axis \(0<\epsilon<\textrm{min}\left\{\pi,\;\pi p,\;\left|\mathfrak{Im}(c_{k})\right|\;\forall \; k\right\}\) whose parts at \(\pm\infty\) do not contribute because of the vanishing of \(\dot{\varphi}_{1}(\vartheta)\) for very large \(\vartheta\): this explains why it is possible to calculate the contour integral as a sum of two integrals from \(-\infty\) to \(+\infty\). With the help of simple manipulations the derivative of the Eq. (\ref{eq:147}) may be rewritten as

\begin{align}
&\int_{-\infty}^{+\infty}\textrm{d}x\left[\delta(x)\dot{Z}_{N}(\vartheta-x)+K(\vartheta-x)\dot{Z}_{N}(x)\right]=\nonumber \\
&=N\left[\dot{\varphi}_{1}\left(\vartheta+\Theta\right)+\dot{\varphi}_{1}\left(\vartheta-\Theta\right)\right]+\dot{\varphi}_{1}(\vartheta)+\dot{\varphi}_{1}(2\vartheta)\nonumber \\
&+\sum_{k=-N_{H}}^{N_{H}}\dot{\varphi}_{1}(\vartheta-h_{k})+\dot{\varphi}_{H_{+}/2}(\vartheta)+\dot{\varphi}_{H_{-}/2}(\vartheta)\nonumber \\
&-\sum_{k=-M_{C}}^{M_{C}}\dot{\varphi}_{1}(\vartheta-c_{k})-\sum_{k=-M_{W}}^{M_{W}}\dot{\varphi}_{1}(\vartheta-c_{k})\nonumber \\
&+\int_{-\infty}^{+\infty}\frac{\textrm{d}x}{2\pi i}\dot{\varphi}_{1}\left(\vartheta-x+i\epsilon\right)\frac{-ie^{iZ_{N}(x-i\epsilon)}\dot{Z}_{N}(x-i\epsilon)}{1-e^{iZ_{N}(x-i\epsilon)}}\nonumber \\
&-\int_{-\infty}^{+\infty}\frac{\textrm{d}x}{2\pi i}\dot{\varphi}_{1}\left(\vartheta-x-i\epsilon\right)\frac{-ie^{iZ_{N}(x+i\epsilon)}\dot{Z}_{N}(x+i\epsilon)}{1-e^{iZ_{N}(x+i\epsilon)}} \label{eq:149} 
\end{align}
with

\[
K(x)\equiv\frac{\dot{\varphi}_{1}(x)}{2\pi}=\frac{1}{\pi}\frac{\sin 2\gamma}{\cosh 2x-\cos 2\gamma} \]
Finally, using Fourier transform, one gets an integral equation satisfied by \(\dot{Z}_{N}(\vartheta)\)
 
\begin{align}
\dot{Z}_{N}(\vartheta) & =N\left[\frac{1}{\cosh\left(\vartheta+\Theta\right)}+\frac{1}{\cosh\left(\vartheta-\Theta\right)}\right]\nonumber \\
&+2\pi\sum_{k=-N_{H}}^{N_{H}}G\left(\vartheta-h_{k}\right)+2\pi B\left(x\left|H_{+},H_{-}\right.\right)\nonumber \\
&-2\pi\sum_{k=-M_{C}}^{M_{C}}G\left(\vartheta-c_{k}\right)-2\pi\sum_{k=-M_{W}}^{M_{W}}G_{II}\left(\vartheta-w_{k}\right)\nonumber \\
&-2\pi\sum_{k=-N_{S}}^{N_{S}}\left[G\left(\vartheta-s_{k}+i\epsilon\right)+G\left(\vartheta-s_{k}-i\epsilon\right)\right]\nonumber \\
&-i\int \textrm{d}xG(\vartheta-x-i\epsilon)\frac{\textrm{d}}{\textrm{d}x}\log\left[1-(-1)^{M_{SC}}e^{iZ_{N}(x+i\epsilon)}\right]\nonumber \\
&+i\int \textrm{d}xG(\vartheta-x+i\epsilon)\frac{\textrm{d}}{\textrm{d}x}\log\left[1-(-1)^{M_{SC}}e^{iZ_{N}(x-i\epsilon)}\right]\label{eq:150}
\end{align}
where the logarithms in the last two lines are considered in their fundamental determination and come from the identification

\begin{align}
&\frac{\pm (-1)^{M_{SC}}ie^{\pm iZ_{N}(x\pm i\epsilon)}\dot{Z}_{N}(x\pm i\epsilon)}{1-e^{\pm iZ_{N}(x\pm i\epsilon)}}=\nonumber \\
&\;\;\;\;\;=\frac{\textrm{d}}{\textrm{d}x}\log\left[1-(-1)^{M_{SC}}e^{\pm iZ_{N}(x\pm i\epsilon)}\right]\mp 2\pi i \delta\left(x-s_{k}\right)\nonumber
\end{align}
with \(s_{k}\) denoting the position of special roots which take into account that, if the function \(1-e^{iZ_{N}(x+i\epsilon)}\) crosses the cut of the logarithm, the function \(\log\left[1-e^{iZ_{N}(x+i\epsilon)}\right]\) has a jump by \(\pm 2\pi i\) at the exact real point corresponding to a special root. The following notations have been introduced as well:

\begin{equation}
B\left(x\left|H_{+},H_{-}\right.\right)=F\left(x,H_{+}\right)+F\left(x,H_{-}\right)+J(x)+G(x)\label{eq:151}
\end{equation}

\begin{align}
F\left(x,H\right) & \equiv\int_{-\infty}^{+\infty}\frac{\textrm{d}k}{2\pi}e^{ikx}\textrm{Sign}(H)\frac{\sinh\frac{\pi}{2}\left(p+1-\left|H\right|\right)k}{\sinh\frac{\pi}{2}pk \cosh\frac{\pi}{2}k}\qquad\qquad\nonumber \\
&\qquad\qquad\qquad\qquad\qquad\qquad\qquad\textrm{for}\;\;\;\left|\mathfrak{Im} (x)\right|<\frac{\pi}{2}\left|H\right|\label{eq:152}
\end{align}

\begin{align}
J(x) & \equiv\int_{-\infty}^{+\infty}\frac{\textrm{d}k}{2\pi}e^{ikx}\frac{\sinh\frac{\pi}{4}(p-1)k\cosh\frac{\pi}{4}(p+1)k}{\sinh\frac{\pi}{2}pk \cosh\frac{\pi}{2}k}\qquad\qquad\nonumber \\
&\qquad\qquad\qquad\qquad\qquad\qquad\textrm{for}\;\;\;\left|\mathfrak{Im} (x)\right|<\frac{\pi}{2}\textrm{min}(1,p)\label{eq:153}
\end{align}

\begin{align}
G(x) & \equiv\int_{-\infty}^{+\infty}\frac{\textrm{d}k}{2\pi}e^{ikx}\frac{\sinh\frac{\pi}{2}(p-1)k}{2\sinh\frac{\pi}{2}pk \cosh\frac{\pi}{2}k}\qquad\qquad\nonumber \\
&\qquad\qquad\qquad\qquad\qquad\qquad\textrm{for}\;\;\;\left|\mathfrak{Im} (x)\right|<\pi\textrm{min}(1,p)\label{eq:154}
\end{align}

\begin{equation}
G_{II}(\vartheta)=\left\{\begin{array}{ccc}
\frac{i}{2\pi p}\left[\coth \frac{\vartheta}{p}\textrm{Sign}\mathfrak{Im} (\vartheta)+\coth\frac{(i\pi-\vartheta)\textrm{Sign}\mathfrak{Im} (\vartheta)}{p}\right] & \textrm{if} & p>1 \\
i\textrm{Sign}\mathfrak{Im} (\vartheta)\left[\frac{1}{\sinh\vartheta}+\frac{1}{\sinh\left(\vartheta-i\pi\textrm{Sign}\mathfrak{Im} (\vartheta)\right)}\right] & \textrm{if} & p<1\end{array}\right.\label{eq:155}
\end{equation}
The contribution of wide roots \(G_{II}(\vartheta)\) is different from that of close roots, because of a pole of \(G(\vartheta)\) at \(\vartheta=\pm i\pi\). To obtain the function \(Z_{N}(\vartheta)\) an integration in \(\vartheta\) is needed. If one defines

\begin{equation}
\chi(\vartheta)=2\pi\int_{0}^{\vartheta}\textrm{d}x\; G(x)\label{eq:156}
\end{equation}

\begin{equation}
P\left(\vartheta\left|H_{+},H_{-}\right.\right)=2\pi\int_{0}^{\vartheta}\textrm{d}x\;B\left(x\left|H_{+},H_{-}\right.\right),\label{eq:157}
\end{equation}
the final form of the counting function is

\begin{align}
Z_{N}(\vartheta) & =2N\arctan \frac{\sinh\vartheta}{\cosh\Theta}+ P\left(\vartheta\left|H_{+},H_{-}\right.\right)+g\left(\vartheta\left|\left\{\vartheta_{k}\right\}\right.\right)\nonumber \\
&\;+i\int \textrm{d}xG(\vartheta-x+i\epsilon)\log\left[1-(-1)^{M_{SC}}e^{iZ_{N}(x-i\epsilon)}\right]\nonumber \\
&\;-i\int \textrm{d}xG(\vartheta-x-i\epsilon)\log\left[1-(-1)^{M_{SC}}e^{iZ_{N}(x+i\epsilon)}\right]+\omega\label{eq:158}
\end{align}
with \(\omega\) an integration constant and

\begin{equation}
g\left(\vartheta\left|\left\{\vartheta_{k}\right\}\right.\right)=\sum_{k}c_{k}\left[\chi_{(k)}\left(\vartheta-\vartheta_{k}\right)+\chi_{(k)}\left(\vartheta+\vartheta_{k}\right)\right]\label{eq:159}
\end{equation}
where \(\left\{\vartheta_{k}\right\}\) is the set of positions of the various objects (holes, close roots, wide and special roots) characterizing a certain Bethe state, the coefficients \(c_{k}\) are given by

\[
c_{k}=\begin{cases} +1 & \textrm{for holes} \\
                           -1 & \textrm{otherwise}\end{cases} \]
and for any function \(f(\vartheta)\) we define

\[
f_{(k)}(\vartheta)=\begin{cases} f_{II}(\vartheta) & \textrm{for wide roots} \\
f(\vartheta+i\epsilon)+f(\vartheta-i\epsilon) & \textrm{for specials} \\
f(\vartheta) & \textrm{otherwise}\end{cases}\]
where \(f_{II}(\vartheta)\) is defined as 

\[
f_{II}(\vartheta)=\left\{\begin{array}{ccc} f(\vartheta)+f\left(\vartheta-i\pi\textrm{Sign}\mathfrak{Im} (\vartheta)\right) & \textrm{if} & p>1 \\
f(\vartheta)-f\left(\vartheta-i\pi p\textrm{Sign}\mathfrak{Im} (\vartheta)\right) & \textrm{if} & p<1\end{array}\right. \]
Note that the previous definition (\ref{eq:155}) of \(G_{II}(\vartheta)\) is in agreement with this more general one. From the analysis given in the periodic case \cite{ddv3,frt2,frt4}, it follows that the number \(N_{H}-2N_{S}\) remains constant along the volume size, therefore, one can refer to this number as the effective number of holes \(N_{eff}\).

Notice that for real \(x\) and \(\epsilon\rightarrow 0\)

\begin{align}
Q_{N}(x) & \equiv \lim_{\epsilon\rightarrow 0}2\mathfrak{Im}\log\left[1-e^{iZ_{N}(x+i\epsilon)}\right]=\nonumber \\
&=\lim_{\epsilon\rightarrow 0}\frac{1}{i}\left(\frac{1-e^{iZ_{N}(x+i\epsilon)}}{1-e^{-iZ_{N}(x+i\epsilon)}}\right)=Z_{N}(x)+\pi\textrm{mod}\;2\pi\;.\label{eq:160}
\end{align}
It is possible to compute exactly the integration constant \(\omega\) as well. In fact, one has to take into account the asymptotic values of functions \(\chi(\vartheta)\)

\begin{equation}
\chi_{\infty}\equiv\chi(+\infty)=\frac{\pi}{2}\frac{\pi-2\gamma}{\pi-\gamma}\label{eq:161}
\end{equation}
and the asymptotics of the boundary term \(P\left(\vartheta\left|H_{+},H_{-}\right.\right)\)

\begin{equation}
P_{\infty}\equiv P\left(+\infty\left|H_{+},H_{-}\right.\right)=\frac{\pi}{2}\frac{4\pi-\gamma\left(H_{+}+H_{-}-4\right)}{\pi-\gamma}. \label{eq:162}
\end{equation}
This allows to calculate the aymptotic value of \(Z_{N}(\vartheta)\) and, making use of the counting equation (\ref{eq:146}), it follows

\begin{align}
\omega & =(\pi-2\gamma)S+4\pi-\gamma\left(H_{+}+H_{-}+4\right)+\pi\textrm{Sign}(p-1)M_{W}\nonumber \\
&-2\chi_{\infty}S-P_{\infty}-4\chi_{\infty}+\frac{\chi_{\infty}\gamma}{\pi}\left(H_{+}+H_{-}\right)+\frac{2\chi_{\infty}\gamma}{\pi}(S+2)=\nonumber \\
&= \pi\textrm{Sign}(p-1)M_{W}\;.\label{eq:163}
\end{align}
The last term is a multiple of \(2\pi\) if \(M_{W}\) is even, which is the case unless there are an odd number of self-conjugated roots. \(Z_{N}(\vartheta)\) does not change if one adds or subtracts \(2\pi\) because of the quantization rule. Without loss of generality one can take

\begin{equation}
\omega=\pi\left(M_{SC}\;\textrm{mod}\;2\right)\label{eq:164}
\end{equation}
therefore, it is convenient to include this constant into the logarithmic term of the convolution, by modifying it to \(\log\left[1-(-1)^{\delta}e^{iZ_{N}(x)}\right]\), with \(\delta=M_{SC}\;\textrm{mod}\;2\). The quantization rule of any state becomes

\begin{equation}
Z_{N}\left(\vartheta_{j}\right)=2\pi I_{j}\;\;\;,\;\;\;I_{j}\in\mathbb{Z}+\frac{\delta}{2}\;.\label{eq:165}
\end{equation}
Similar considerations can be done to express the energy in terms of the counting function and the positions of the various roots. The expression which one derives depends on the number of lattice sites \(N\) and the lattice spacing \(a\), it has the form

\begin{align}
 2aE^{(N)} & =\sum_{k}c_{k}\left(\frac{1}{\cosh\left(\Theta-\vartheta_{k}\right)}+\frac{1}{\cosh\left(\Theta+\vartheta_{k}\right)}\right)\nonumber \\
&-\int\frac{\textrm{d}x}{2\pi}\left[\frac{\sinh\left(\Theta-x\right)}{\cosh^{2}\left(\Theta-x\right)}+\frac{\sinh\left(-\Theta-x\right)}{\cosh^{2}\left(-\Theta-x\right)}\right]Q_{N}(x)\nonumber \\
&-N\int\frac{\textrm{d}x}{2\pi}\Phi\left(\Theta,x\right)\left(\frac{1}{\cosh\left(\Theta-x\right)}+\frac{1}{\cosh\left(\Theta+x\right)}\right)\nonumber \\
&-\int\frac{\textrm{d}x}{\pi}\Phi\left(\Theta,x\right)\left[P\left(x\left|H_{+},H_{-}\right.\right)-2\pi\delta(x)\right]\label{eq:166}
\end{align}

where

\[
\Phi\left(\Theta,x\right)\equiv\dot{\varphi}_{1/2}\left(\Theta-x\right)+\dot{\varphi}_{1/2}\left(\Theta+x\right) \]

\subsection{Continuum limit} 
At this point one can calculate the continuum limit of the energy, i.e. an expression valid for \(N\rightarrow +\infty\), \(a\rightarrow 0\) in such a way that the finite volume \(L=Na\) remains fixed. One can try to do that by computing first the continuum limit of the counting function

\begin{equation}
Z(\vartheta)\equiv\lim_{N\rightarrow +\infty}Z_{N}(\vartheta).\label{eq:167}
\end{equation}
One has to set

\begin{equation}
\Theta\sim\log\frac{2N}{\mathcal{M}L}\label{eq:168}
\end{equation}
with \(\mathcal{M}\) a mass scale. The source term \(g\left(\vartheta\left|\left\{\vartheta_{k}\right\}\right.\right)\) and the convolution integral do not present any dependence on \(N\), which is instead crucial in the driving term. As a result, one can define the \emph{Nonlinear Integral Equation} (NLIE) 

\begin{align}
Z(\vartheta) & =2\mathcal{M}L\sinh\vartheta+g\left(\vartheta\left|\left\{\vartheta_{k}\right\}\right.\right)+P\left(\vartheta\left|H_{+},H_{-}\right.\right)\nonumber \\
&\qquad-2i\mathfrak{Im} \int\textrm{d}xG(\vartheta-x-i\epsilon)\log\left[1-(-1)^{\delta}e^{iZ(x+i\epsilon)}\right]\;.\label{eq:169}
\end{align}
For the vacuum state containing real roots only, this equation coincides with the one found some years ago in \cite{lmss}. The quantization relation

\begin{equation}
Z\left(\vartheta_{j}\right)=2\pi I_{j}\;\;\;,\;\;\;I_{j}\in\mathbb{Z}+\frac{\delta}{2}\label{eq:170}
\end{equation}
holds as well and fixes the positions of holes and complex roots determining the source term \(g\left(\vartheta\left|\left\{\vartheta_{k}\right\}\right.\right)\) characterizing the Bethe excited states. The term \(P\left(x\left|H_{+},H_{-}\right.\right)\) takes into account the effects of boundaries and, in particular, it depends on the parameters \(H_{\pm}\) which can be related to the DSG boundary parameters \(\phi_{\pm}\). We will be considering this point in the next Chapter. 

In order to compute the continuum limit of the expression (\ref{eq:166}) for the energy, let us just say that the following asymptotic behaviours hold for \(N\rightarrow +\infty\)

\[
\frac{N}{L}\arctan \sinh\left(\Theta-\vartheta\right)\sim\frac{N\pi}{L}-\mathcal{M}e^{+\vartheta} \]

\[
\frac{N}{L}\int\frac{\textrm{d}x}{2\pi}\frac{Q_{N}(x)}{\cosh\left(\pm\Theta-x\right)}\sim\frac{\mathcal{M}}{2}\int\frac{\textrm{d}x}{2\pi}e^{\pm x}Q(x) \]
The last formula is the contribution of the integral term in the second line of Eq. (\ref{eq:166}) with \(Q(x)\equiv\lim_{N\rightarrow\infty}Q_{N}(x)\). The contribution of the third line of Eq. (\ref{eq:166}) can be evaluated by considering that it takes the form

\begin{equation}
E^{(N)}_{\textrm{bulk}}=-\frac{2N^{2}}{L}\int\frac{\textrm{d}x}{\pi}\frac{\textrm{d}}{\textrm{d}\Theta}\left(\frac{\varphi_{1/2}(x)}{\cosh\left(2\Theta-x\right)}\right).\label{eq:171}
\end{equation}
It is, evidently, divergent for \(N\rightarrow +\infty\). The contribution of the boundaries to the energy is given by the last line of (\ref{eq:166}) and can be expressed, making use of Fourier representation, as

\begin{align}
E^{(N)}_{B} & \equiv \frac{2N}{L}\int\frac{\textrm{d}k}{2\pi}\frac{\cos k\Theta}{\cosh\frac{\pi}{2}k}\left[\frac{\sinh\frac{\pi}{2}\left(p+1-H_{+}\right)k}{\sinh\frac{\pi}{2}(p+1)k}+\right.\nonumber \\
&\qquad\qquad+\left.\frac{\sinh\frac{\pi}{2}\left(p+1-H_{-}\right)k}{\sinh\frac{\pi}{2}(p+1)k}-\frac{\sinh\frac{\pi}{4}(p-1)k}{\sinh\frac{\pi}{4}(p+1)k}+1 \right]\label{eq:172}
\end{align}
This term, after the shift \(k\rightarrow ik\), splits into two terms. The first one diverges with \(N\), let us call it \(E_{B}^{\textrm{bulk}}\). The second one takes the form

\begin{align}
E_{B}\equiv & -\frac{\mathcal{M}}{2}\left(\frac{\sin\frac{\pi}{2}\left(p+1-H_{+}\right)}{\sin\frac{\pi}{2}(p+1)}+\right.\nonumber \\
&\qquad\qquad\qquad\left.+\frac{\sin\frac{\pi}{2}\left(p+1-H_{-}\right)}{\sin\frac{\pi}{2}(p+1)}-\cot\frac{\pi}{4}(p+1)-1\right)\label{eq:173}
\end{align}
At this point it is allowed to renormalize the energy subtracting all the divergent terms, i.e. shifting the energy \(E\rightarrow E-E_{B}^{\textrm{bulk}}-E_{\textrm{bulk}}^{(N)}\). The final result is

\begin{equation}
E=E_{B}+\mathcal{M}\sum_{k}c_{k}\cosh_{(k)}\vartheta_{k}-\mathcal{M}\int\frac{\textrm{d}x}{2\pi}\sinh(x)\;Q(x)\label{eq:174}
\end{equation}
Eqs. (\ref{eq:169}) and (\ref{eq:174}) are the most important results of this Chapter. Indeed, they allows to express in a continuum framework the energy, i.e. the eigenvalues of the Hamiltonian (\ref{eq:126quater}) in the formalism of Chapter 3, as a function of Bethe roots and boundary parameters. The energy (\ref{eq:174}) is the perfect candidate to define the scaling functions of the model and to relate different states in the IR and UV limits introduced in the next Chapter.

\chapter{The continuum theory}
Once the \(Z(\vartheta)\) and the energy are known, in the IR limit \(\mathcal{M}L\rightarrow +\infty\) one can make contact between the NLIE excited states and the scattering theory of the underlying field theory. On the other hand, in the UV limit \(\mathcal{M}L\rightarrow 0\), one finds contact with the CFT at the UV regime of the DSG model.

\section{IR limit} 
The IR limit is expected to reproduce a field theory in a large volume \(L\). As a scale parameter it is convenient to choose the dimensionless size parameter \(l=\mathcal{M}L\) with \(\mathcal{M}\) playing the role of a mass. Therefore, the very large \(l\) regime can be interpreted both as infinite large size, that reproduces the conditions of a field theory in an infinite Minkowski space-time, or large mass scale, compared to energies in play, that corresponds to an infra-red point in the RG flow. There are many possibilities to give an interpretation of the physical content of the model described by the NLIE. The main question is to find a correspondence between the physics of the IR model and the different states of the NLIE which one can deal with. Let us notice that the infinite volume limit is not a thermodynamic limit, because the number of the various objects in the NLIE, apart the roots on the real axis, remains fixed. So, it is expected that the energy \(E\) should approach a finite limit depending only on a finite number of roots, reproducing thus a free massive spectrum. In order to prove this conjecture, it is sufficient to notice that for \(l\) large enough the integrals in Eqs. (\ref{eq:169}) and (\ref{eq:174}) become small of order \(\mathcal{O}\left(e^{-l}\right)\). They can be dropped. Hence

\begin{equation}
Z(\vartheta)=2l\sinh \vartheta +g\left(\left.\vartheta\right|\left\{\vartheta_{k}\right\}\right)+P\left(\vartheta\left|H_{+},H_{-}\right.\right)\;\;\;\textrm{for}\;\;\;l\rightarrow +\infty\label{eq:175}
\end{equation}
which can be seen as a constraint on the asymptotic states. Similarly, in the energy formula the integral term drops and

\begin{equation}
E=E_{B}+\mathcal{M}\sum_{k} c_{k}\cosh_{(k)} \vartheta_{k}\label{eq:176}
\end{equation}

\subsection{Holes-Solitons correspondence}
The important thing is to understand what type of constraints Eqs. (\ref{eq:175}) and (\ref{eq:176}) fix and what they mean from a physical point of view. Consider the state without complex roots \( \left(M_{C}=M_{W}=0\right) \). Thus one can take into account two different situations. The simplest one is that corresponding to the presence of real roots only. It is evident that in this case the energy reduces to \(E=E_{B}\) because neither the NLIE nor the energy depend explicitly on these roots, i.e. one has a state with a fixed energy value in which from the constraint (\ref{eq:175}) it does not come any informations. One could associate this NLIE state to the vacuum of the theory. This assumption is also supported by the fact that for \(l\rightarrow +\infty\) the number of real roots goes to infinity forming a vacuum Dirac sea and that, from the Eq. (\ref{eq:146}), the total spin of the system is \(S=0\). This is exaclty true only if \(\left|h_{\pm}\right|<h_{c}\) where the quantity \(h_{c}\) has been defined in Chapter 3, otherwise the behaviour of energy presents some special effects and, in order to have \(S=0\), one should consider the presence of some different type of roots. We will be treating this situation later, so far let us consider only the range \(\left|h_{\pm}\right|<h_{c}\).  

The other possibility is to have a NLIE state defined by \( N_{H}\) holes in the positive real axis, say at positions \( \vartheta_{k},\;k=1,2,\;.\;.\;.\;N_{H}\). For simplicity take just two holes at positions \(\vartheta_{1}\) and \(\vartheta_{-1}\). Due to the fact that \(\left\{\vartheta_{-k}\right\}=\left\{-\vartheta_{k}\right\}\;\forall\;k=1,2,\;.\;.\;.\) and to the antisymmetric property of the NLIE, for any hole of type \(\vartheta_{-k}\) the same equation holds and one can consider only objects with positive real part: each of them implicitly describes also the contribution of those with negative real part. From now on we will consider objects with positive real part and use the notation \(N_{A}\) with \(A=H,C,W,\;.\;.\;.\) to refer to their numbers. The NLIE valued at \(\vartheta_{1}\) becomes 

\begin{align}
Z\left(\vartheta_{1}\right)&=2l\sinh\vartheta_{1}+\chi\left(2\vartheta_{1}\right)+P\left(\vartheta_{1}\left|H_{+},H_{-}\right.\right)\nonumber \\
&=2l\sinh\vartheta_{1}+\mathcal{F}\left(\vartheta_{1},H_{+}\right)+\mathcal{F}\left(\vartheta_{1},H_{-}\right)=2\pi I_{1}\label{eq:177}
\end{align}
where

\begin{equation}
\mathcal{F}\left(\vartheta,H\right)=2\pi\int_{0}^{\vartheta}\textrm{d}x\left[F\left(x,H\right)+\int_{-\infty}^{+\infty}\frac{\textrm{d}k}{2\pi}e^{ikx}\frac{\sinh \frac{3\pi}{2}k\sinh\frac{\pi}{2}(p-1)k}{\sinh\frac{\pi}{4}pk\sinh \pi k}\right].\label{eq:178}
\end{equation}
Eq. (\ref{eq:177}) may be interpreted as a quantization rule for the momentum of a particle of energy \(\mathcal{M}\cosh\vartheta_{1}\). In fact, taking the exponent

\begin{equation}
e^{2il\sinh\vartheta_{1}}e^{i\left[\mathcal{F}\left(\vartheta_{1},H_{+}\right)+\mathcal{F}\left(\vartheta_{1},H_{-}\right)\right]}=1\;,\label{eq:179}
\end{equation}
after some mathematical manipulations \cite{fensal}, one can prove that 

\begin{equation}
\mathcal{F}\left(\vartheta,H\right)=-i\log\mathcal{R}^{+}\left(\vartheta,H\right)\label{eq:180}
\end{equation}
where \(\mathcal{R}^{+}\left(\vartheta,H\right) \), which has the same structure of the function \(\mathcal{P}_{+}(u)\) introduced in Eq. (\ref{eq:111}), is exactly the soliton boundary reflection matrix as found in \cite{ghoshzam}; therefore,

\begin{equation}
e^{2il\sinh\vartheta_{1}}\mathcal{R}^{+}\left(\vartheta_{1},H_{+}\right)\mathcal{R}^{+}\left(\vartheta_{1},H_{-}\right)=1\label{eq:181}
\end{equation}
which corresponds to the quantization rule of a particle moving in a volume \(L\) with momentum \(\mathcal{M}\sinh\vartheta_{1}\) towards two boundaries and interacting off them with reflection amplitudes \(\mathcal{R}^{+}\left(\vartheta_{1},H_{+}\right)\) and \(\mathcal{R}^{+}\left(\vartheta_{1},H_{-}\right)\).

The parameters of reflection matrices are related to the Dirichlet boundaries \(\phi_{\pm}\) in the boundary sine-Gordon Action and the boundary magnetic fields of the XXZ based construction, then, taking into account that \(\phi_{\pm}\) are always positive parameters while the parameters \( H_{\pm}\) and \(h_{\pm}\) take values in positive and negative ranges, one gets

\begin{equation}
H_{\pm}=\left\{ \begin{array}{ccc} p\left(1-\frac{8}{\beta}\phi_{\pm}\right) & \textrm{if} & h_{\pm}\ge 0 \\
 p\left(1+\frac{8}{\beta}\phi_{\pm}\right) & \textrm{if} & h_{\pm}<0\end{array}\right.\label{eq:182}
\end{equation}
In particular, from our choice of boundary magnetic fields (\(h_{-}\ge 0\) and \(h_{+}\le 0\)), it follows \(H_{-}=p\left(1-\frac{8}{\beta}\phi_{-}\right)\) and \(H_{+}=p\left(1+\frac{8}{\beta}\phi_{-}\right)\). It is thus tempting to say that the one hole NLIE state corresponds to a soliton of DSG. This interpretation is also supported by the fact that the energy of this state is

\begin{equation}
E=E_{B}+\mathcal{M}\cosh \vartheta_{1}\label{eq:183}
\end{equation}
that looks like a particle excitation over the vacuum (remember the parametrization of particle energy in terms of rapidity given in Chapter 2). Moreover, from the counting equation it follows that the state has spin \(S=\frac{1}{2}\), i.e. an excitation over the vacuum spin \(S=0\). We conclude that the physical meaning of a hole state is a soliton with energy \(\mathcal{M}\cosh\vartheta_{1}\) scattering off the boundaries via the Ghoshal-Zamolodchikov reflection amplitudes \cite{io}. 

Consider, now, a more complicated state, say the one with a generic number \(N_{H}\) of holes at positions \(\vartheta_{k},\;k=1,2,\;.\;.\;.\;N_{H}\). Due to (\ref{eq:106bis}) and to the definitions (\ref{eq:154}) and (\ref{eq:156}), the function \(\chi(\vartheta)\) can be written as

\begin{equation}
\chi(\vartheta)=-i\log a(\vartheta)= -i\log S_{++}^{++}(\vartheta)\label{eq:184}
\end{equation}
where \(S_{++}^{++}(\vartheta)\) is the sine-Gordon soliton-soliton scattering amplitude. Therefore, for \(N_{H}\) holes one has that the spin of the state is \(S=\frac{N_{H}}{2}\), the energy is

\begin{equation}
E=E_{B}+\mathcal{M}\sum_{k}^{N_{H}}\cosh\vartheta_{k}\label{eq:185}
\end{equation}
and for any hole \(\vartheta_{k}\) the following equation holds

\begin{align}
&Z\left(\vartheta_{k}\right) =2l\sinh\vartheta_{k}+\sum_{j=1}^{N_{H}}\left[\chi\left(\vartheta_{k}-\vartheta_{j}\right)+\chi\left(\vartheta_{k}+\vartheta_{j}\right)\right]+P\left(\vartheta_{k}\left|H_{+},H_{-}\right.\right)=\nonumber \\
&=2l\sinh\vartheta_{k}+\sum_{j\ne k}^{N_{H}}\left[\chi\left(\vartheta_{k}-\vartheta_{j}\right)+\chi\left(\vartheta_{k}+\vartheta_{j}\right)\right]+\mathcal{F}\left(\vartheta_{k},H_{+}\right)+\mathcal{F}\left(\vartheta_{k},H_{-}\right).\label{eq:186}
\end{align}
After an exponentiation, Eq. (\ref{eq:186}) becomes
   
\begin{equation}
e^{2il\sinh \vartheta_{k}} \prod_{j\ne k}^{N_{H}}\mathcal{S}\left(\vartheta_{k}\left|\vartheta_{j}\right.\right) \mathcal{R}^{+}\left(\vartheta_{k},H_{+}\right)\mathcal{R}^{+}\left(\vartheta_{k},H_{-}\right)=1\label{eq:187}
\end{equation}
where, for simplicity, we have defined

\begin{equation}
\mathcal{S}\left(\vartheta\left|\vartheta_{j}\right.\right)\equiv S_{++}^{++}\left(\vartheta-\vartheta_{j}\right)S_{++}^{++}\left(\vartheta+\vartheta_{j}\right)\label{eq:187bis}
\end{equation}
Eq. (\ref{eq:187}) is the quantization rule of a soliton interacting on two boundaries via \(\mathcal{R}^{+}\left(\vartheta_{k},H_{+}\right)\) and \(\mathcal{R}^{+}\left(\vartheta_{k},H_{-}\right)\) and with \(N_{H}-1\) solitons via the soliton S matrix \(S_{++}^{++}\left(\vartheta_{k}\pm\vartheta_{j}\right)\). Notice that the signs \(\pm\) refer to the fact that the soliton \(\vartheta_{k}\) can scatter with the other solitons either before or after they have respectively scattered off one of the two boundaries.

\subsection{The antisoliton}

According to the symmetry \(\mathcal{C}\) introduced in Chapter 3 we can deal with states of negative spin just reversing both magnetic fields \(h_{\pm}\). In terms of DSG model this symmetry requires some care. It corresponds to reversing the topological charge \(Q\rightarrow -Q\), but, in order to achieve the correct result it is not enough to send \(\phi_{\pm}\rightarrow -\phi_{\pm}\). In fact, an exact definition of the DSG theory describing a state of topological charge \(-Q\) in the same ranges of boundary fields, comes from the successive application of the symmetries \(\phi\rightarrow -\phi\), \(\phi\rightarrow \phi+\frac{2\pi}{\beta}\) and \(\phi_{\pm}\rightarrow -\phi_{\pm}\), \(\phi_{\pm}\rightarrow \phi_{\pm}+\frac{2\pi}{\beta}\) simultaneously because one has fixed the boundary fields to be always positive. This leads to the same description in terms of NLIE states for physical states with charge \(-Q\): a state with \( N_{H}\) holes describes \( N_{H}\) particles of mass \(\mathcal{M}\) scattering with S-matrix given by the \( S_{++}^{++}(\vartheta)=S_{--}^{--}(\vartheta)\) sine-Gordon amplitude. Therefore, one espects to get a theory describing corresponding conjugated states of those treated previously, for instance antisoliton states instead of soliton states. From Eqs. (\ref{eq:182}) it follows that, due to the transformations of fields, in all formulae one has to substitute 

\begin{equation}
H_{\pm}+1\rightarrow -H_{\pm}-1\;\;,\;\;\forall\; H_{\pm}\; :\; -p-1<H_{\pm}<p+1\;.\label{eq:188}
\end{equation}
As a consequence, the soliton reflection factor goes into the antisoliton reflection factor \(\mathcal{R}^{-}\left(\vartheta,H_{\pm}\right)\) found in \cite{ghoshzam}. Under these transformations the pure holes state can be interpreted as a state of \(N_{H}\) antisolitons of sine-Gordon with Dirichlet boundaries \(\phi_{\pm}\)

\begin{equation}
e^{2il\sinh \vartheta_{k}} \prod_{j\ne k}^{N_{H}} \mathcal{S}\left(\vartheta_{k}\left|\vartheta_{j}\right.\right) \mathcal{R}^{-}\left(\vartheta_{k},H_{+}\right)\mathcal{R}^{-}\left(\vartheta_{k},H_{-}\right)=1\label{eq:189}
\end{equation}
where \(\mathcal{S}\left(\vartheta_{k}\left|\vartheta_{j}\right.\right)\) has been defined in (\ref{eq:187bis}). 

\subsection{Boundary bound states}
From the analysis given in \cite{ss} of the Bethe Ansatz equations there is evidence that boundary bound states should be described by purely imaginary objects. Therefore, in the continuum NLIE we search for solutions characterized by the presence of one or more roots on the imaginary axis. Such roots give contributions \(\chi_{(k)}\left(\vartheta-i\alpha_{k}\right)\) to the NLIE source term \(g\left(\vartheta\left|\left\{\vartheta_{k}\right\}\right.\right)\) if their position is at \(\vartheta_{k}=i\alpha_{k}\). Moreover, in the large \(l\) limit from the oddity of the \(Z(\vartheta)\) it follows that any such a root implies

\begin{equation}
\begin{cases} \mathfrak{Re} Z\left(i\alpha_{k}\right)=2\pi I_{\alpha_{k}}=0 \\
\mathfrak{Im} Z\left(i\alpha_{k}\right)=2l\sin\alpha_{k}+\mathfrak{Im} \chi\left(2i\alpha_{k}\right)+\mathfrak{Im} P\left(i\alpha_{k}\left|H_{+},H_{-}\right.\right)=0\end{cases}\label{eq:190}
\end{equation}
The exploding term \(2l\sin\alpha_{k}\) for \(l\rightarrow +\infty\) has to be compensated by some singularity coming from the other two terms. This may happen when one considers poles in the boundary reflection amplitude. In the function \(P\left(\vartheta\left|H_{+},H_{-}\right.\right)\) the boundary content is encoded in functions \(F\left(\vartheta,H_{\pm}\right)\), hence, only poles coming from these terms and depending on boundary parameters \(H_{\pm}\) can be taken into account, any other poles has to be interpreted in the bulk theory. In the region \(0<H_{\pm}<p+1\) there are no poles in functions \(F\left(\vartheta,H_{\pm}\right)\). All possible poles in the integral terms of the NLIE may be interpreted in the bulk theory. Hence, taking into account the range of definition of boundary magnetic fields \(h_{\pm}\) which we have chosen, the poles which one is interested in have to be found in the region \(-1<H_{-}<0\) for the boundary parameter \(H_{-}\) and \(-p-1<H_{+}<-1\) for \(H_{+}\), this means that \(h_{-}>h_{c}\) and \(h_{+}<-h_{c}\) and clarifies why in the previous section, where \(\left|h_{\pm}\right|<h_{c}\), we have not treated boundary bound states. To find such poles one has to consider the large \(\left|k\right|\) limit in the corresponding Fourier transform \(\hat{F}\left(k,H\right)\). It follows

\begin{equation}
\hat{F}\left(k,H\right)\sim e^{-\frac{\pi}{2}\left|H\right|\left| k\right|}\;\;\;\textrm{for}\;\;\;\left|k\right|\rightarrow +\infty\label{eq:191}
\end{equation}
therefore, in \(F\left(\vartheta,H\right)\) there appear singularities at 

\begin{equation}
\mathfrak{Re} (\vartheta)=0\;\;\;,\;\;\;\mathfrak{Im} (\vartheta)=\pm\frac{\pi}{2}\left(H+2pl\right)\;\;\;\textrm{with}\;\;\;l\in \mathbb{Z}\;.\label{eq:192}
\end{equation}
These singularities correspond exactly to the first set of poles (\ref{eq:115}) parametrized by Dorey and Mattsson in \cite{dm} if \(H_{\pm}\) take values in the range (\ref{eq:182}): \(-1<H_{-}<0\) and \(-p-1<H_{+}<-1\). The sets of these poles can be denoted as

\begin{align}
\mathbb{A}^{\pm} &=\left\{i\alpha_{n}^{(\pm)}=-i\frac{\pi}{2}\left[H_{\pm}\pm p +p(2n+1)\right]:\right.\nonumber \\
&\qquad\qquad \left.\frac{\pi}{2}>\left|\alpha_{0}^{(\pm)}\right|>\left|\alpha_{1}^{(\pm)}\right|\;.\;.\;.\;>0,\;n=0,1,\;.\;.\;.\;\le n^{\pm}\right\}\label{eq:193}
\end{align}
where \(n^{\pm}\) are integers defined such that \(0<\left|H_{\pm}\pm p+p\left(2n^{\pm}+1\right)\right|<1\) and the roots \(i\alpha_{n}^{(\pm)}=i\frac{\pi}{2}\left[H_{\pm}\pm p+p(2n+1)\right]\) may be discarded because of the properties of the Bethe equations. So one can consider the boundary bound states as states in which combinations of purely imaginary roots with imaginary part in the set (\ref{eq:193}) appear. 

Let us deal with the simplest case with only one excited boundary, say the one at the left. In terms of NLIE this means the presence of a pure imaginary root \(i\alpha_{0}^{(-)}=-i\frac{\pi}{2}H_{-}\) which fixes the expression of the IR energy to be

\begin{equation}
E=E_{B}-\mathcal{M}\cos\alpha_{0}^{(-)}\label{eq:195}
\end{equation}
where, due to range of definition of the boundary parameter \(H_{-}\), the term \(\mathcal{M}\cos\alpha_{0}^{(-)}\) is always positive, i.e. the energy becomes smaller than the vacuum energy \(E_{B}\). In general, the whole set of pure imaginary roots contributes a negative amount in the energy and these roots cannot be considered as excitations. This means that the vacuum which we have been working with is instable for \(h_{-}>h_{c}\) and \(h_{+}<-h_{c}\). The vacuum decreases as the value of boundary parameters take values grater than the critical one. This could appear as a non-sense, but remember that the vacuum energy has been defined up to a bulk term, therefore, in principle, one could redefine that by shifting by a finite amount. In this sense any root on the imaginary axis corresponding to a pole of \(F\left(\vartheta,H_{\pm}\right)\) defines lower levels of vacuum energy. To cure this situation one has to define a new ground state by adding up to the Dirac sea made of real roots all the pure imaginary roots in the set \(\mathbb{A}^{\pm}\) defined in (\ref{eq:193}). The vacuum energy becomes 

\begin{equation}
E_{B}^{\textrm{bbs}}\equiv E_{B}-\mathcal{M}\sum_{k=0}^{n^{+}}\cos\alpha_{k}^{(+)}-\mathcal{M}\sum_{k=0}^{n^{-}}\cos\alpha_{k}^{(-)}\;\;\;\;,\;\;\;\;\alpha_{k}^{(\pm)}\in \mathbb{A}^{\pm}\label{eq:196}
\end{equation}
For reasons of consistency, one expects the total spin associated to the vacuum state to be \(S=0\). This is possible only by redefining the formula (\ref{eq:138}) as \(S=\frac{N}{2}-M=\frac{N}{2}-M_{R}-M_{I}\), where \(M_{R}\) is the number of real roots, as usual, and \(M_{I}\) the number of pure imaginary roots in the set \(\mathbb{A}^{\pm}\). This new definition of the total spin might have some consequences on the counting equation. Indeed, it has been derived taking into account the relation among the spin, the number of lattice sites and the number of total roots. Notice that from the first of Eqs. (\ref{eq:190}) it follows that the quantization numbers of pure imaginary roots are always zero. This means that the definition of \(I_{\textrm{max}}\), the greatest integer smaller than \(Z_{N}(+\infty)/2\pi\), and \(I_{\textrm{min}}\), the smallest integer greater than \(Z_{N}(-\infty)/2\pi\), needs some care. The quantization numbers of pure imaginary roots, due to the fact that they are zero, do not give any amount neither to the sum of all quantum numbers nor to the greatest or the smallest among them. Therefore, \(I_{\textrm{max}}\) and \(I_{\textrm{min}}\) need redefining as \(I_{\textrm{max}}=2S+M-M_{I}\), \(I_{\textrm{min}}=-2S-M+M_{I}\). Following the same deduction, the number \(N_{I}\) of pure imaginary holes does not depend on the integers \(I_{\textrm{max}}\) and \(I_{\textrm{min}}\), hence \(I_{\textrm{max}}-I_{\textrm{min}}=2\left(M_{R}+N_{H}-2N_{S}\right)\) as in (\ref{eq:145}). As a result, the counting equation remains the same and the numbers of pure imaginary roots and holes are not involved in it. In particular, one can choose their number independently from the total spin and the numbers of all other objects.

In this scheme boundary excitations are obtained by successive removing of such imaginary roots, i.e. by creating holes \(\vartheta_{k}\) with \(\mathfrak{Re}\left(\vartheta_{k}\right)=0\) and \(i\left(\mathfrak{Im} \left(\vartheta_{k}\right)\right)\in \mathbb{A}^{\pm}\). The creation of a hole at position, say, \(i\alpha_{0}^{(-)}=-i\frac{\pi}{2}H_{-}\) describes the excited boundary with energy coming from the Eq. (\ref{eq:176})

\begin{equation}
E=E_{B}^{\textrm{bbs}}+\mathcal{M}\cos \alpha_{0}^{(-)}\label{eq:197}
\end{equation}
This is the first boundary bound state at the boundary on the left. Notice that it is exactly an excited state over the new vacuum \(E_{B}^{\textrm{bbs}}\). In particular, the state described by a number of imaginary holes equivalent to the number of imaginary roots continues to be an excited state although its energy becomes quantitatively equivalent to \(E_{B}\), in fact, for \(\left|h_{\pm}\right|>h_{c}\) the system is completely different from the case \(\left|h_{\pm}\right|<h_{c}\) and the quantity \(E_{B}\) has not longer the same meaning.

One also expects that a boundary bound state implies some consequence on the scattering theory, in the sense that particles scattering off the excited boundary have to be affected by a different behaviour from that of free particles in not excited boundary system. Let us, thus, consider the simple state consisting of an excited boundary (an imaginary hole at position \(-i\frac{\pi}{2}H_{-}\) and a real hole at position \(\vartheta_{1}\)). Substituting in the NLIE, considering its antisymmetric properties and the quantization equation, it follows that the position \(\vartheta_{1}\) of the hole fulfils a quantization rule of a particle of energy \(\mathcal{M}\cosh\vartheta_{1}\) in a finite domain such that

\begin{equation}
e^{2il\sinh \vartheta_{1}}\mathcal{P}^{+}_{(-,0)}\left(\vartheta_{1},H_{-}\right)\mathcal{R}^{+}\left(\vartheta_{1},H_{+}\right)=1\label{eq:198}
\end{equation}
where, making use of the function (\ref{eq:187bis}), the notation

\begin{equation}
\mathcal{P}^{+}_{(\pm,0)}\left(\vartheta,H\right)\equiv \mathcal{S}\left(\vartheta\left|i\alpha_{0}^{(\pm)}\right.\right)\mathcal{R}^{+}\left(\vartheta,H\right)\label{eq:199}
\end{equation}
has been introduced with \(\alpha_{0}^{\pm}\in\mathbb{A}^{\pm}\). It corresponds to the first boundary bound state reflection amplitude given in \cite{dm}. From a physical point of view one can look at this result as a soliton in the boundary system scattering on the left excited boundary by the amplitude \(\mathcal{P}^{+}_{(-,0)}\left(\vartheta,H_{-}\right)\) and  on the right one by the unexcited amplitude \(\mathcal{R}^{+}\left(\vartheta,H_{+}\right)\). The energy of this state turns out to be

\begin{equation}
E=E_{B}^{\textrm{bbs}}+\mathcal{M}\cos\alpha_{0}^{(-)}+\mathcal{M}\cosh \vartheta_{1}\;.\label{eq:200}
\end{equation}
From now on we will refer to the amplitude \(\mathcal{P}^{+}_{(\pm,0)}\left(\vartheta,H\right)\) as the solitonic first excited boundary reflection factor.

Boundary bound states may occour at the right boundary simultaneously as well. In this case, evidently, a first pole in \(F\left(\vartheta,H_{+}\right)\) may be found at position \(i\alpha_{0}^{(+)}=-i\frac{\pi}{2}\left(H_{+}+2p\right)\). The two excited boundaries state is described by creating two pure imaginary holes at the exact position of the two imaginary roots \(i\alpha_{0}^{(-)}\), \(i\alpha_{0}^{(+)}\) and

\begin{equation}   
e^{2il\sinh\vartheta_{1}}\mathcal{P}^{+}_{(-,0)}\left(\vartheta_{1},H_{-}\right)\mathcal{P}^{+}_{(+,0)}\left(\vartheta_{1},H_{+}\right)=1\label{eq:201}
\end{equation}
with energy

\begin{equation}
E=E_{B}^{\textrm{bbs}}+\mathcal{M}\cos\alpha_{0}^{(-)}+\mathcal{M}\cos\alpha_{0}^{(+)}+\mathcal{M}\cosh\vartheta_{1}\;.\label{eq:202}
\end{equation}
In this case the reflection amplitude at the right boundary is excited as well. Obviously, when \(\left|h_{\pm}\right|>h_{c}\) any real hole can be regarded as in the previous section such that \(N_{H}\) of them describe \(N_{H}\) solitons reflected back of the boundaries and scattering each others by the sine-Gordon $S$ matrix. If the boundaries are unexcited they are reflected by the usual soliton boundary amplitude \(\mathcal{R}^{\pm}\left(\vartheta,H_{\pm}\right)\) through the Eq. (\ref{eq:181}) and a reference ground state energy \(E_{B}^{\textrm{bbs}}\), otherwise by \(\mathcal{P}^{+}_{(\pm,0)}\left(\vartheta,H_{\pm}\right)\) from the first excited boundary. Considering such a state, it is simple to find that, after an exponentiation, the NLIE becomes for any real hole

\begin{equation}
e^{2il\sinh\vartheta_{k}}\prod_{j\ne k}^{N_{H}}\mathcal{S}\left(\vartheta_{k}\left|\vartheta_{j}\right.\right)\mathcal{P}^{+}_{(-,0)}\left(\vartheta_{k},H_{-}\right)\mathcal{P}^{+}_{(+,0)}\left(\vartheta_{k},H_{+}\right)=1\;.\label{eq:203}
\end{equation}
According to the values of \(H_{\pm}\) and \(p\), it can happen to have more than one pure imaginary hole on the left and on the right boundaries. In this case, considering a number \(n\) of pure imaginary holes in the set \(\mathbb{A}^{-}\) and \(n^{\prime}\) in the set \(\mathbb{A}^{+}\), indeed these numbers are in general different because of the strong dependence of positions of pure imaginary objects on parameters \(H_{\pm}\), any soliton with energy \(\mathcal{M}\cosh\vartheta_{k}\) should scatter by the two excited boundaries via the equation 

\begin{equation}
e^{2il\sinh\vartheta_{k}}\prod_{j\ne k}^{N_{H}}\mathcal{S}\left(\vartheta_{k}\left|\vartheta_{j}\right.\right)\mathcal{P}^{+}_{(-,n)}\left(\vartheta_{k},H_{-}\right)\mathcal{P}^{+}_{(+,n^{\prime})}\left(\vartheta_{k},H_{+}\right)=1\label{eq:206}
\end{equation}
where the excited boundary reflection factors \(\mathcal{P}^{+}_{(-,n)}\left(\vartheta_{k},H_{-}\right)\) and \(\mathcal{P}^{+}_{(+,n^{\prime})}\left(\vartheta_{k},H_{+}\right)\) on the left and on the right have the form 

\begin{equation}
\mathcal{P}_{(\pm,q)}^{+}\left(\vartheta,H_{\pm}\right)=\prod_{i=0}^{q}\mathcal{S}\left(\vartheta\left|i\alpha^{(\pm)}_{i}\right.\right)\mathcal{R}^{+}\left(\vartheta,H_{\pm}\right)\label{eq:204}
\end{equation}
and the terms \(\mathcal{S}\left(\vartheta_{k}\left|\vartheta_{j}\right.\right)\) take into account the presence of other solitons (real holes). 

In order to study the antisolitonic excited boundary reflection factors one has to consider the symmetry which changes states of topological charge \(Q\) into states of charge \(-Q\). It has been stated that this corresponds to send \(H_{\pm}+1\rightarrow -H_{\pm}-1\) with the consequences that \(\mathcal{R}^{+}\left(\vartheta,H\right)\rightarrow \mathcal{R}^{-}\left(\vartheta,H\right)\) and one deals with antisolitonic states. Observe that the antisolitonic reflection factor \(\mathcal{R}^{-}\left(\vartheta,H\right)\) has simple poles at position \(i\alpha_{-n}^{(\pm)}\) as the solitonic reflection amplitude. In addition it presents simple poles at positions

\begin{equation}
\mathfrak{Re} (\vartheta)=0\;\;\;,\;\;\;\mathfrak{Im} (\vartheta)=\mp\frac{\pi}{2}\left(2+H+2pj\right)\;\;\;\textrm{with}\;\;\;j\in\mathbb{Z}\label{eq:207}
\end{equation}
which represent a second set characterizing new boundary bound states

\begin{align}
\mathbb{B}^{\pm} &=\left\{i\beta_{n}^{(\pm)}=i\frac{\pi}{2}\left[2+H_{\pm}\pm p-p(2m-1)\right]:\right.\nonumber \\
&\qquad\qquad\left.\frac{\pi}{2}>\left|\beta_{0}^{(\pm)}\right|>\left|\beta_{1}^{(\pm)}\right|\;.\;.\;.\;>0,\;m=1,2,\;.\;.\;.\;.\le m^{\pm}\right\}\label{eq:194}
\end{align}
where \(m^{\pm}\) are integers defined such that \(0<\left|2+H_{\pm}\pm p-p\left(2m^{\pm}-1\right)\right|<1\). According to \cite{dm} these poles represent the second set of boundary bound states (\ref{eq:116}) describing antisolitonic boundary bound states. Therefore, one can repeat the same derivation given above for solitonic states and find similar equations. The most general case with \(N_{H}\) antisolitons and boundary bound states at left and right boundaries characterized by \(m\) and \(m^{\prime}\) pure imaginary holes in sets \(\mathbb{B}^{\pm}\) can be expressed through the following expressions for the quantized scattering of a particle of energy \(\mathcal{M}\cosh\vartheta_{k}\) and for the total energy

\begin{equation}
e^{2il\sinh\vartheta_{k}}\prod_{j\ne k}^{N_{H}}\mathcal{S}\left(\vartheta_{k}\left|\vartheta_{j}\right.\right)\mathcal{P}^{-}_{(-,m)}\left(\vartheta_{k},H_{-}\right)\mathcal{P}^{-}_{(+,m^{\prime})}\left(\vartheta_{k},H_{+}\right)=1\label{eq:207bis}
\end{equation}

\begin{equation}
E=E_{B}^{\textrm{bbs}}+\mathcal{M}\sum_{i=1}^{m^{\prime}}\cos\beta_{i}^{(+)}+\mathcal{M}\sum_{i=1}^{m}\cos\beta_{i}^{(-)}+\mathcal{M}\sum_{j=1}^{N_{H}}\cosh\vartheta_{j}\label{eq:207tris}
\end{equation}
where

\begin{equation}
\mathcal{P}_{(\pm,r)}^{-}\left(\vartheta,H_{\pm}\right)=\prod_{i=0}^{r}\mathcal{S}\left(\vartheta\left|i\beta^{(\pm)}_{i}\right.\right)\mathcal{R}^{-}\left(\vartheta,H_{\pm}\right)\label{eq:207quater}
\end{equation}
with \(r=m\) for the left reflection amplitude and \(r=m^{\prime}\) for the right reflection amplitude. The reference ground state is redefined as

\begin{equation}
E_{B}^{\textrm{bbs}}=E_{B}-\mathcal{M}\sum_{i=1}^{m^{+}}\cos\beta_{i}^{(+)}-\mathcal{M}\sum_{i=1}^{m^{-}}\cos\beta_{i}^{(-)}
\end{equation}

\section{UV limit} 
Now we want to give a computation of the \(l\rightarrow 0\) limit in order to describe the system in the UV limit. Starting points are the NLIE and the expression for the energy (\ref{eq:174}).

It is known from boundary perturbation theory that the scaling behaviour of the energy for small volume size is

\begin{equation}
E(L)\sim\frac{\pi}{L}\left(\Delta-\frac{c}{24}\right)\label{eq:208}
\end{equation}
as sketched in Chapter 1 and Chapter 2. In this limit the roots and holes may rescale to infinity or stay in a finite region. Their behaviour is

\begin{equation}
\vartheta_{k}=\vartheta_{k}^{\pm}\pm\log\frac{1}{l}\;\;\;\;\textrm{or}\;\;\;\;\vartheta_{k}=\vartheta_{k}^{0}\label{eq:209}
\end{equation}
where thus we classify these objects into three types denoted by the indices "\(\pm,\;0\)"

\begin{equation}
\left\{\vartheta_{k}^{\pm}(l)\right\}=\left\{\vartheta_{k}^{\pm}\pm\log\frac{1}{l}\right\}\;\;\;\;,\;\;\;\;\left\{\vartheta_{k}^{0}(l)\right\}=\left\{\vartheta_{k}^{0}\right\}.\label{eq:210}
\end{equation}
Accordingly the NLIE splits into left and right kink equations. Define, thus, the kink NLIE as

\begin{equation}
Z_{\pm}(\vartheta)=\lim_{l\rightarrow 0}Z\left(\left.\vartheta\pm\log\frac{1}{l}\right|l\right)\;\;\;,\;\;\;Z_{0}(\vartheta)=\lim_{l\rightarrow 0}Z\left(\vartheta\left|l\right.\right)\label{eq:211}
\end{equation}
and introduce the scaling function

\begin{equation}
E_{i}(l)=E_{\textrm{boundary}}-\frac{\pi}{24L}c_{i}(l)\;\;\;\textrm{with}\;\;\;\lim_{l\rightarrow 0}c_{i}(l)=c-24\Delta_{i}\label{eq:212}
\end{equation}
where \(E_{\textrm{boundary}}\) refers to any contribution coming from boundaries and not depending on the space size \(L\), that is in the limit \(l\rightarrow 0\) the boundary bulk energy terms can be ignored because they are finite quantities.

Due to the antisymmetry of the NLIE it is enough to consider just the right kink equation and the objects with positive real parts. All the others have to be described by the same equations and one gets for their contributions the same results. Observe that the source terms corresponding to the various roots and holes may have different behaviour according to their set of reference (\ref{eq:210}). For source in \(\left\{\vartheta_{k}^{+}(l)\right\}\)

\begin{equation}
\lim_{l\rightarrow 0}\chi\left(\vartheta+\log\frac{1}{l}-\vartheta^{+}_{k}(l)\right)=\chi\left(\vartheta-\vartheta_{k}^{+}\right)\label{eq:213}
\end{equation}
and for sources in \(\left\{\vartheta_{k}^{0}(l)\right\}\)

\begin{equation}
\lim_{l\rightarrow 0}\chi\left(\vartheta+\log\frac{1}{l}-\vartheta^{0}_{k}(l)\right)=\chi\left(\vartheta+\log\frac{1}{l}-\vartheta_{k}^{0}\right)=\chi_{\infty}\label{eq:214}
\end{equation}
where \(\chi_{\infty}\) has been defined in (\ref{eq:161}). Therefore, given e.g. \(N_{H}^{+}\) holes of type "+", there will be in the NLIE \(N_{H}-N_{H}^{+}\) contributions of term \(\chi_{\infty}\). One can introduce for the numbers of objects in the sets (\ref{eq:210}) the notation \(N_{H}^{+,0},M_{C}^{+,0},\;.\;.\;.\) and expect that counting equations similar to that given in (\ref{eq:146}) hold for the spin \(S^{+}\) and \(S^{0}\) according to the condition that \(N_{H}=2N_{H}^{+}+2N_{H}^{0},\; M_{C}=2M_{C}^{+}+2M_{C}^{0},\;.\;.\;.\) and \(S=2S^{+}+2S^{0}\). Here the factor 2 take into account that the number of objects with negative real part is equal to that of objects with positive real part and that each of them carries the same spin. Hence, one considers only sources with positive real part and take care about the factor 2 to have the total contribution of any source. Notice that there might be \(N_{I}\) contributions \(\chi_{\infty}\) as well, due to pure imaginary holes. It has been stated that they do not count in the counting equation and cannot be related to the spin, but they have to be taken into account. After few manipulations one has, for \(l\rightarrow 0\),

\begin{equation}
Z_{+}(\vartheta)=e^{+\vartheta}+g_{+}\left(\vartheta\left|\left\{\vartheta^{+}_{k}\right\}\right.\right)+P\left(\vartheta\left|H_{+},H_{-}\right.\right)+\int\textrm{d}xG(\vartheta-x)Q_{+}(x)\label{eq:215}
\end{equation}
where

\begin{align}
g_{+}\left(\vartheta\left|\left\{\vartheta^{+}_{k}\right\}\right.\right) & =\lim_{l\rightarrow 0}g\left(\left.\vartheta+\log\frac{1}{l}\right|\left\{\vartheta_{k}(l)\right\}\right)=\nonumber \\
&=2\chi_{\infty}\left(S+\frac{N_{I}}{2}-S^{+}\right)+\sum_{k}^{M^{+}}c_{k}\chi_{(k)}\left(\vartheta-\vartheta^{+}_{k}\right)+\pi L_{W}^{+}\label{eq:216}
\end{align}
where \(L_{W}^{+}=\textrm{Sign}(\pi-2\gamma)\left(M_{W}-M_{W}^{+}\right)\) and \(M^{+}=N_{H}^{+}+M_{C}^{+}+M_{W}^{+}+N_{S}^{+}\). Similarly for sources that remain in a finite region

\begin{equation}
Z_{0}(\vartheta)=g_{0}\left(\vartheta\left|\left\{\vartheta^{0}_{k}\right\}\right.\right)+P\left(\vartheta\left|H_{+},H_{-}\right.\right)+\int\textrm{d}xG(\vartheta-x)Q_{0}(x)\label{eq:217}
\end{equation}
where

\begin{align}
g_{0}\left(\vartheta\left|\left\{\vartheta^{0}_{k}\right\}\right.\right) & =\lim_{l\rightarrow 0}g\left(\left.\vartheta\right|\left\{\vartheta_{k}(l)\right\}\right)=\nonumber \\
&=2\chi_{\infty}\left(S-S^{0}\right)+\sum_{k}^{M^{0}}c_{k}\chi_{(k)}\left(\vartheta-\vartheta^{0}_{k}\right)\label{eq:218}
\end{align}
The functions \(Q_{+,0}(x)\) are related to \(Z_{+,0}(x)\) as \(Z(x)\) to \(Q(x)\), \(M^{0}=N_{H}^{0}+M_{C}^{0}+M_{W}^{0}+N_{S}^{0}+N_{I}\). In order to calculate the UV of the energy (\ref{eq:174}), some asymptotics for \(\vartheta\rightarrow +\infty\) are required:

\begin{equation}
Z_{+}(+\infty)=+\infty\;\;\;\;,\;\;\;\;Q_{+}(+\infty)=0\label{eq:218bis}
\end{equation}

\begin{equation}
g_{+}\left(-\infty\left|\left\{\vartheta^{+}_{k}\right\}\right.\right)=2\chi_{\infty}\left(S+\frac{N_{I}}{2}-2S^{+}\right)+2\pi K_{W}^{+}\label{eq:218tris} 
\end{equation}

\begin{equation}
Z_{+}(-\infty)=g_{+}\left(-\infty\left|\left\{\vartheta^{+}_{k}\right\}\right.\right)+P_{\infty}+\frac{\chi_{\infty}}{\pi}Q_{+}(-\infty)\label{eq:218quater} 
\end{equation}
with \(K_{W}^{+}\in\mathbb{Z}\) and \(P_{\infty}\) defined in formula (\ref{eq:162}). By definition (see Eq. (\ref{eq:160})), for any \(K\in\mathbb{Z}\), it holds

\begin{equation}
Z_{+}(\pm\infty)=Q_{+}(\pm \infty)+\pi+2\pi K\label{eq:219}
\end{equation}
therefore, making use of the asymptotics (\ref{eq:218bis}) and (\ref{eq:218tris}) and comparing Eqs. (\ref{eq:218quater}) and (\ref{eq:219}), 

\begin{equation}
Q_{+}(-\infty)=\frac{\pi\left[2\chi_{\infty}\left(S+\frac{N_{I}}{2}-2S^{+}\right)+P_{\infty}-\pi-2\pi K+2\pi K_{W}^{+}\right]}{\pi-\chi_{\infty}}\label{eq:220}
\end{equation}
and finally, observing that for \(l\rightarrow 0\) the terms in the energy behave as

\[
\frac{\mathcal{M}}{2}\int\frac{\textrm{d}x}{2\pi}\sinh\;x\;Q(x)\sim \frac{1}{L}\int\frac{\textrm{d}x}{2\pi}e^{+x}Q_{+}(x) \]

\[
\mathcal{M}\cosh \vartheta_{k}\sim \frac{1}{L}e^{+\vartheta_{k}^{+}}, \]
one can rearrange the formula (\ref{eq:208}) as

\begin{equation}
\Delta=\frac{c}{24}+\frac{1}{2\pi}\left[\sum_{k}^{M^{+}}c_{k}e^{+\vartheta_{k}^{+}}-\int\frac{\textrm{d}x}{2\pi}e^{+x}Q_{+}(x)\right]+\frac{\mathcal{M}L}{24\pi}\sum_{k}^{N_{I}}\cos \vartheta_{k}\label{eq:221}
\end{equation}
where the last term takes into account the whole set of pure imaginary holes \(i\vartheta_{k}\) in the sets \(\mathbb{A}^{\pm}\) and \(\mathbb{B}^{\pm}\) describing in the IR limit the boundary bound states. For \(l\rightarrow 0\) this term goes to zero. At this point, making use of the notations introduced above, defining the sums over all quantum numbers for any type of sources as \(I_{A}^{+}=\sum_{j}I_{A,j}^{+}\) with \(A=H,C,W,\;.\;.\;.\) and following the scheme given in \cite{ddv3}, one gets the following expression

\begin{align}
\Delta= & \frac{p}{p+1}\left[\frac{1}{2}\left(\frac{H_{+}+H_{-}}{2p}-1+\right.\right.\nonumber \\
&\left.\left.-2\left(S+\frac{N_{I}}{2}-S^{+}+k\right)\right)+\frac{p+1}{2p}\left(S+\frac{N_{I}}{2}\right)\right]^{2}+N\label{eq:222}
\end{align}
where

\[
N=I_{H}^{+}-I_{C}^{+}-I_{W}^{+}-2I_{S}^{+}+L_{W}^{+}S^{+}+2k-S^{+}-2\left(S^{+}\right)^{2}\;\in\mathbb{Z}_{+} \]
and \(k=K-K_{W}^{+}\). This is an exact computation of the scaling limit of the scaling function (\ref{eq:212}). 

\subsection{UV limit as a BCFT}
\begin{figure}
\begin{center}
\includegraphics[angle=-90, width=0.7\textwidth]{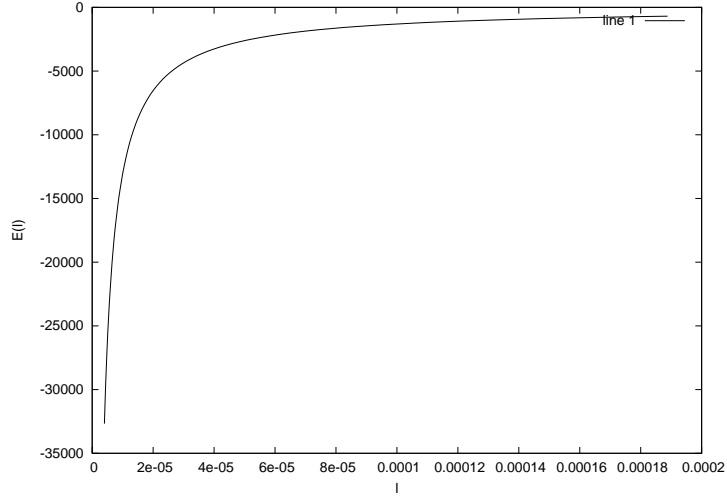}
\caption{\emph{The behaviour of the vacuum energy as a function of the scale parameter \(l\) for the case: \(p=\frac{2}{5}\), \(H_{-}=0.3642\), \(H_{+}=0.4238\). According to} Eq. (\ref{eq:212}), \emph{\(E_{B}\) can be discarded. Therefore \(E(l)\rightarrow 0\) for \(l\rightarrow +\infty\)}.\label{fig:12}}
\end{center}
\end{figure}

\begin{figure}
\begin{center}
\includegraphics[angle=-90, width=0.7\textwidth]{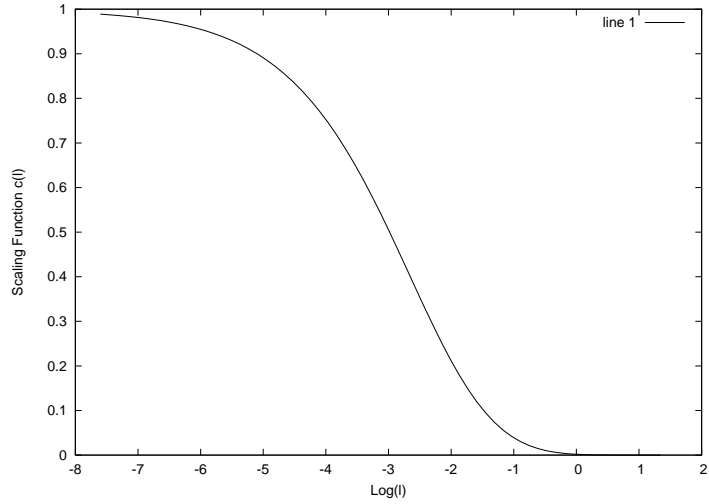}
\caption{\emph{The scaling function \(c(l)\) for the vacuum state with: \(p=\frac{2}{5}\), \(H_{-}=0.3642\) and \(H_{+}=0.4238\).}\label{fig:13}}
\end{center}
\end{figure}

\begin{figure}
\begin{center}
\includegraphics[angle=-90, width=0.7\textwidth]{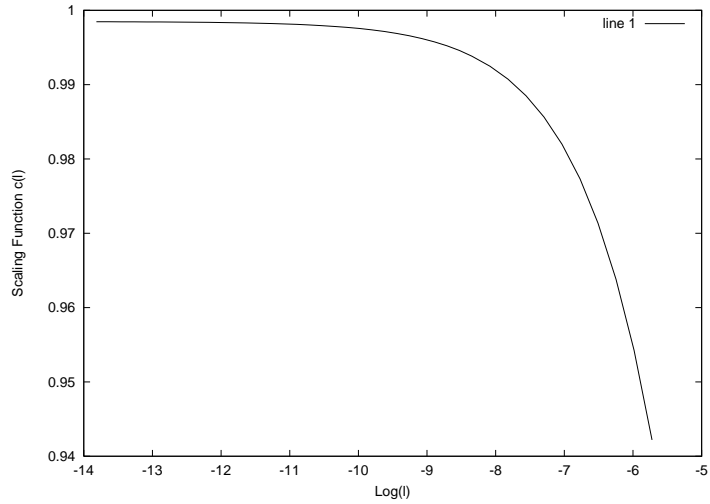}
\caption{\emph{Particular of Fig.}(\ref{fig:13}). \emph{According to analytic formulae, one expects an UV central charge \(c=0.999618\;.\;.\;.\), here it is reproduced with a precision \(10^{-6}\)}.\label{fig:13bis}}
\end{center}
\end{figure}

The UV limit allows one to make contact with a \(c=1\) boundary CFT. The scaling function as defined in (\ref{eq:212}) takes the form

\begin{equation}
c_{i}=1-24\Delta_{i}\label{eq:223}
\end{equation}
and, in general, for a theory with non trivial boundaries \(\Delta_{i}\ne 0\) for the conformal vacuum too. In our case the main aim is to relate NLIE states of the IR limit to conformal states. In the sections devoted to the large volume limit, we have established that, for \(\left|h_{\pm}\right|<h_{c}\), the vacuum state is characterized by the NLIE state without any source and with spin \(S=0\). In the language of UV limit this corresponds to \(S=S^{+}=S^{0}=0\), therefore

\begin{equation}
c_{i}^{\textrm{vac}}=1-\frac{6p}{p+1}\left(1-\frac{H_{+}+H_{-}}{2p}\right)^{2}\;.\label{eq:224}
\end{equation}
It is the effective central charge in the UV fixed point, in particular \(c_{i}\le 1\). Eq. (\ref{eq:224}) describes, thus, the scaling of the vacuum of the theory and reproduces exactly the formula given some years ago in \cite{ss}. These few considerations confirm that the NLIE allows to control the complete flow of the vacuum scaling functions from the IR to the UV limits. To be more precise, if the vacuum of the scattering theory is fixed to be \(E_{B}\) as in (\ref{eq:173}), which is a quantity strongly dependent on the boundary parameters \(H_{\pm}\), the correspondent conformal state has an energy \(E_{i}=-\frac{\pi}{24L}c_{i}^{\textrm{vac}}\) according to Eq. (\ref{eq:212}). It is evident that the boundaries contribute some energy to the Casimir effect due to the finite size. The complete flow of the vacuum energy \(E_{0}(l)\) from zero to infinity can be seen in Fig. (\ref{fig:12}). In Figs. (\ref{fig:13}) and (\ref{fig:13bis}) the behaviour of the corresponding scaling function \(c_{0}(l)\) is displaied. In order to obtain such pictorial representations, the NLIE has been integrated numerically at several values of \(l\).
 
In a more general situation, one would be able to treat the excited states as well. For semplicity let us introduce the notations

\[
R=\sqrt{\frac{p+1}{2p}}=\frac{\sqrt{4\pi}}{\beta}\;\;\;\;,\;\;\;\;m=2S+N_{I}\in\mathbb{Z},
\]
where \(S\) denotes the total spin carried by sources at positive real position, i.e. the total spin of the system is exactly twice. The conformal dimensions of any conformal state can be parametrized in terms of relevant quantities of NLIE states as

\begin{equation}
\Delta=\frac{1}{2}\left[\frac{\phi_{+}-\phi_{-}}{\sqrt{\pi}}-\frac{S^{0}+k}{R}+\frac{1}{2}mR\right]^{2}+N\label{eq:225}
\end{equation}
where \(\phi_{\pm}\) are the boundary values of DSG related to the parameters \(H_{\pm}\) as in Eq. (\ref{eq:182}). Eq. (\ref{eq:225}) (see \cite{io}) reproduces exactly the conformal dimensions in the partition function (\ref{eq:73}) for all states with \(S^{0}+k=0\). We will see that any physical state satisfies this condition. Therefore, without loss of generality, we can rewrite for physical states 

\begin{equation}
\Delta=\frac{1}{2}\left[\frac{\phi_{+}-\phi_{-}}{\sqrt{\pi}}+\frac{1}{2}mR\right]^{2}+N\label{eq:226}
\end{equation}
This result can be exactly compared with the \(c=1\) CFT of a boson with Dirichlet conformal boundary conditions which is compactified on a circle of radius \(R\) \cite{s,affl}. 

\subsection{Classification of BCFT states}
The Hilbert space \(\mathcal{H}\) of the \(c=1\) CFT (free boson) with Dirichlet boundary conditions is composed of Heisenberg algebra representations \(\mathcal{H}_{m}\) of \(c=1\) CFT whose primary states \(\left|m\right\rangle\) are created from the vacuum by radial ordered vertex operators \(e^{i\kappa\phi}\) of \(U(1)\) charge and corresponding conformal dimensions 

\begin{equation}
\kappa=\frac{1}{2}\left(\frac{\phi_{+}-\phi_{-}}{\sqrt{\pi}}+\frac{1}{2}mR\right)\;\;\;\;,\;\;\;\;\Delta_{m}=2\kappa^{2}.\label{eq:227}
\end{equation}
The whole set of states of a representation \(\mathcal{H}_{m}\) is created by applying repeatedly the creation operators (\ref{eq:59}) in such a way that

\begin{equation}
\mathcal{H}_{m}=\left\{\alpha_{-k_{1}}\;.\;.\;.\;\alpha_{-k_{q}}\left|m\right\rangle\;\left|\;k_{1},\;.\;.\;.\;,k_{q}\in\mathbb{Z}_{+}\right.\right\}\label{eq:228}
\end{equation}
Notice that only creation operators in the holomorphic sector need applying. The Hilbert space is, thus, a linear combination \(\mathcal{H}=\bigoplus_{m}\mathcal{N}_{\Delta_{m}}\mathcal{H}_{m}\) of sets (\ref{eq:228}). For any state \(\left|i\right\rangle\in\mathcal{H}_{m}\) the energy is given by

\begin{equation}
E_{i}=\frac{\pi}{L}\left(\Delta_{m}+N_{i}-\frac{1}{24}\right)\;,\;\;\;\;N_{i}=\sum_{j=1}^{q}k_{j}\in\mathbb{Z}_{+}\label{eq:229}
\end{equation}
It is evident that the value in the expression (\ref{eq:226}) is a good candidate to classify several states in relation to NLIE states, one has just to identify \(\Delta=\Delta_{m}+N_{i}\). For instance the conformal vacuum takes the form

\begin{equation}
\Delta^{\textrm{vac}}=\frac{1}{2\pi}\left[\phi_{+}-\phi_{-}\right]^{2}=\Delta_{0}\;.\label{eq:230}
\end{equation} 

\subsubsection{Solitonic primary states}
\begin{figure}
\begin{center}
\includegraphics[angle=0, width=0.8\textwidth]{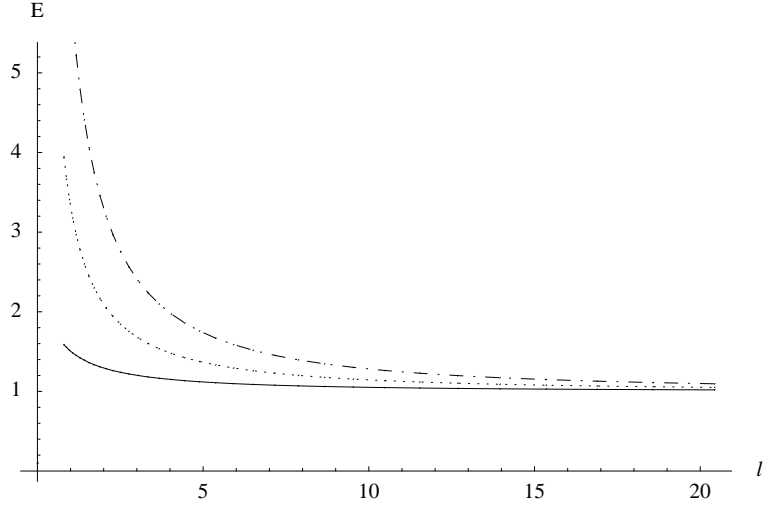}
\caption{\emph{One soliton state energy as a function of \(l\). \(p=\frac{2}{5}\), \(H_{-}=0.3642\), \(H_{+}=0.4238\). For \(l\rightarrow +\infty\) the energy goes to a positive finite value corresponding to the soliton rest energy. For simplicity it has been set \(\mathcal{M}=1\). The filled line represents the case with \(I_{1}=1\), the dotted line \(I_{1}=2\) and the dashed line \(I_{1}=3\)}.\label{fig:14}}
\end{center}
\end{figure}

The first interesting conformal primary states which one can consider are those correponding to solitonic IR states. It has been shown that sine-Gordon solitons can be represented in the scattering theory limit as pure NLIE holes states. Take e.g. the state of one hole at generic position \(\vartheta_{1}\) in the positive real axis and take the minimal quantum number \(I_{1}=1\). In the IR theory it describes a soliton of energy \(\mathcal{M}\cosh\vartheta_{1}\) scattering off the left and right boundaries. Suppose that when \(l\rightarrow 0\) the hole becomes an object in the set \(\left\{\vartheta_{j}^{+}(l)\right\}\), i.e. its position rescales to infinity. The spin is \(S=\frac{1}{2}\) and, in particular, the spin has to be \(S=S^{+}=\frac{1}{2}\) and the quantum number \(I_{1}^{+}=I_{1}=1\). Hence, the associated conformal state in the UV limit is

\begin{equation}
\Delta=\frac{1}{2}\left[\frac{\phi_{+}-\phi_{-}}{\sqrt{\pi}}+\frac{1}{2}R\right]^{2}=\Delta_{1}\label{eq:231}
\end{equation}
One is thus dealing with the conformal primary state of the Heisenberg representation \(\mathcal{H}_{1}\). Observe that in this situation one is forced to choose \(I_{1}=1\). Any other choice should not be compatible with a primary state. In fact, any quantum number greater than 1 should imply \(N>0\), i.e. a descendant state with energy grater than \(E_{1}\). Therefore, the quantum number is crucial to have exactly a primary state. The behaviour of the one soliton energy is represented in Fig. (\ref{fig:14}) for cases with \(I_{1}=1\), \(I_{1}=2\) and \(I_{1}=3\).

It is possible to take more complicated states. For instance the state with a generic number \(N_{H}=N_{H}^{+}\), that is a state with \(N_{H}\) holes in the positive real axis lying in the set \(\left\{\vartheta^{+}_{j}\right\}\) for \(l\rightarrow 0\). According to our notations

\[
S=S^{+}=\frac{N_{H}}{2}\;\;\;\;,\;\;\;\;m=N_{H}\;\;\;\;\textrm{and}\;\;\;\;N=I_{H}^{+}-\frac{1}{2}N_{H}\left(N_{H}+1\right) \]
therefore

\begin{equation}
\Delta=\frac{1}{2}\left[\frac{\phi_{+}-\phi_{-}}{\sqrt{\pi}}+N_{H}\frac{R}{2}\right]^{2}+I_{H}^{+}-\frac{1}{2}N_{H}\left(N_{H}+1\right)\label{eq:232}
\end{equation}
and, observing that for the monotoniticity of the \(Z(\vartheta)\) one has to choose linearly increasing integers as quantization numbers, hence \(I_{H}^{+}=\sum_{j=1}^{N_{H}}I_{j}^{+}=\frac{1}{2}N_{H}\left(N_{H}+1\right)\), it follows: \(\Delta=\Delta_{m}\). This means that the generic \(m\) holes NLIE state characterized by minimal quantum numbers, i.e. the \(m\) solitons sine-Gordon state, corresponds to the primary state of the \(\left[\phi_{m}\right]\) conformal family in the \(c=1\) BCFT.

\subsubsection{About the topological charge}
In formula (\ref{eq:109}) the topological charge of any DSG state has been introduced. We set

\begin{equation}
Q=\frac{\beta}{2\pi}\left(\phi_{+}-\phi_{-}\right)+m_{+}-m_{-}\;\;\;,\;\;\;m_{\pm}\in\mathbb{Z}\label{eq:233}
\end{equation}
i.e. the topological charge consists of a combination of a term depending on boundary  and an integer. It is evident that as far as the formula (\ref{eq:226}) is concerned, one could expect to give an expression of conformal dimensions in terms of the topological charge of the corresponding DSG state. Identifying the topological charge of the vacuum with \(Q_{0}=\frac{\beta}{2\pi}\left(\phi_{+}-\phi_{-}\right)\) and the topological charge of a soliton as \(Q_{1}=m=1\), the formula (\ref{eq:231}) for the solitonic conformal dimension becomes

\begin{equation}
\Delta=\frac{2\pi}{\beta^{2}}\left[Q_{0}+\frac{1}{2}\right]^{2}\label{eq:234}
\end{equation}
and, in general, for any excited solitonic state over the vacuum with topological charge \(Q_{s}=m=2S\)

\begin{equation}
\Delta=\frac{2\pi}{\beta^{2}}\left[Q_{0}+\frac{m}{2}\right]^{2}+N\;.\label{eq:235}
\end{equation}
This means that, once the topological charges fixed, one has directly the corresponding conformal state in the UV limit. In particular, for any anti-solitonic state the topological charge comes from applying the symmetry \(\phi\rightarrow -\phi\) and \(\phi_{\pm}\rightarrow -\phi_{\pm}\) simultaneously. Therefore, one can generalize the Eq. (\ref{eq:235}) for all antisolitonic states, which have been defined in the scattering theory as pure holes states under the symmetry \(\mathcal{C}:\;h_{\pm}\rightarrow -h_{\pm}\;,\;S\rightarrow -S\), as

\begin{equation}
\Delta=\frac{2\pi}{\beta^{2}}\left[-Q_{0}+\frac{Q_{-s}}{2}\right]^{2}+N\label{eq:236}
\end{equation}
where \(Q_{-s}=-Q_{s}=-m\). Of course, due to the structure of conformal dimensions expression, solitonic and antisolitonic states have the same dimensions. This is not surprising, in fact, the symmetries of the model makes irrelevant the signs of the boundary fields and of the total spin: any state is equivalent to the \(\mathcal{C}\)-transformed state. This property has to be encoded also in the conformal limit.

\subsubsection{Boundary bound states as conformal states} 
Among all other possible states the most interesting ones are those corresponding to the boundary bound states described in the scattering theory. In order to have such boundary states one has to choose \(\left|h_{\pm}\right|>h_{c}\), i.e. the DSG boundary fields have to be in the ranges \(\frac{\beta}{8}<\phi_{-}<\frac{\beta(p+1)}{8p}\), \(\frac{\beta(p+1)}{8p}<\phi_{+}<\frac{\beta(2p+1)}{8p}\), according to the relations (\ref{eq:182}). In these ranges of definition, the conformal dimension of the vacuum state turns out to be \(\Delta=\frac{1}{2\pi}\left[\phi_{+}-\phi_{-}\right]^{2}=\Delta_{0}\), where \(\Delta_{m}\) refers to the conformal dimension of the primary state in the represenatation \(\mathcal{H}_{m}\). This conformal dimension has the same structure of dimension (\ref{eq:230}), but it is quantitatively different due to different values of \(\phi_{\pm}\). At this point one may distinguish several situations: 

\begin{itemize}
\item the case without boundary bound states can be described as before for all allowed states, the only difference is the value of boundary fields and, hence, the value of vacuum conformal dimension; 

\item the case with boundary bound states which implicitely contains a variation of boundary conditions. The description of such states is the following. 
\end{itemize}
The first excited state over the vacuum is described, in the large \(l\) NLIE, by an imaginary hole at position \(i\alpha_{0}^{(-)}=-i\frac{\pi}{2}H_{-}\) (see Eqs. (\ref{eq:190}) and (\ref{eq:197})) with quantum number \(I_{\alpha_{0}^{(-)}}=0\) and energy \(E=E_{B}^{\textrm{bbs}}+\mathcal{M}\cos\alpha_{0}^{(-)}\) (notice that in the UV limit the constant term \(E_{B}^{\textrm{bbs}}\) can be ignored because of Eq. (\ref{eq:212})). It corresponds to a conformal state with \(N_{I}=1\) and conformal dimension

\begin{equation}
\Delta=\frac{1}{2}\left[\frac{\phi_{+}-\phi_{-}}{\sqrt{\pi}}+\frac{1}{2}R\right]^{2}\label{eq:237}
\end{equation}
A soliton in the IR limit is described by a real hole \(\vartheta_{1}\) state (\(N_{H}=1\)) with energy \(\mathcal{M}\cosh\vartheta_{1}\) over the first excited state, i.e. over the boundary bound state energy. This means that a real hole is no more the first excited state, in fact it has to describe a soliton scattering on an excited boundary, i.e. a free particle in a system with changed boundary states. Taking a boundary bound state means fixing new physical boundary states, any particle has to be an excitation over this reference state. The conformal state corresponding to a soliton in an excited boundary system has \(S^{+}=S=\frac{1}{2}\) and \(I_{H}^{+}=1\), then

\begin{equation}
\Delta=\frac{1}{2}\left[\frac{\phi_{+}-\phi_{-}}{\sqrt{\pi}}+R\right]^{2}=\Delta_{2}\label{eq:238}
\end{equation}
In general, the \(N_{H}\) solitons state in a boundary bound state system corresponds to the conformal state with conformal dimensions

\begin{equation}
\Delta=\frac{1}{2}\left[\frac{\phi_{+}-\phi_{-}}{\sqrt{\pi}}+\frac{1}{2}\left(N_{H}+N_{I}\right)R\right]^{2}=\Delta_{N_{H}+N_{I}}\label{eq:239}
\end{equation}
As a consequence of the boundary bound states the conformal dimensions of the one soliton state appear to be in the Heisenberg representation \(\mathcal{H}_{1+N_{I}}\) instead of \(\mathcal{H}_{1}\). This is not surprising because a NLIE boundary bound state has to correspond to a physically relevant change of boundary state conditions and then to a change of the partition function and of the conformal theory. In particular, an expression for the conformal partition function taking into account different boundary state conditions might be

\begin{equation}
\mathcal{Z}_{DD}\propto\frac{1}{\eta(q)}\sum q^{\frac{1}{2}\left[\frac{\phi_{+}-\phi_{-}}{\sqrt{\pi}}+\frac{1}{2}\left(2S+N_{I}\right)R\right]^{2}}\label{eq:240} 
\end{equation}
where \(\phi_{\pm}\) together with \(N_{I}\) takes into account the boundary contributions, in the sense that not only the value of boundary fields but also the specific boundary states, i.e. the number of boundary bound states in the NLIE, define the various Heisenberg representations which the conformal states are organized into and the partition function of the theory. It is evident that there are as many partition functions, therefore distinct conformal theories, as boundary states. Due to the NLIE, one can classify these conformal theories with the number \(N_{I}\) of boundary bound states. In formula (\ref{eq:240}) the number of boundary bound states \(N_{I}\) has to be fixed \emph{a priori}, it depends, actually, on the fact that the boundary state has to be fixed before the study and classification of the conformal states in the representations \(\mathcal{H}_{m}\). Let us observe, finally, that the integer \(n\) in the sum (\ref{eq:73}) has to be associated to the combination \(n\equiv 2S+N_{I}\) of the third component of the spin and the number of boundary bound states.    

\section{Conclusions}
In this thesis it has been shown how the Nonlinear Integral Equation deduced from the XXZ spin chain with boundary magnetic fields (\ref{eq:117}) describes the finite size effects of the sine-Gordon theory with Dirichlet boundary conditions (\ref{eq:107}). A summary of most important results is the following:

\begin{enumerate}
\item we have presented a derivation of the NLIE (\ref{eq:169}) governing the finite size effects for sine-Gordon theory with two boundaries, each with an independent Dirichlet boundary condition, as a continuum limit of the Bethe ansatz equations (\ref{eq:128}) of the alternating inhomogeneous XXZ spin chain with magnetic fields at the boundaries for the case of antiparallel fields. Excited states have been considered as well;

\item by examining the infrared limit of the NLIE (\ref{eq:175}) it has been shown that it gives strong evidence that the underlying quantum field theory is actually the Dirichlet sine-Gordon model. The relation among NLIE boundary parameters \(H_{\pm}\) and boundary fields \(\phi_{\pm}\) of Dirichlet sine-Gordon has been fixed;

\item the correct two particle sine-Gordon \(S\) matrices are reproduced (\ref{eq:184}) leading to a coherent description of NLIE states describing scattering solitonic and antisolitonic particles. In particular, the correct solitonic and antisolitonic Ghoshal-Zamolodchikov boundary reflection amplitudes (\ref{eq:180}) are obtained;

\item by analyzing several infrared states, a relation among sine-Gordon boundary bound states and NLIE states has been proposed. The correct excited boundary reflection amplitudes (\ref{eq:199}) and (\ref{eq:204}) are reproduced as well and the scattering of particles with them is under control;

\item by computing the ultraviolet conformal weights (\ref{eq:222}) from the NLIE it has been shown that the energy spectrum is consistent with the spectrum of a \(c=1\) boundary conformal field theory (\ref{eq:229});

\item conformal dimensions of solitonic and antisolitonic states and boundary bound states can be reproduced and it has been shown that they belong into the correct representations (\ref{eq:228}), even if the states which we have considered are not enough to describe the whole sine-Gordon spectrum, for example we have not taken into account breather states;

\item we argued a general formula to relate NLIE states to boundary conformal states described by the partition function (\ref{eq:240}).
\end{enumerate}
The understanding of the Dirichlet sine-Gordon NLIE and of the finite size behaviour of the continuum theory is not quite complete. Indeed, some open questions have not been answered yet. Interesting questions need further investigation:

\begin{enumerate}
\item the whole set of NLIE solutions describing the finite volume behaviour of scaling functions is not completely understood. The main difficulty is the problem of counting, i.e. the complete knowledge of the structure of solutions that are highly dependent on the value of the coupling constant;

\item there is an unresolved technical difficulty too. Apart the first simple states with only real and pure imaginary sources describing solitonic and antisolitonic states and boundary bound states, in order to study configurations of sources with non-zero imaginary parts, the understanding of the NLIE boundary term over the fundamental branches of definition of functions (\ref{eq:152}) and (\ref{eq:157}) is required. This is crucial for a complete comprehension of other NLIE excited states as those which should describe breathers scattering off unexcited or excited boundaries;

\item although the vacuum state NLIE obtained here reproduces some known facts (see \cite{lmss,ss,s}), the excited states interpretation and their scaling behaviour should need to be numerically compared to results coming from other nonperturbative methods, namely the Truncated Conformal Space approach. The agreement of NLIE and TCS approach predictions should confirm the validity of the NLIE approach for the boundary sine-Gordon model. 
  
\end{enumerate}
Finally, to conclude this thesis, let us mention that, although many progresses on the knowledge of general integrable boundary conditions for the sine-Gordon theory have been done \cite{bk1,bk2,bk3}, from the point of view of the NLIE approach the understanding of the Dirichlet boundary conditions case should be a first step towards the investigation of the general boundary conditions case: first attempts have been recentely presented (see, for example, \cite{ahn,nep1,nep2,nep3,caolin}), but a complete description has not been done yet. Nevertheless, we think that further investigation could throw some light on the general structure of IQFTs and finite size effects which, at this point, few people might consider as a pure academic or even theoretical game, even if theories such as sine-Gordon model are usually called toy models.

\Bibliography

\end{document}